\documentclass[twocolumn,twocolappendix]{aastex701}
\usepackage{amsmath}

\newcommand\teff{$T_{\rm eff}$}
\newcommand\logg{$\log g$}

\newcommand\vsini{$v\sin i$}

\defcitealias{kounkel2018a}{Paper I}
\usepackage{listings}

\begin{document}
\title{ABYSS. IV. Identifying signatures of stellar youth in APOGEE spectra}

\author[0009-0009-2841-1091]{Valentina Bonilla Villalobos}
\affil{Department of Physics and Astronomy, University of North Florida, 1 UNF Dr, Jacksonville, FL, 32224, USA}
\email{n01569033@unf.edu}
\author[0000-0002-5365-1267]{Marina Kounkel}
\affil{Department of Physics and Astronomy, University of North Florida, 1 UNF Dr, Jacksonville, FL, 32224, USA}
\email[show]{marina.kounkel@unf.edu}
\author[0000-0002-1650-2764]{Joseph Mullen}
\affil{Department of Physics and Astronomy, University of North Florida, 1 UNF Dr, Jacksonville, FL, 32224, USA}
\email{j.p.mullen@unf.edu}
\author[0000-0003-3769-8812]{Eleonora Zari}
\affiliation{Dipartimento di Fisica e Astronomia, Università degli Studi di Firenze, Via G. Sansone 1, I-50019, Sesto F.no (Firenze), Italy}
\email{eleonoramaria.zari@unifi.it}
\author[0000-0002-1379-4204]{Alexandre Roman-Lopes}
\affil{Department of Astronomy, Universidad de La Serena, Av. Raul Bitran \#1302, La Serena, 170000, Chile}
\email{Alexandre Roman Lopes <aroman@userena.cl>}
\author[0000-0002-3481-9052]{Keivan Stassun}
\affil{Department of Physics and Astronomy, Vanderbilt University, VU Station 1807, Nashville, TN 37235, USA}
\email{keivan.stassun@vanderbilt.Edu}
\author[0000-0001-8600-4798]{Carlos Román-Zúñiga}
\affiliation{Universidad Nacional Aut\'onoma de M\'exico, Instituto de Astronom\'ia, AP 106,  Ensenada 22800, BC, M\'exico}
\email{croman@astro.unam.mx}
\author[0000-0002-7795-0018]{Ricardo L\'opez-Valdivia}
\affiliation{Universidad Nacional Aut\'onoma de M\'exico, Instituto de Astronom\'ia, AP 106,  Ensenada 22800, BC, M\'exico}
\email{rlopezv@astro.unam.mx}
\author[0000-0001-6072-9344]{Jinyoung Serena Kim}
\affiliation{Steward Observatory, University of Arizona, 933 N. Cherry Ave., Tucson, AZ 85721-0065, USA}
\email{serena00@arizona.edu}
\author[0000-0001-8341-3940]{Mojgan Aghakhanloo}
\affil{Department of Astronomy, University of Virginia, 530 McCormick Rd, Charlottesville, VA 22904, USA}
\affiliation{Virginia Institute of Theoretical Astronomy, University of Virginia, Charlottesville, VA 22904, USA}
\email{mvy4at@virginia.edu}
\author[0000-0002-4128-7867]{Facundo Perez Paolino}
\affil{Department of Astronomy, California Institute of Technology, Pasadena, CA 91125, USA}
\email{fperezpa@caltech.edu}

\author[0000-0002-3389-9142]{Jonathan C. Tan}
\affil{Department of Space, Earth \& Environment, Chalmers University of Technology, SE-412 96 Gothenburg, Sweden}
\email{jct6e@virginia.edu}

\author[0000-0001-9797-5661]{Jes\'us Hern\'andez}
\affiliation{Universidad Nacional Aut\'onoma de M\'exico, Instituto de Astronom\'ia, AP 106,  Ensenada 22800, BC, M\'exico}
\email{hernandj@astro.unam.mx}

\begin{abstract}

We develop a convolutional neural network classifier that performs spectroscopic identification of stellar youth ($<$40 Myr) in APOGEE spectra. This classifier is sensitive to several youth-related features, including rotational broadening, which is common in younger stars, and to several discrete lines that appear to be indicative of a very spotted photosphere. The model is successful at identifying youth across a wide range of stars, achieving its strongest performance at discriminating young M and K dwarfs, while maintaining useful discriminatory power for hotter stars as well. This work enables more robust separation of pre-main-sequence stars from more evolved sources in the field even when they have comparable \teff\ and \logg, providing a reliable means to substantially reduce contamination in photometrically selected YSO candidates. In addition to constructing a classifier for APOGEE spectra, we also perform classification of young stars in optical BOSS spectra.
\end{abstract}


\section{Introduction}

In many respects, young pre-main-sequence stars resemble their more evolved counterparts, making them non-trivial to identify in the field among more evolved stars \citep[and references therein]{wright2023}. For example, despite their larger radii and higher luminosities, young stars have lower $\log g$ and photometrically overlap with binary-sequence stars and reddened red giants on the HR diagram \citep[e.g.,][]{baraffe2015}. Spectroscopically, their identification at the low-mass end is somewhat easier, as \logg\ has fewer contaminants than luminosity; nonetheless, young stars overlap with evolved sub-giants in parameter space. Despite these difficulties, young stars possess peculiar features that gradually weaken over time. Reliable identification of young stars is therefore essential for studies of star formation, early stellar evolution, disk evolution, and the spatial and kinematic structure of young Galactic populations \citep{wright2023,haisch2001,gagne2024,kounkel2023}.

The most notable feature of youth in stellar spectra is the Li I 6707 \AA\ line in low-mass stars \citep[e.g.,][]{briceno1997, jeffries2017, zerjal2019, armstrong2025}, which is depleted over a few to a few tens of Myr. In low-mass stars, a strong Li I absorption line is therefore commonly interpreted as evidence of incomplete lithium depletion and, hence, of stellar youth. Another indicator of youth is H$\alpha$ emission. During the T-Tauri phase in pre-main-sequence stars, this line is dominated by the accretion shock from the protoplanetary disk, but over time, as the disk depletes, the line is primarily driven by chromospheric activity, which is orders of magnitude stronger in young stars than in more evolved systems \citep[e.g.,][]{white2003,saad2024}.
While these optical features are not particularly prominent individually, when considered together, they can provide a more unambiguous confirmation of youth and a spectroscopic basis for identifying young low-mass stars \citep{briceno1997,jeffries2017,white2003,saad2024}.
These commonly used youth features are predominantly observed in the optical regime. Nevertheless, recent advances in large spectroscopic surveys now make it possible to obtain spectra of $\gtrsim10^5$ candidate young stellar objects (YSOs), and because young stars are predominantly found along the Galactic plane and extinction towards them can be high, many of these spectroscopic observations are conducted in the near-IR. This creates an important observational asymmetry: the most widely used spectroscopic diagnostics of youth lie in the optical, whereas many of the largest and most efficient spectroscopic surveys of embedded and reddened young populations operate in the near-IR \citep{lada2003,kounkel2023}. In particular, the Apache Point Observatory Galactic Evolution Experiment (APOGEE) is a near-IR spectrograph operated by the Sloan Digital Sky Survey (SDSS) that has conducted a number of observations of star forming regions in the past \citep[e.g.,][]{cottaar2014,da-rio2016} and it is currently performing the largest spectroscopic census of the young Galaxy to date \citep{kounkel2023}. As such, it is critical to devise criteria to confirm the youth of the observed candidates. At the same time, the spectroscopic signatures of youth in the near-IR remain less well established than their optical counterparts, and no comparably mature set of diagnostics exists for APOGEE spectra \citep{wright2023,kounkel2023}.

In this context, machine-learning methods are particularly well suited to the problem, because they can combine multiple weak and distributed spectral signatures that may not be diagnostic individually but become informative when considered jointly across the full spectrum \citep{kang2023, sizemore2024}. In this work, we develop a classifier that operates on optical and near-IR spectra obtained by SDSS using a convolutional neural network (CNN) and determine which spectroscopic features the classifier relies on when making its predictions. This work aims both to provide a practical tool for identifying likely young stars in APOGEE data and to investigate the physical origin of the spectral features used by the classifier, placing the APOGEE results in the broader context of youth classification in SDSS spectroscopy through comparison with BOSS data. In Section \ref{sec:data}, we present an overview of the data used in this analysis. In Section \ref{sec:methods}, we describe the development of the classifier and the construction of average spectra for young and field stars, which are used to assess the classifier's ability to separate the two populations. In Section \ref{sec:results}, we present our results. In Section \ref{sec:discussion}, we discuss the features of youth in near-IR spectra. Finally, in Section \ref{sec:conclusions}, we present our conclusions.

\section{Data}\label{sec:data}

The fifth iteration of the Sloan Digital Sky Survey \citep[SDSS-V,][]{kollmeier2026} is operated using two facilities: Apache Point Observatory \citep[APO,][]{gunn2006} in the northern hemisphere, and Las Campanas Observatory \citep[LCO,][]{bowen1973} in the southern hemisphere. Both observatories are equipped with two spectrographs: APOGEE, which obtains spectra in the H band (1.5--1.7 $\mu$m) with a resolution of $R\sim22,500$ \citep{wilson2019}, and the Baryon Oscillation Spectroscopic Survey (BOSS), which covers the entire optical range and has a significantly lower resolution of $R\sim1800$ \citep{smee2013}.

One of the programs supported by SDSS-V is the APOGEE \& BOSS Young Star Survey \citep[ABYSS,][]{kounkel2023}, which aims to observe $\sim$200,000 candidate young stars across the Milky Way with both spectrographs. These candidates were selected using a variety of techniques, including infrared excess, membership in nearby moving groups, and optical variability. To date, more than 100,000 of these candidates have been observed as of the Internal Product Launch (IPL) 4, which includes all data observed up to the Modified Julian Date (MJD) $<60708$ (February 2025). Approximately half of these data were available in IPL-3 (MJD $<60130$, July 2023, APO only). When we began this project, only IPL-3 data were available; as such, we primarily use these data for training, but we use the full IPL-4 dataset for evaluation.

Several pipelines have been developed for these data. In particular, APOGEE Net and BOSS Net \citep{sizemore2024} provide effective temperature \teff\ (with a precision of $<0.005$ dex), surface gravity \logg\ (with a precision of $<0.1$ dex), and metallicity [Fe/H] (with a precision of $<0.1$ dex). These two pipelines can process all stars observed by SDSS and are particularly calibrated for young stars. In particular, these \logg\ values have sufficient accuracy to serve as independent proxies for the ages of pre-main-sequence stars, consistent with ages derived from photometry.

In addition to extracting fundamental spectroscopic parameters, the pipeline measures equivalent widths of various youth-sensitive lines in optical BOSS spectra, including Li I and H$\alpha$ \citep{saad2024}. Using these equivalent widths, along with \teff, \logg, and optical and near-IR photometry, \citet{saad2024} developed a rudimentary CNN to identify young stars in BOSS spectra. This classifier was trained with TensorFlow \citep{tensorflow} on sources targeted by ABYSS as representative YSO candidates, with some cuts to exclude the most obvious false positives. Through these efforts, it became not only possible to more rigorously select a subset of spectroscopically confirmed bona fide YSOs but also to find YSOs serendipitously observed by SDSS outside of the ABYSS program (for example, as part of the Solar Neighborhood Census, which is a volume-limited program aimed at obtaining spectra of stars within 250 pc).

This classifier was most effective for low-mass pre-main-sequence stars, achieving $\sim$90\% recall. This is to be expected, given that most of the classically recognized features of youth dominate the low-mass regime. However, it has also had some success in recovering solar-type stars with ages $<$20 Myr, which is more surprising given that such sources reach the main sequence quickly, and thus they have few known reliable features of youth. 
Because this classifier relied on only a handful of lines extracted in the optical regime, it is unsuited for APOGEE spectra. Given the absence of well-known discrete features of youth in the H-band, we develop a similar classifier by training the model on the entire spectral range of APOGEE spectra; this process is described in the next section.\par
Furthermore, to improve the accuracy of the BOSS classifier, given both the significantly larger sample that has been produced by SDSS since \citet{saad2024}, and to similarly avoid reliance only on a handful of lines (which at low spectral resolution can be difficult to measure with a high degree of accuracy), we also retrain the classifier for BOSS data; these efforts are described in Appendix \ref{sec:appendix}.

\section{Methodology} \label{sec:methods}
\subsection{Development of the classifier}

\begin{figure*}
\epsscale{1.1}
\plottwo{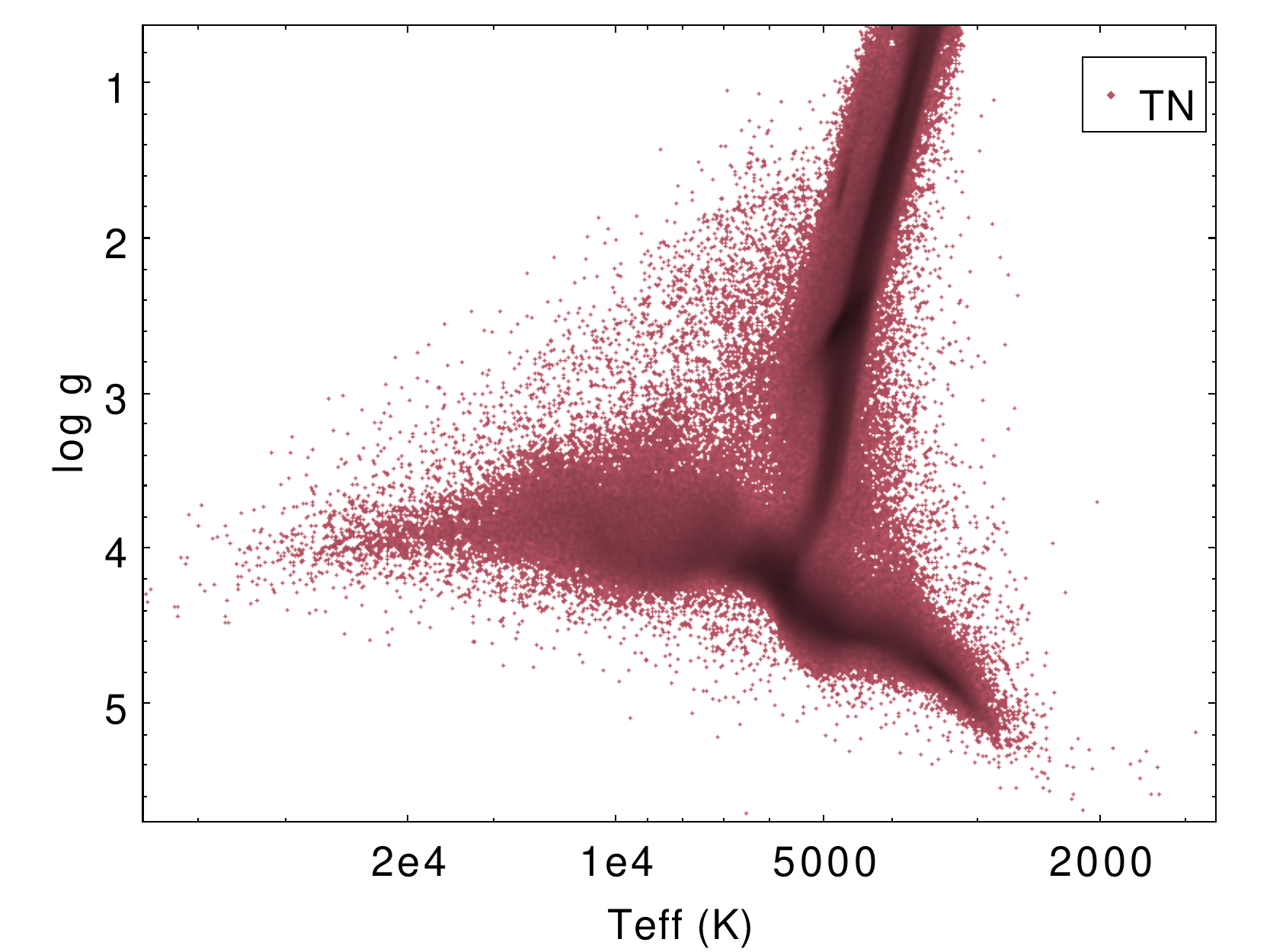}{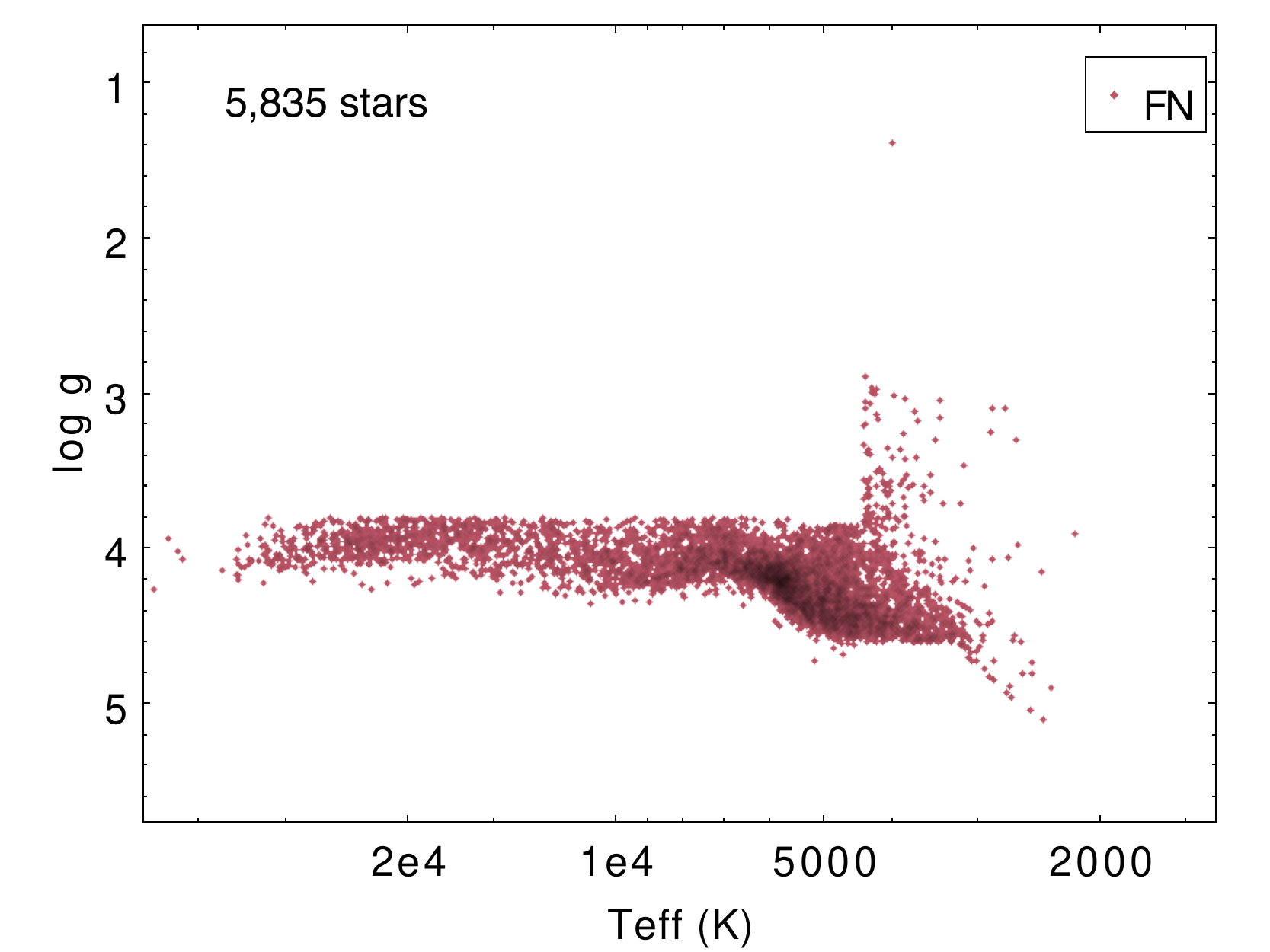}
\plottwo{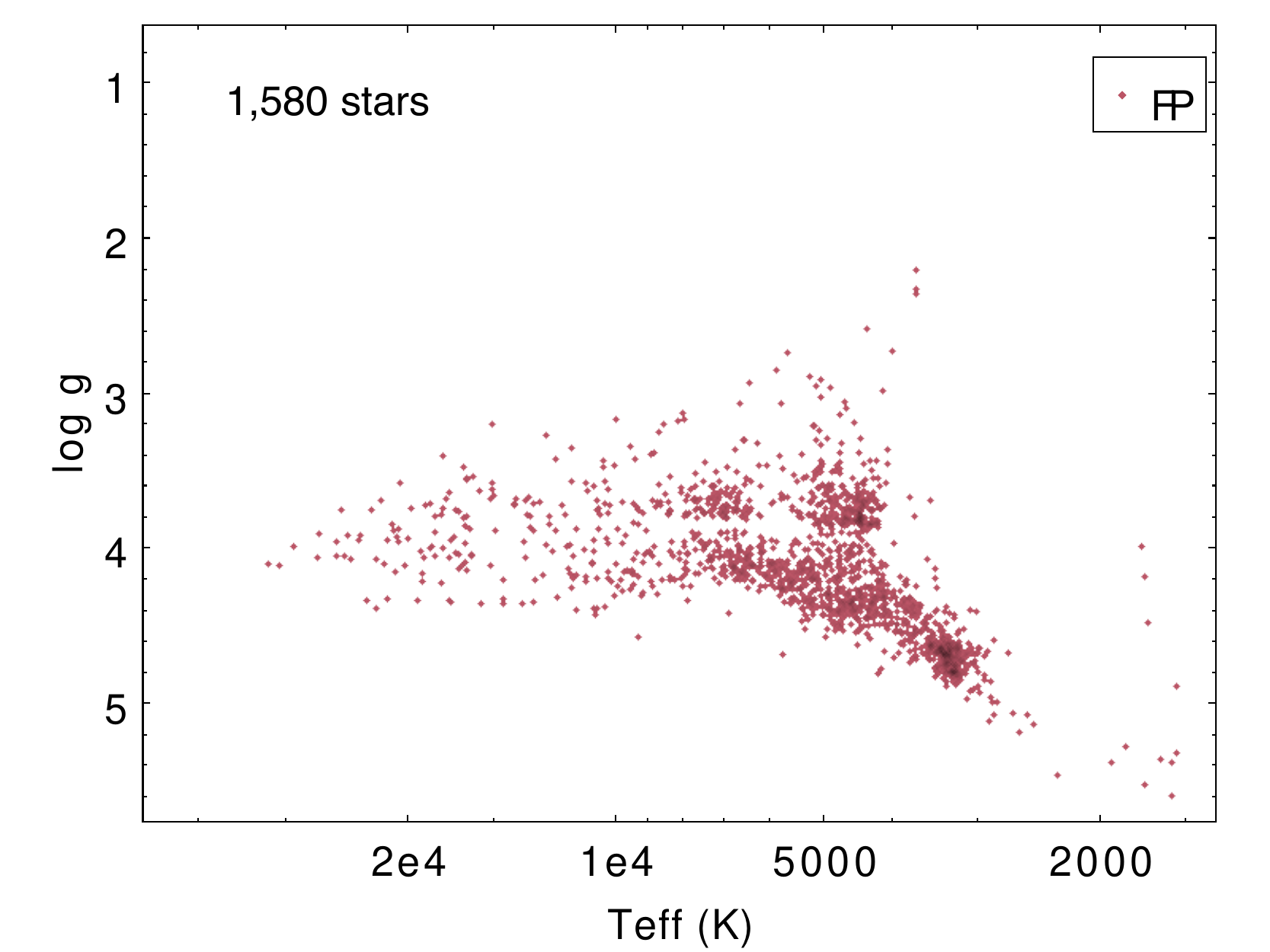}{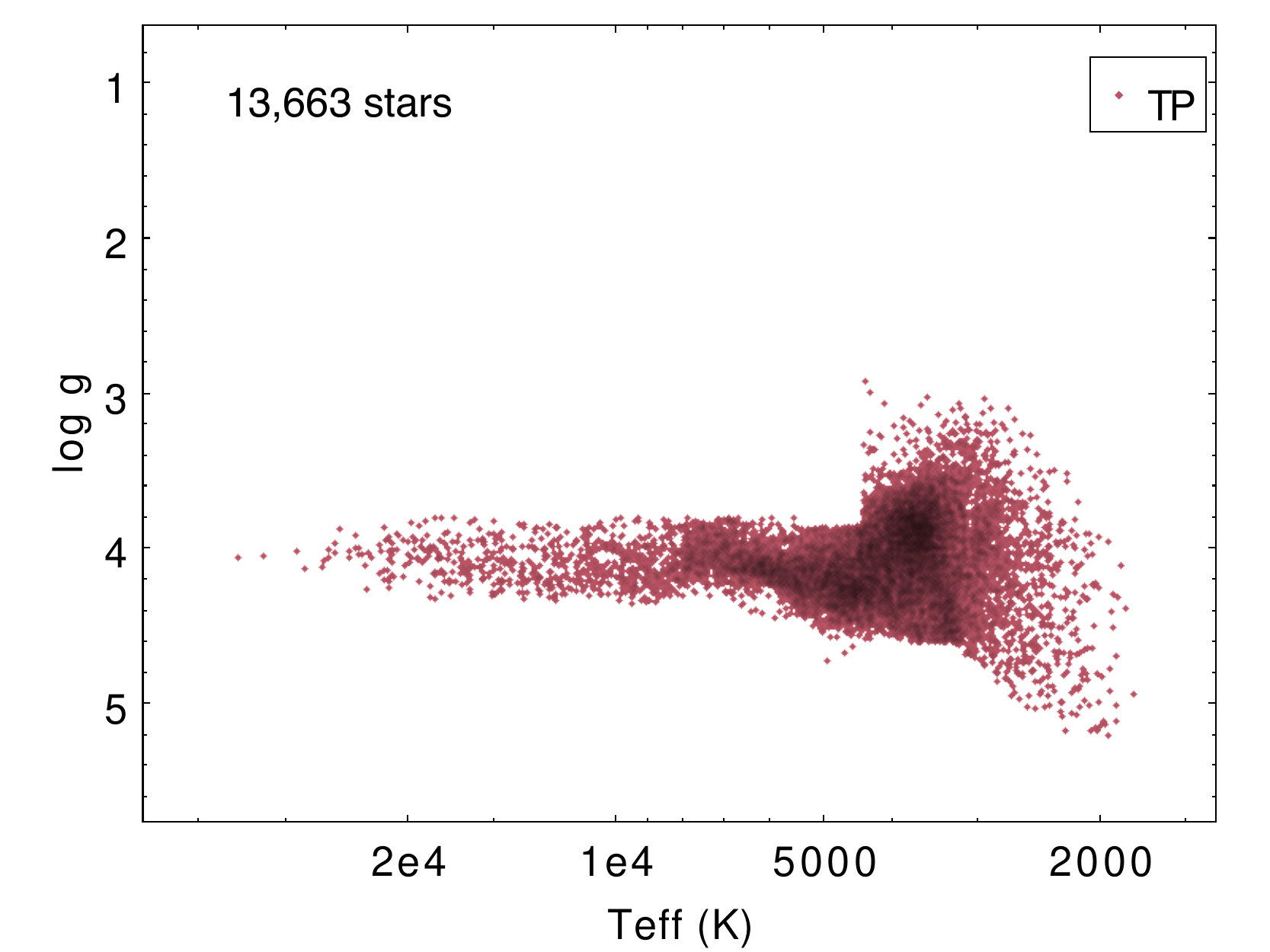}
\caption{Kiel plots of the stars in the various subsets of the labeled sample. Top left: true negatives, i.e., stars labeled as field stars and identified as such by the classifier. Top right: false negatives, i.e., sources labeled as young, but predicted by the classifier to be field stars. Bottom left: false positives, i.e., sources labeled as field stars, but identified by the classifier to be young. Bottom right: true positives, i.e., sources labeled as young and identified as such by the classifier. Note that the sharp horizontal cut in \logg\ is due to attempting to limit contamination from red giants in the training set; thus, some of the bona fide YSOs that were subjected to this cut were recovered as ``false positives''.
\label{fig:kiel}}
\end{figure*}

To perform spectroscopic classification of YSOs in APOGEE spectra, we used a convolutional neural network that takes the entire spectrum as input. Before being fed into the model, each spectrum was normalized by dividing out the median flux, and bad or missing points were handled by setting their values to 1. The architecture of the model consists of an initial 1D convolution, ELU activation function, and Max Pooling for the initial data processing. The results are fed into a series of four Residual Network blocks, each one consisting of two 1D convolutional layers; these blocks are connected with batch normalization and the ELU activation function. Afterward, the data are passed to an adaptive average pooling layer and through six fully connected layers connected by ReLU activation. This architecture was adapted from \citet{sizemore2024}, (see Figure 5 from that paper for the full description), as it has proven to have sufficient complexity for extracting fundamental stellar parameters from SDSS spectra, while still remaining manageable. We did not explore any other model architectures. We implemented this model using TensorFlow \citep{tensorflow}.

Following \citet{saad2024}, we constructed the classifier training set by assigning to the sources targeted by ABYSS the label ``likely young stars'' and to the remaining sources observed by APOGEE the label ``field stars''. Neither of these classes is entirely free of cross-contamination. For example, observations of star-forming regions SDSS-III and -IV with APOGEE \citep[e.g., NGC 1333, IC 348, Orion, and others][]{foster2015,cottaar2015,da-rio2016,roman-zuniga2023} had longer exposure times than what is commonly practiced by SDSS-V; as such, these archival observations include some of the fainter bona fide YSOs that are not a part of ABYSS. Additionally, other SDSS-V programs, for example, the Solar Neighborhood Census, might have serendipitously targeted some YSOs that were not a part of ABYSS. On the other hand, the ABYSS catalog comprises candidate young stars that effectively exhibit signatures of youth, but it suffers from non-negligible field-star contamination because disk-bearing T Tauri stars may have colors similar to those of reddened red giants, while older YSOs may overlap with main-sequence stars.

This cross-contamination can degrade model performance, as inconsistent labeling across similar-looking sources blurs the ability to separate the classes. To address this, we re-labeled a subset of sources. First, most of the obvious evolved stars in the ABYSS sample were excluded on the basis of their \teff\ and \logg, removing everything with \logg$<3.8$ and \teff$>4500$ K (i.e., primarily red giants or supergiants). We also used these spectroscopic parameters to estimate stellar ages using MIST isochrones \citep{choi2016}, excluding any stars that appear older than 40 Myr. Although some bona fide YSOs may remain in these excluded regions, given that YSOs constitute only a small fraction of the overall sample, we prioritized minimizing contamination.

Using this sample, we trained a preliminary model and examined the sources predicted to be young, paying particular attention to those not part of ABYSS (i.e., false positives). In that sample, we searched for signatures of significant clustering (both in the plane of the sky, as well as in parallax and in proper motions), particularly in regions associated with nearby star-forming regions (Orion, Taurus, Perseus, and Ophiuchus). This clustering was visually examined and manually selected in TOPCAT \citep{topcat} if a significant overdensity was visually apparent. The clustered sources were reincorporated into the training set as likely YSOs. This procedure was repeated several times; a new model was retrained each time until no residual clustering was observed outside the excluded parameter space described above.

Such an approach to training does create some uncertainty in classifications due to the absence of a well-defined ``ground truth'' in labeling. Unfortunately, well-characterized YSOs are very biased toward the youngest and lowest-mass sources; as such, they cannot be used to unambiguously evaluate the performance of the classifier with regard to the moving groups that are starting to dissolve. We acknowledge that some uncertainty remains, particularly among stars that reach the main sequence rapidly and therefore exhibit fewer obvious signatures of youth. We expect such sources to represent only a small fraction of the sample. However, we primarily evaluate the performance of the classifier in the qualitative sense, to ensure that stars that are likely to be identified as YSOs through closer examination (both in the labeled and unlabeled data, see Section \ref{sec:unlabeled}) appear to be recovered by the classifier, while excluding most obvious contaminants. However, this also does make the trends seen in the precision and recall metrics more qualitative as well, rather than the ``true'' classification rates.

The training was conducted using APOGEE spectra processed with Astra version 0.6. In the final model, 19,498 sources were labeled as likely YSOs and 889,094 as field stars. Overall, young stars constitute only a small fraction of the full sample. No attempt was made to correct for the class imbalance, as in our prior experience, attempting to even out the numbers of rare candidates results in the model more biased to finding false positives, while preserving the intrinsic imbalance helps to maintain clean sample, being more conservative in its assessment, and forcing to rely on the most unique features to separate out the young stars.

A random subset of 12,800 spectra was withheld prior to training and maintained as a dev set. This sample was used exclusively to monitor model performance and was not used for training. The composition of the dev set remained fixed throughout the analysis. Similarly, sources that became available only in Astra version 0.8 were not used during model development and were instead used solely to evaluate the final classifier. Training was optimized for accuracy using a sparse categorical cross-entropy loss function with a learning rate of $10^{-5}$. We used early stopping, terminating training after the accuracy on the withheld dev data stopped improving for 2 epochs (i.e., two full passes of the data in training). No systematic hyperparameter optimization or cross-validation was performed. The adopted learning rate and architecture were selected based on preliminary experiments and their ability to achieve stable convergence. Outside of early stopping and evaluation on the independent Astra 0.8 sample, no additional regularization methods were employed.

The final model is released on Zenodo \citep{ysoclassifier}.

\subsection{Empirical templates}\label{sec:templates}

To evaluate the model's performance and identify which features it used to make its predictions, we grouped all APOGEE sources into bins with similar \teff, \logg, and [Fe/H]. This binning was performed separately for field stars and for sources identified as likely young. All of the young stars were assumed to have [Fe/H]$\sim$0.

To normalize all spectra, we first excluded the edges of the chips and all of the pixels that were flagged as poor or those that contain strong telluric lines, and applied aggressive smoothing to the resulting data with a Savitzky-Golay filter \citep{savitzky1964} (window size 1000 pixels, polynomial order 1) to obtain a continuum representation. We then fit a line to this continuum. The APOGEE spectra cover the H band, which is relatively narrow, and for most stars this includes the Rayleigh-Jeans tail primarily; as such, the continuum fit does not require a higher-order polynomial.

The resulting fit was divided out of the spectral flux and its uncertainties to preserve the relative ratios between them. We do note that ``true'' continuum normalization is not always trivial to assess, particularly in low-mass stars with strong molecular bands, or for sources with very low signal-to-noise. However, we considered this sufficient for our analysis as this was performed self-consistently for all stars of a given type.

All of the spectra in a given bin were stacked together, and a weighted average flux (weighted by the flux uncertainty) was computed at each wavelength. This average spectrum can be considered representative of all sources in the bin, but it has a higher signal-to-noise ratio than any individual observation.

\section{Results} \label{sec:results}

\begin{deluxetable}{ccl}[!ht]
\tablecaption{Catalog of sources identified as YSOs
\label{tab:catalog}}
\tabletypesize{\scriptsize}
\tablewidth{\linewidth}
\tablehead{
 \colhead{Column} &
 \colhead{Units} &
 \colhead{Description}
 }
\startdata
sdss\_id & & SDSS unique identifier \\
RA & deg & Right ascension in J2000 \\
Dec & deg & Declination in J2000 \\
log \teff & [K] & Effective temperature \\
$\sigma$ log \teff & [K] & uncertainty in log \teff \\
log g &  & Surface gravity \\
$\sigma$ log g & & Uncertainty in log g \\
YSO & & Probability from the classifier \\
Instrument & & APOGEE or BOSS
\enddata
\end{deluxetable}

\begin{figure}
\epsscale{1.2}
\plotone{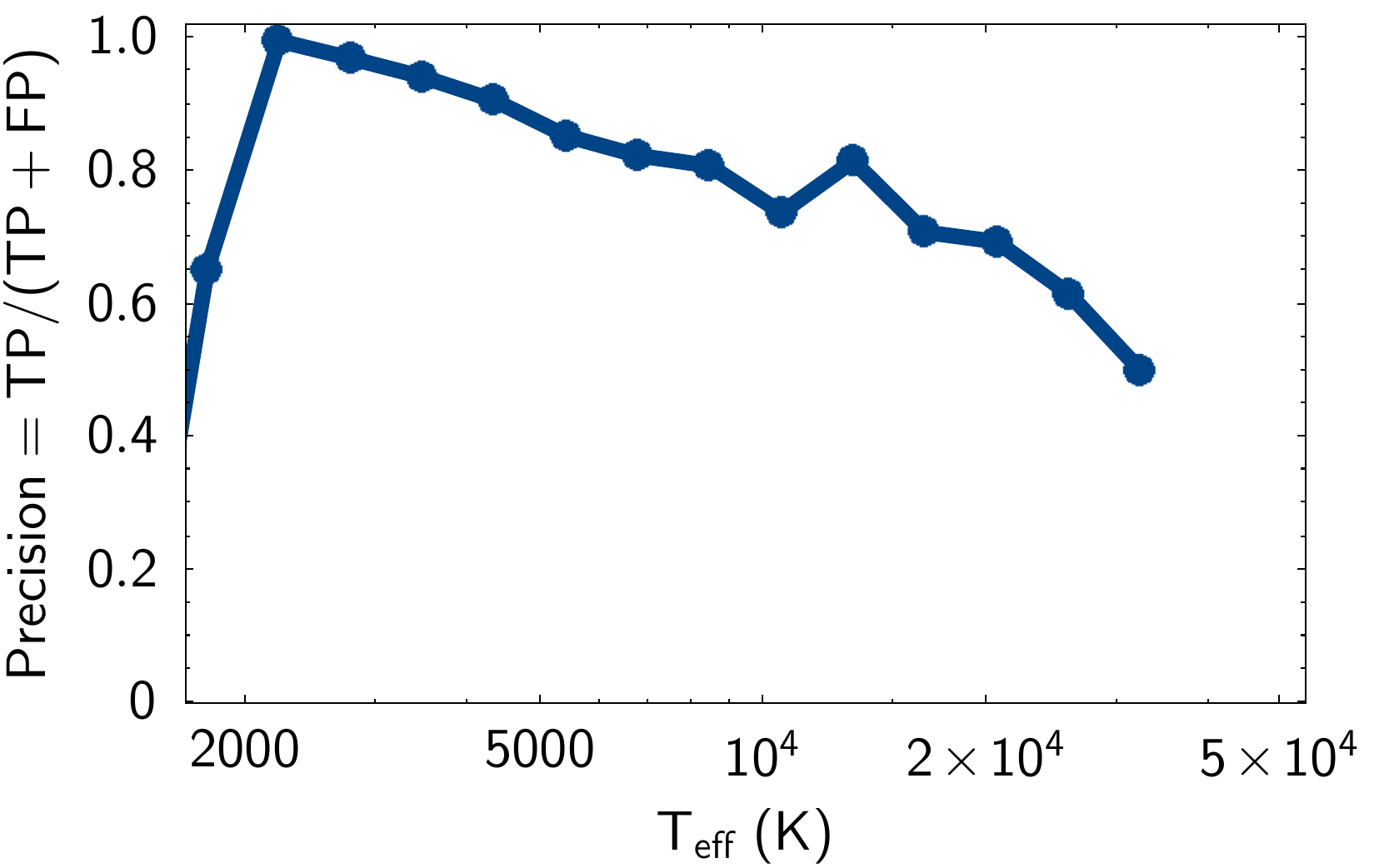}
\plotone{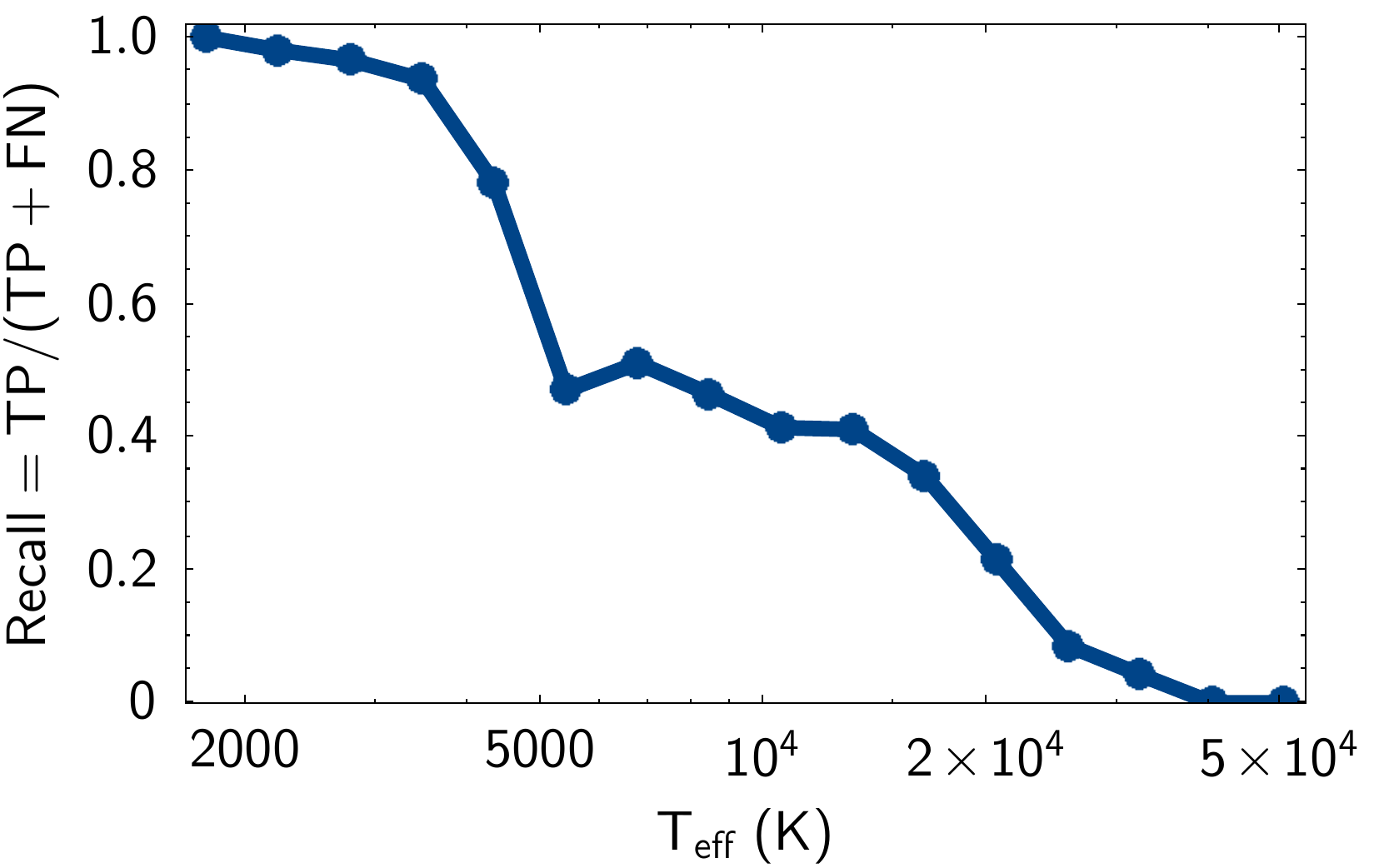}
\caption{Typical precision and recall of the classifier as a function of \teff, computed based on the adopted labels after manual and iterative relabeling process.
\label{fig:pr}}
\end{figure}
\begin{figure}
\epsscale{1.2}
\plotone{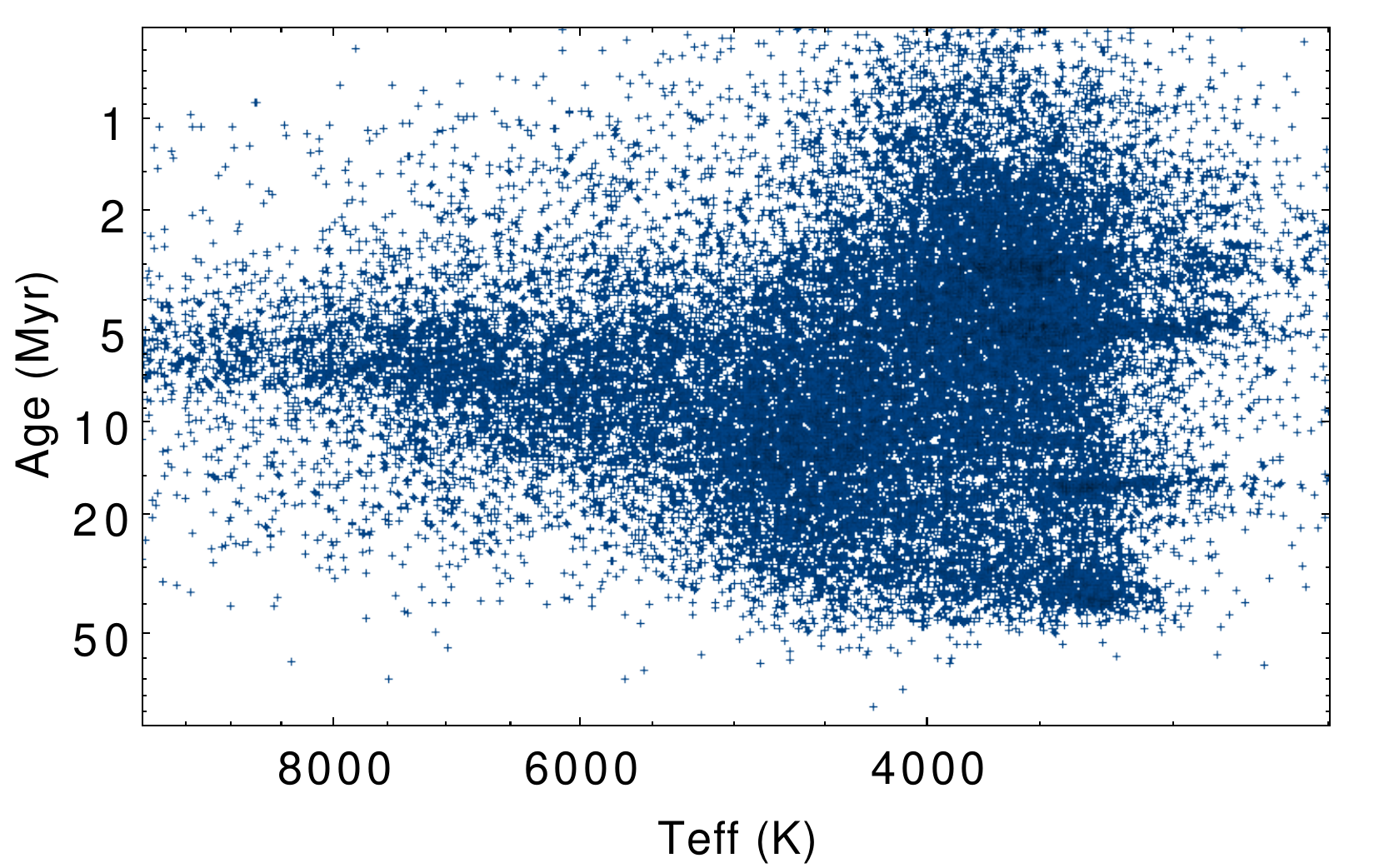}
\plotone{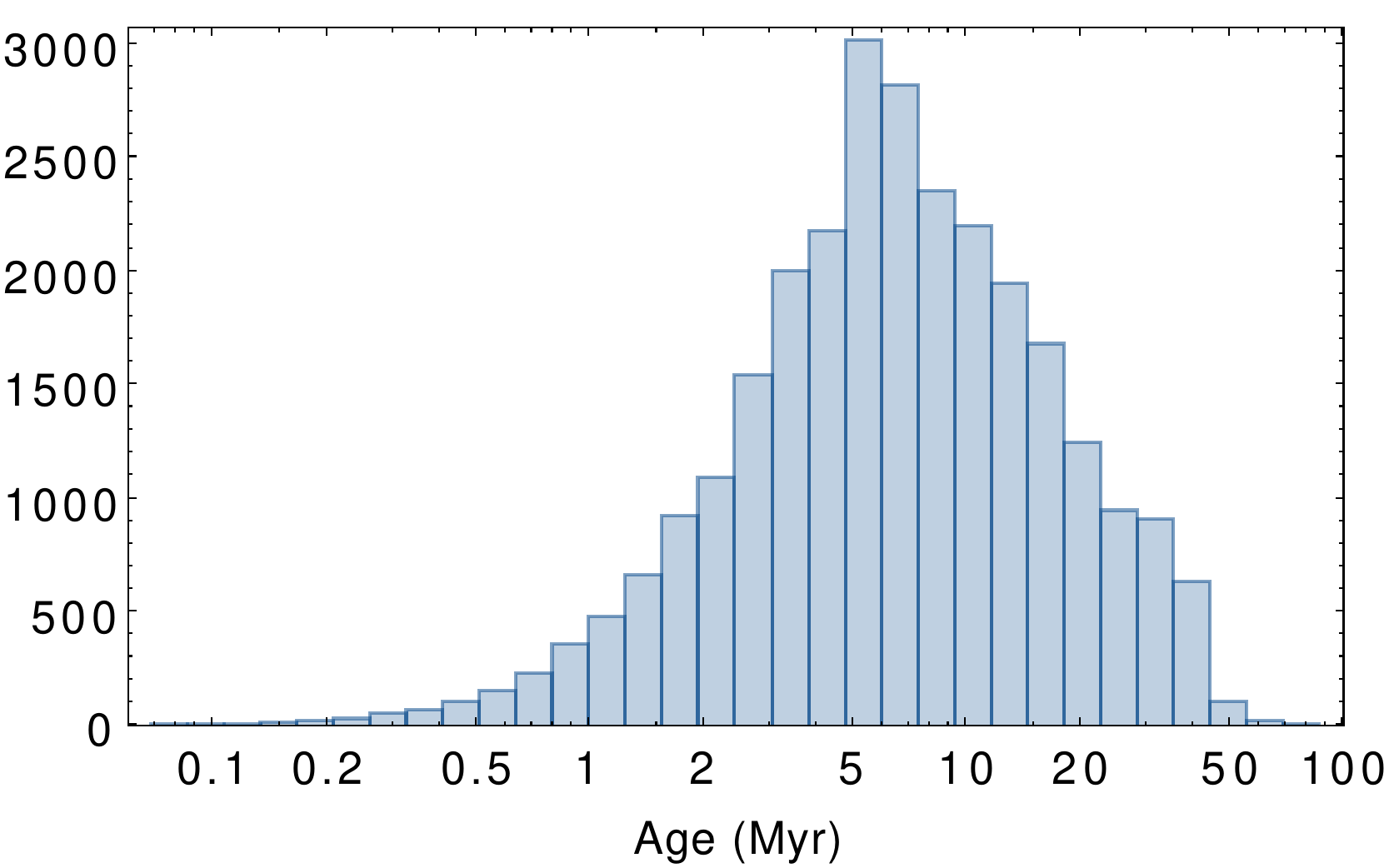}
\caption{Distribution of ages of stars identified by the classifier as young.
\label{fig:age}}
\end{figure}

\begin{figure*}    
\epsscale{1.0}
		\gridline{\fig{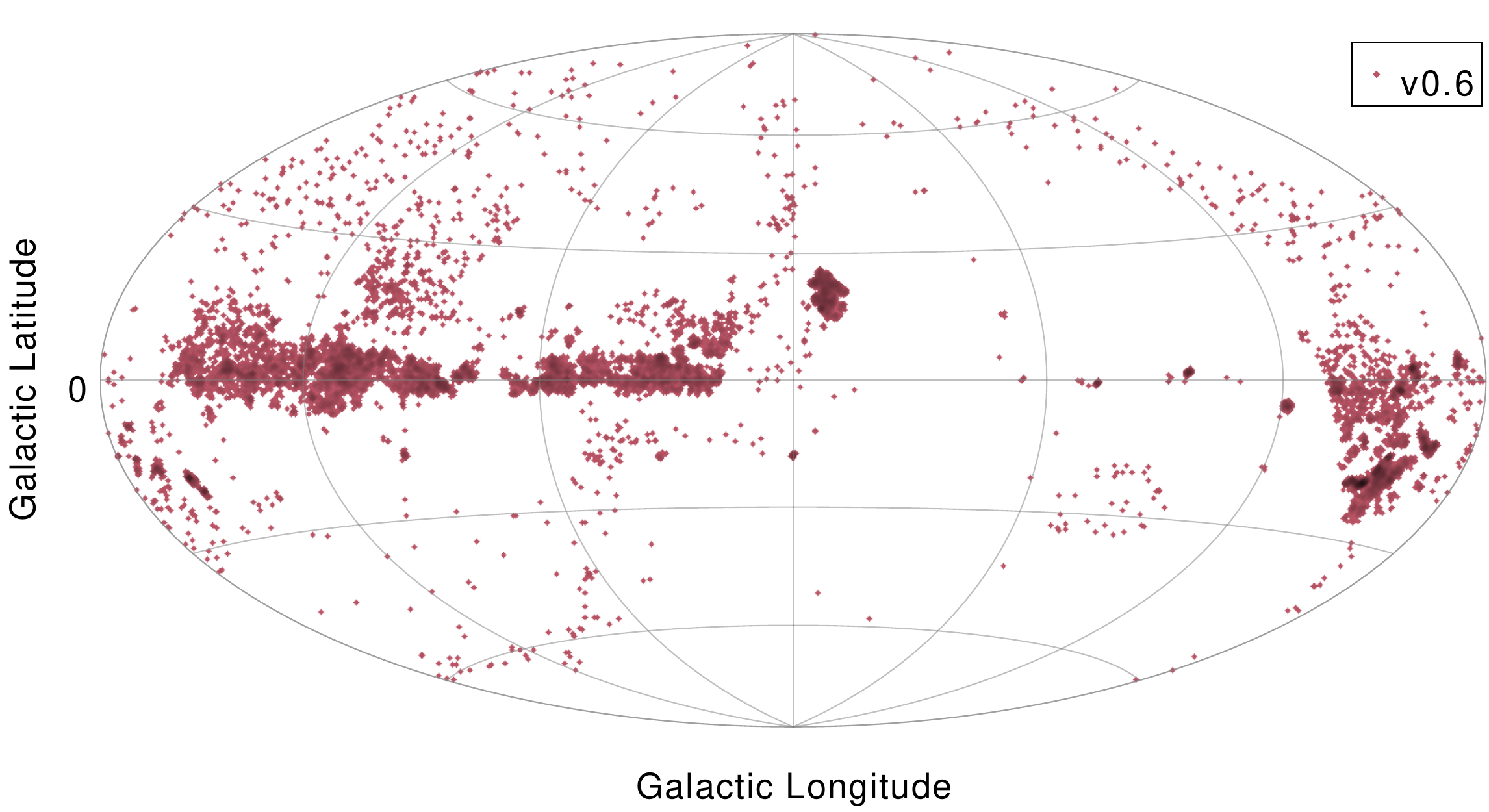}{0.58\textwidth}{}
		          \fig{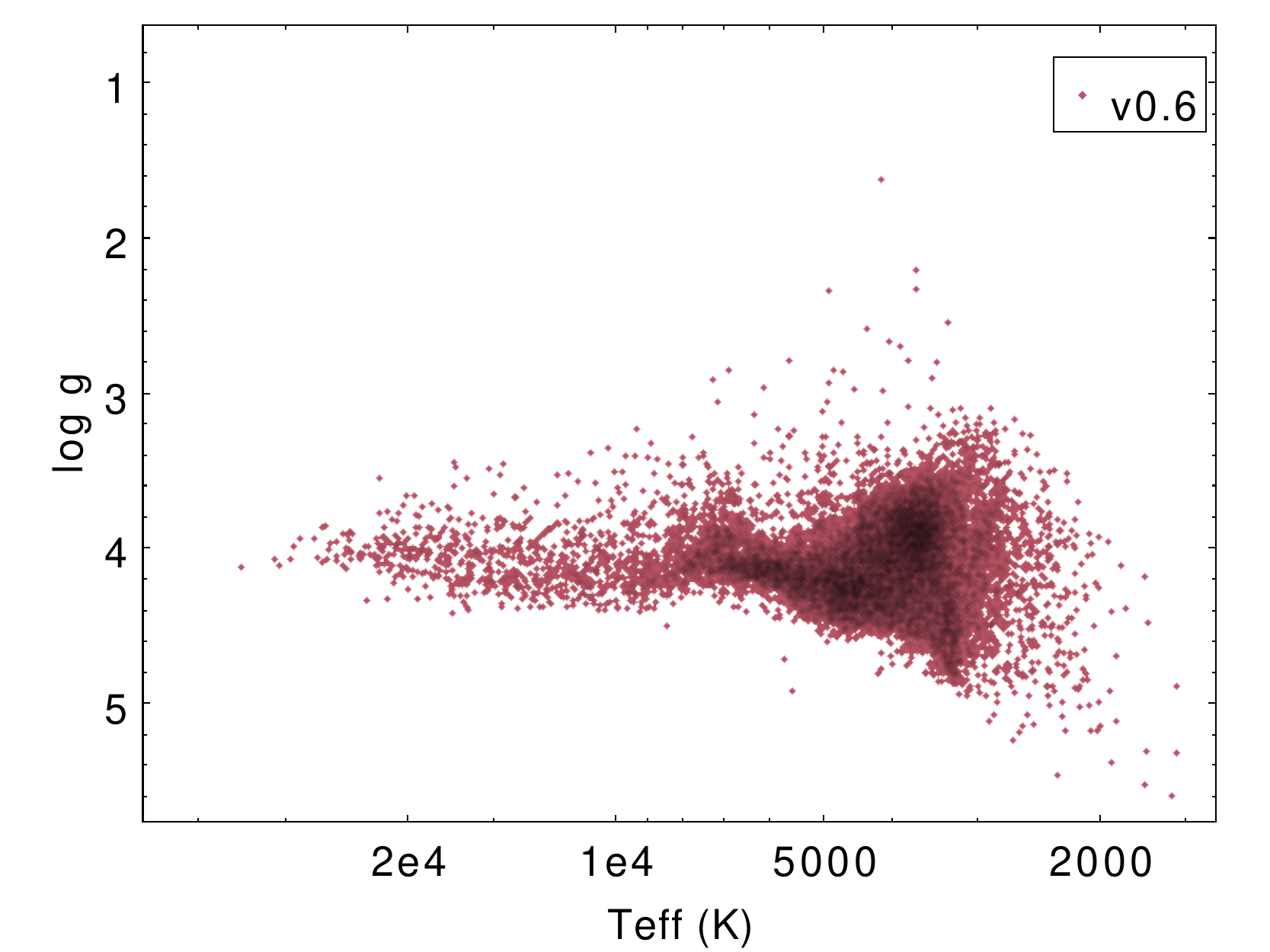}{0.42\textwidth}{}
        }\vspace{-1cm}
		\gridline{\fig{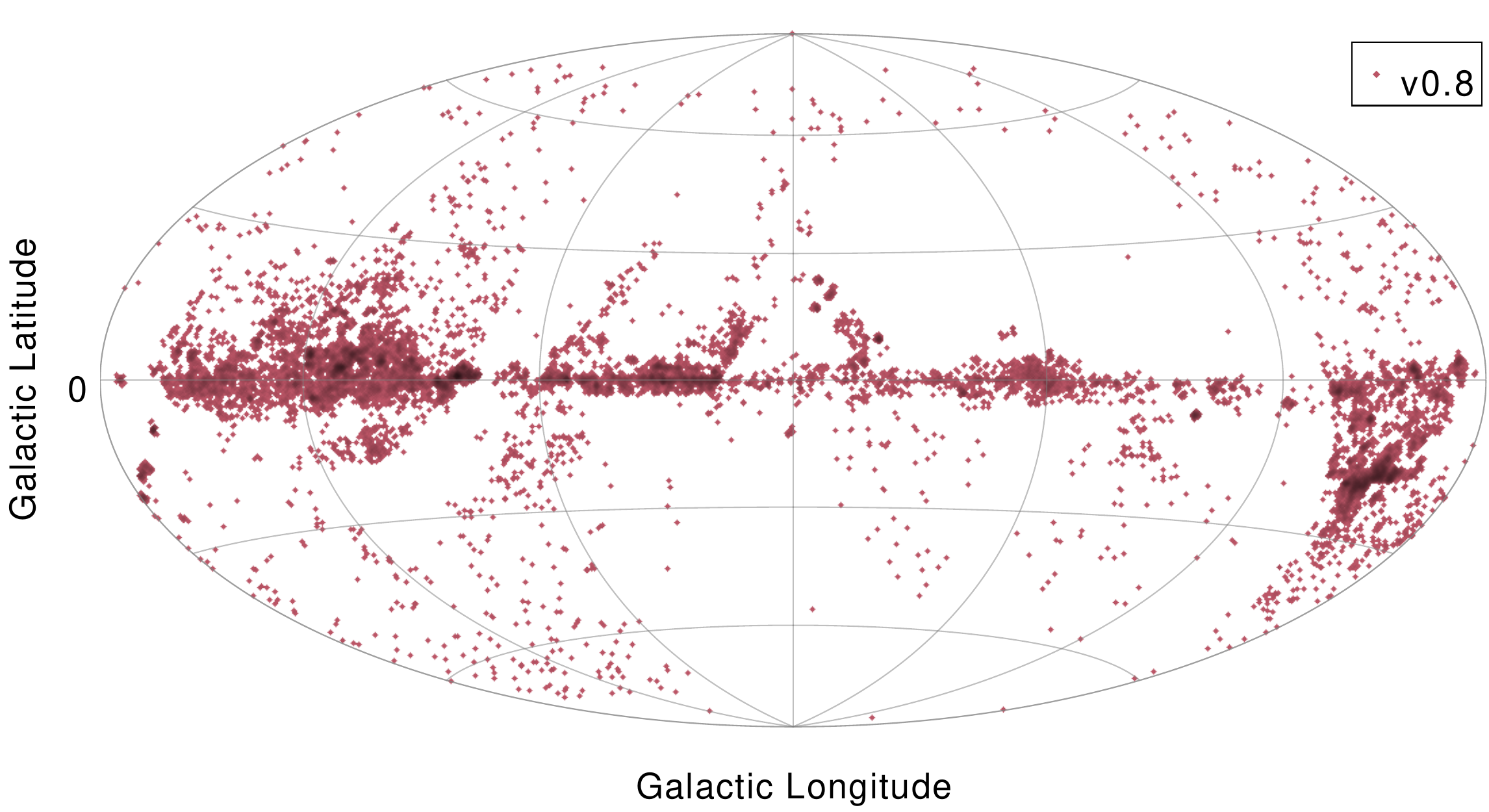}{0.58\textwidth}{}
		          \fig{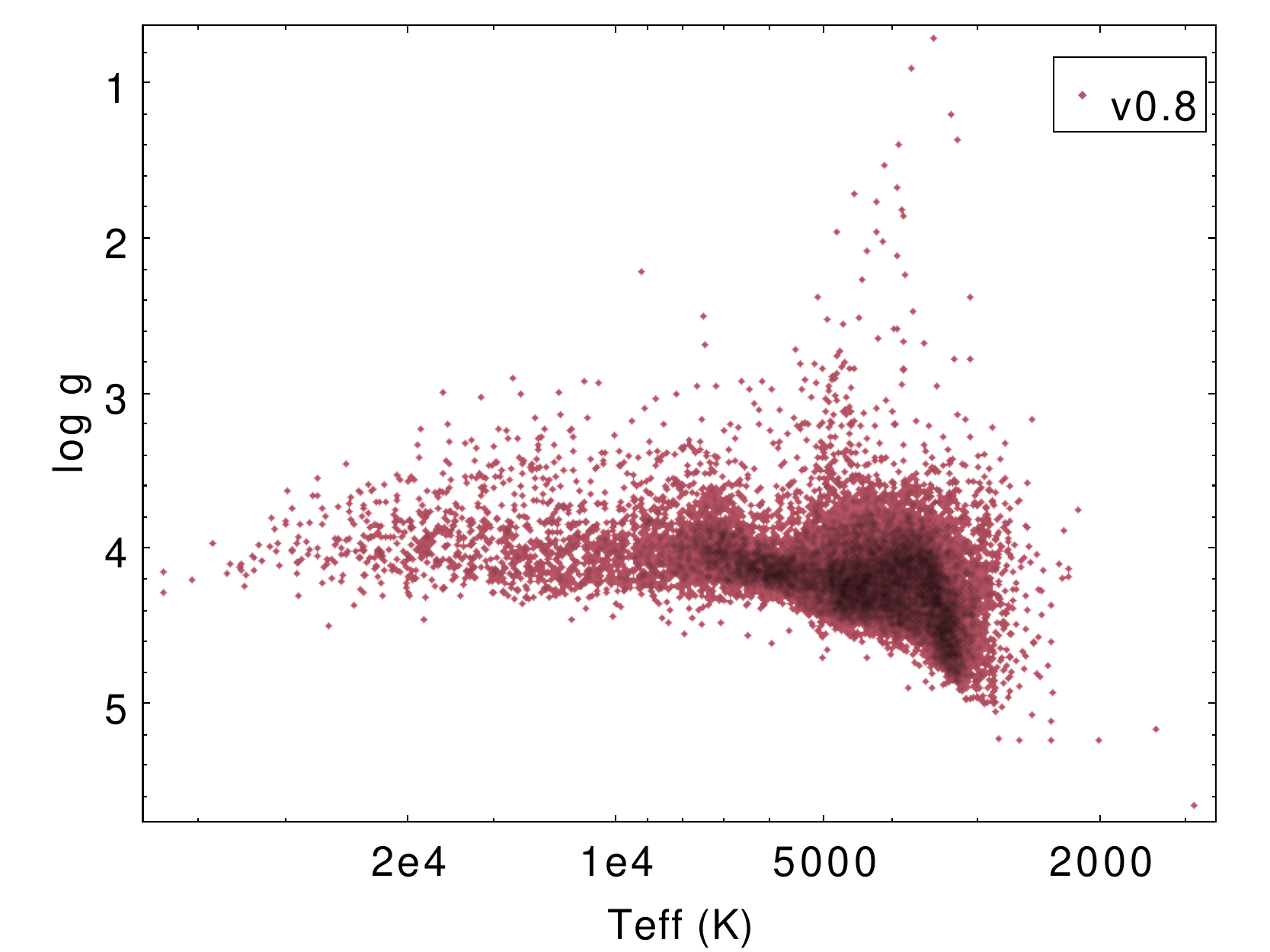}{0.42\textwidth}{}
        }\vspace{-1cm}
\caption{Left: distribution of the likely young stars selected by the classifier on the plane of the sky. Right: Kiel plot of these sources. Top row: sources in IPL-3 processing of the data, which translates to the labeled dataset (train and test). Bottom row: new unlabeled sources observed since then, included in IPL-4 processing. 
\label{fig:labels}}
\end{figure*}

\begin{figure}
\epsscale{1.2}
\plotone{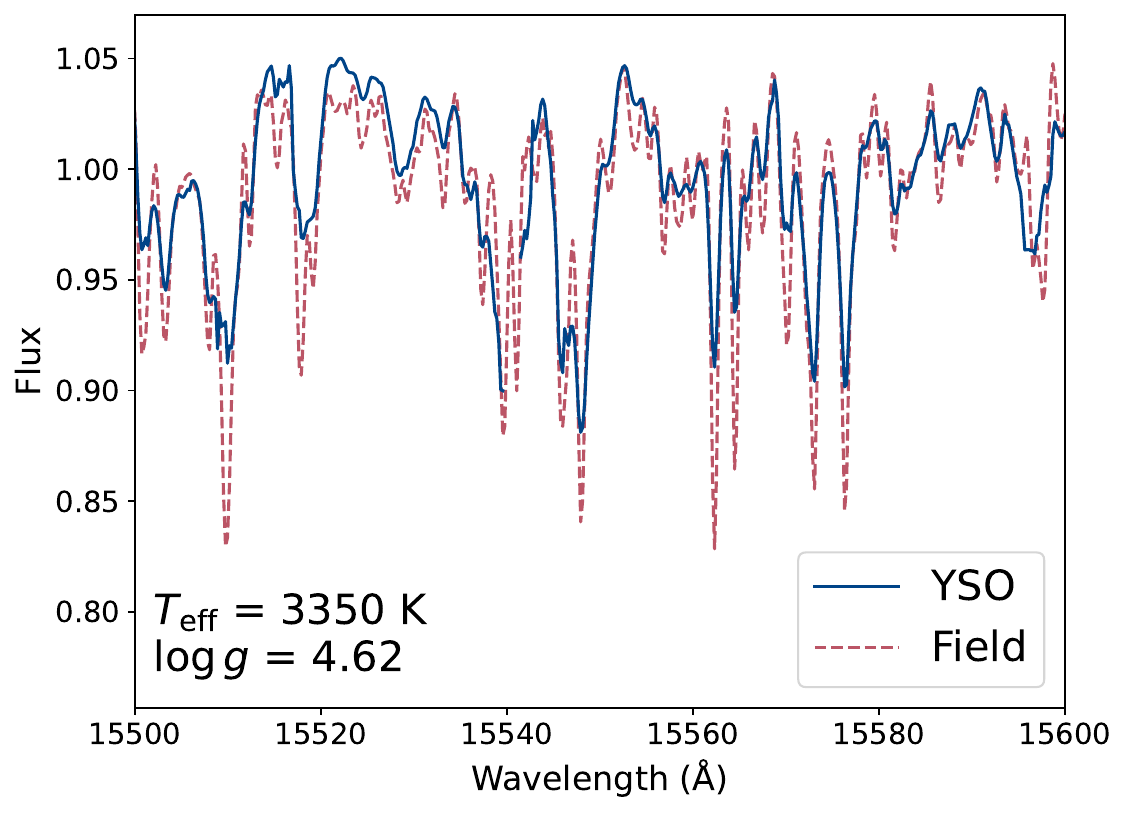}
\caption{Coadded empirical templates within the same $T_{\mathrm{eff}}$, $\log g$, and $[\mathrm{Fe/H}]$ bins, showing the spectrum of a YSO (blue) and a more evolved field star (red). The most prominent difference between the spectra is the stronger line broadening in the YSO, consistent with its typically higher \vsini. Note that while many young stars would have lower \logg, this demonstrates differences between young stars and field stars even in the regime when young stars begin approaching the main sequence.
\label{fig:rot}}
\end{figure}

\begin{figure*}
\epsscale{1.2}
\plotone{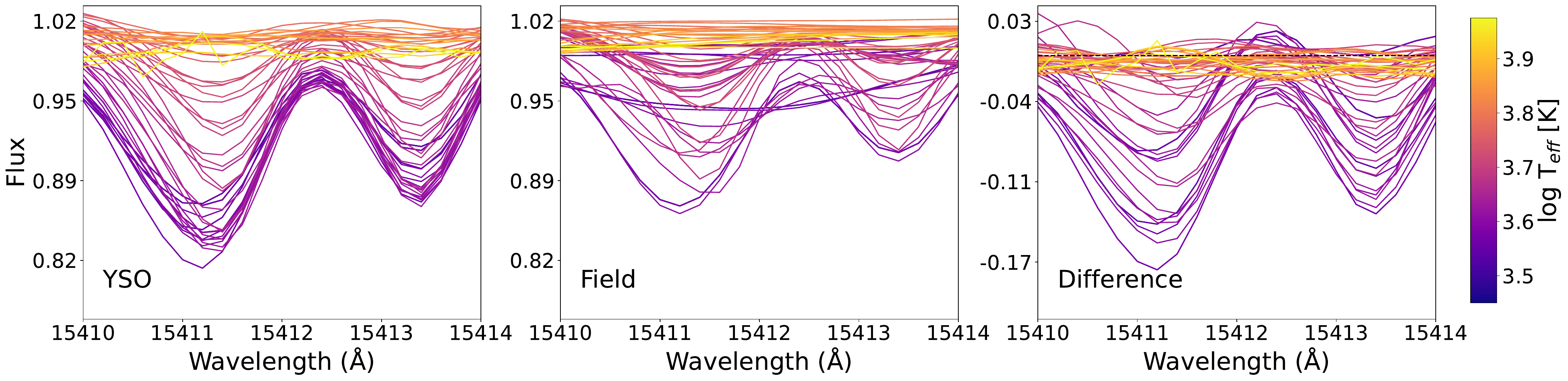}
\plotone{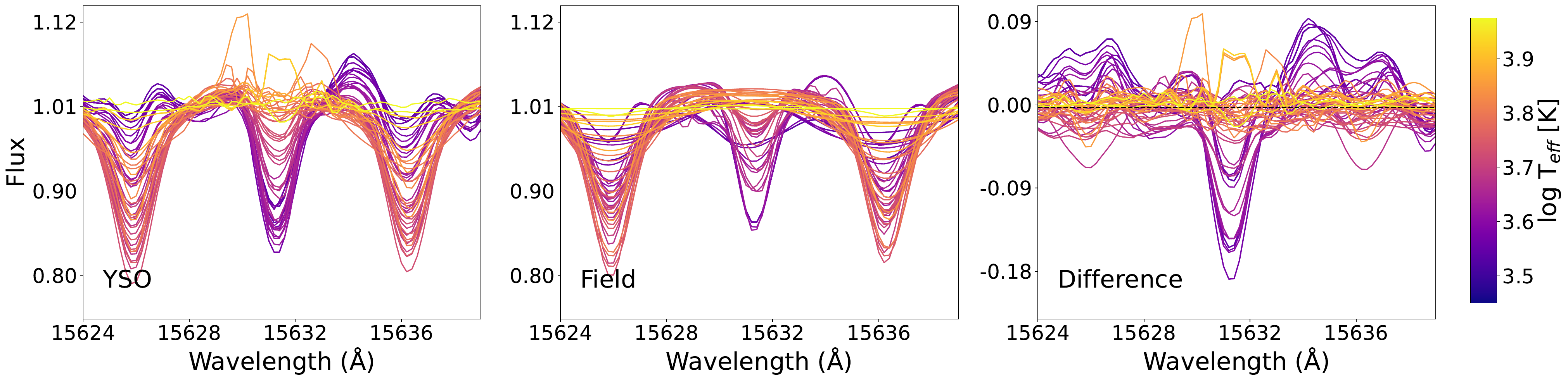}
\plotone{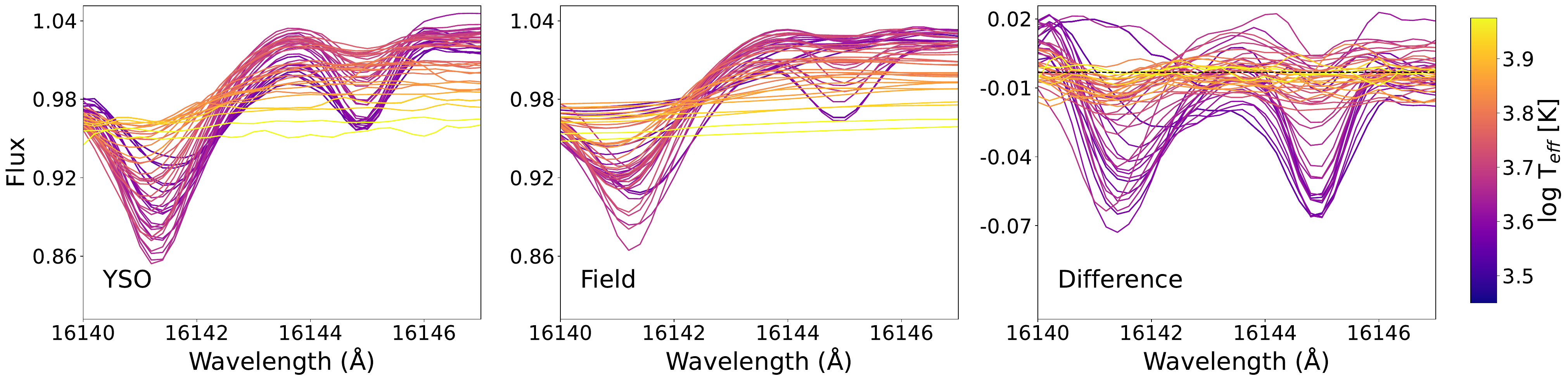}
\plotone{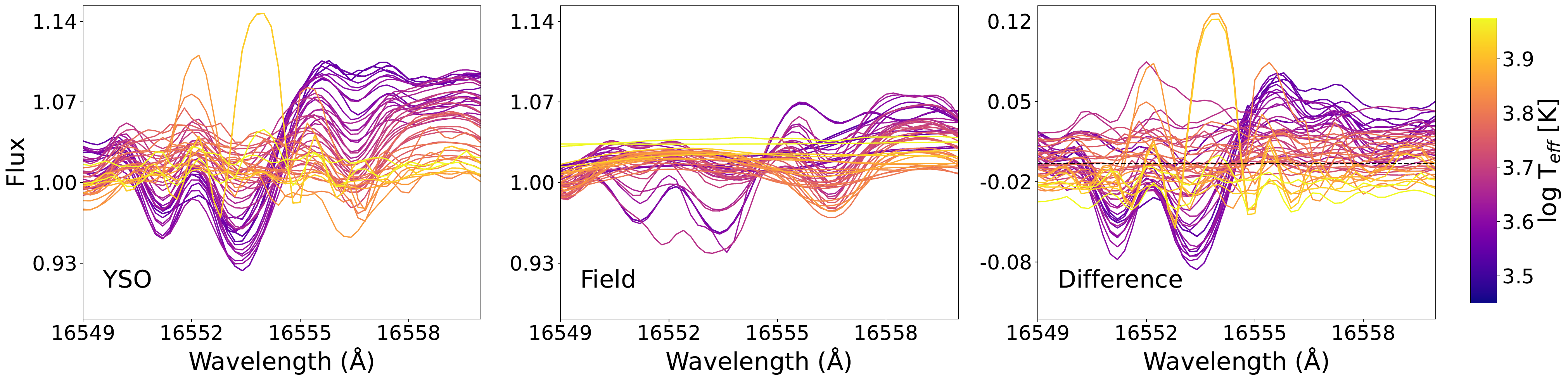}
\plotone{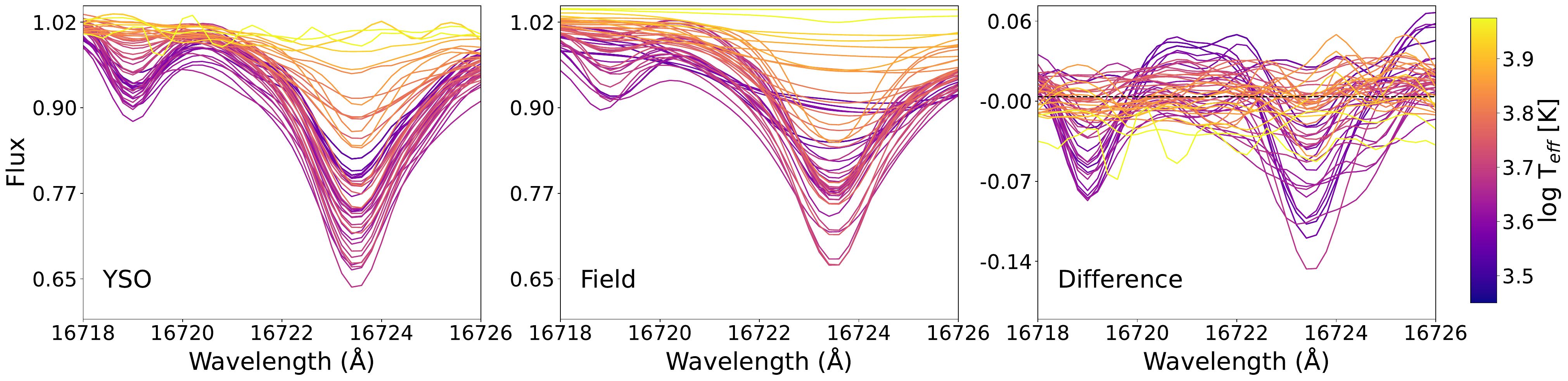}
\caption{Median-stacked APOGEE spectra of young stars near the main sequence identified by the classifier (left column) and more evolved field stars (middle column). Each trace includes spectra within a narrow $T_{\mathrm{eff}}$ bin. The evolved-star spectra are rotationally broadened to match the $v\sin i$ of the YSOs. The right column shows the residuals between the two, highlighting candidate youth-sensitive lines. Traces are color-coded by $T_{\mathrm{eff}}$. The full figure set (29 images) is available in the HTML version.
\label{fig:diff}}
\end{figure*}

\figsetstart
\figsetnum{\ref{fig:diff}}
\figsettitle{Median-stacked APOGEE spectra showing signatures of youth.}

\figsetgrpstart
\figsetgrpnum{\ref{fig:diff}.1}
\figsetgrptitle{15281--15286.2}
\figsetplot{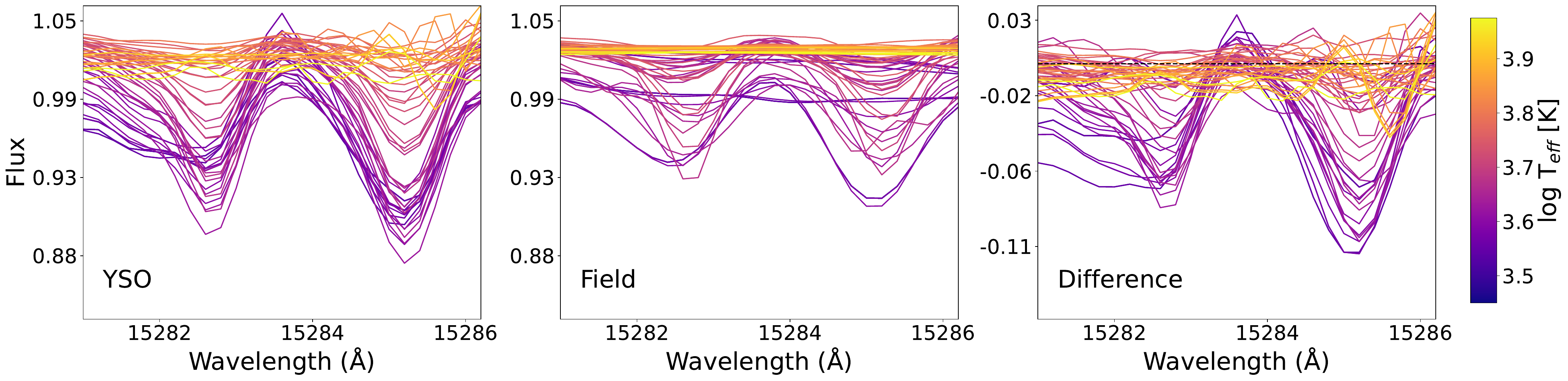}
\figsetgrpnote{Comparison of the mean spectra of younger and older stars showing youth-sensitive features.}
\figsetgrpend

\figsetgrpstart
\figsetgrpnum{\ref{fig:diff}.2}
\figsetgrptitle{15296--15301}
\figsetplot{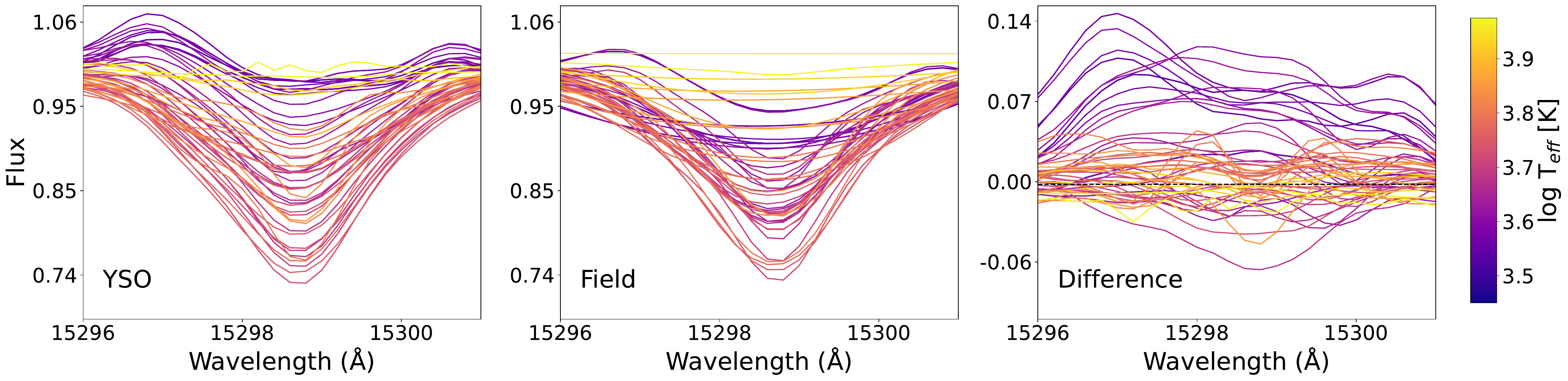}
\figsetgrpnote{Comparison of the mean spectra of younger and older stars showing youth-sensitive features.}
\figsetgrpend

\figsetgrpstart
\figsetgrpnum{\ref{fig:diff}.3}
\figsetgrptitle{15329--15335}
\figsetplot{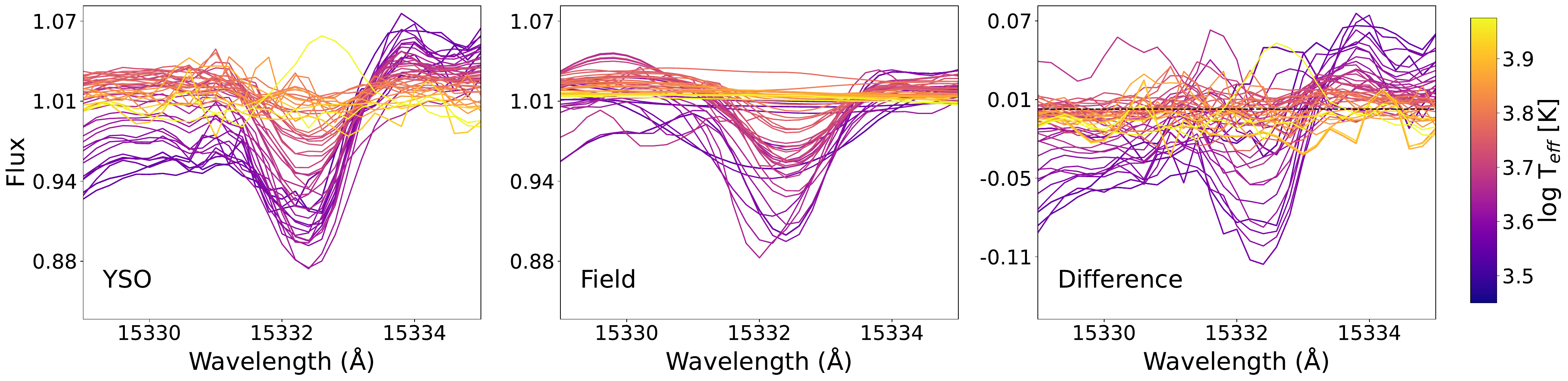}
\figsetgrpend
\figsetgrpnote{Comparison of the mean spectra of younger and older stars showing youth-sensitive features.}
\figsetend

\figsetgrpstart
\figsetgrpnum{\ref{fig:diff}.4}
\figsetgrptitle{15338--15341}
\figsetplot{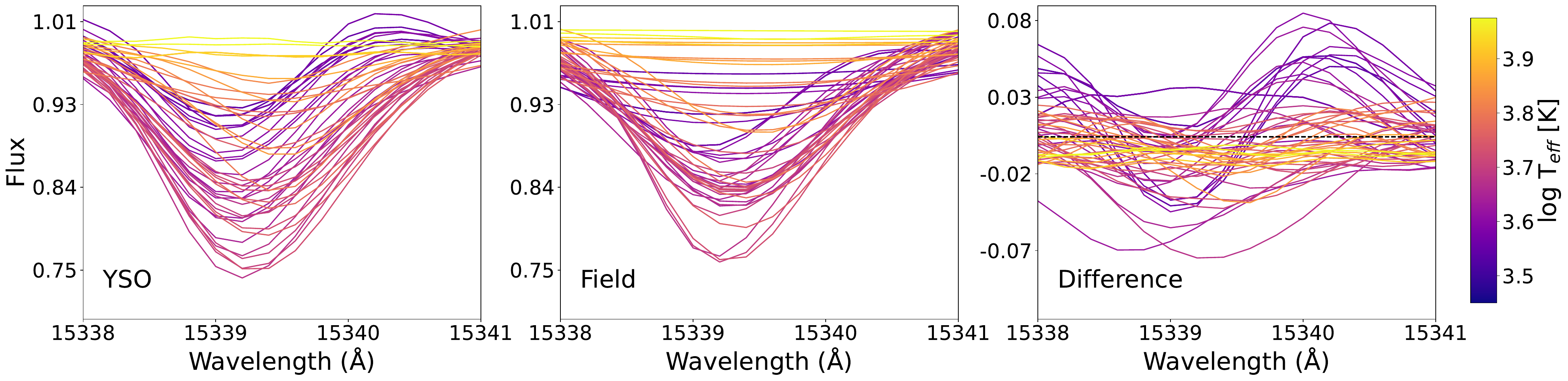}
\figsetgrpend
\figsetgrpnote{Comparison of the mean spectra of younger and older stars showing youth-sensitive features.}
\figsetend

\figsetgrpstart
\figsetgrpnum{\ref{fig:diff}.5}
\figsetgrptitle{15393--15401}
\figsetplot{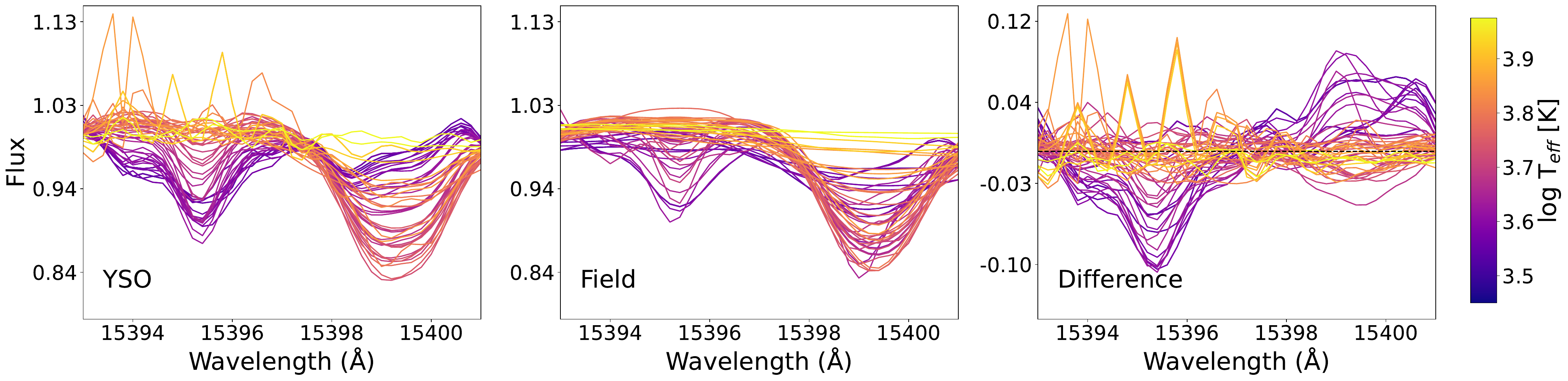}
\figsetgrpend
\figsetgrpnote{Comparison of the mean spectra of younger and older stars showing youth-sensitive features.}
\figsetend

\figsetgrpstart
\figsetgrpnum{\ref{fig:diff}.6}
\figsetgrptitle{15473--15476}
\figsetplot{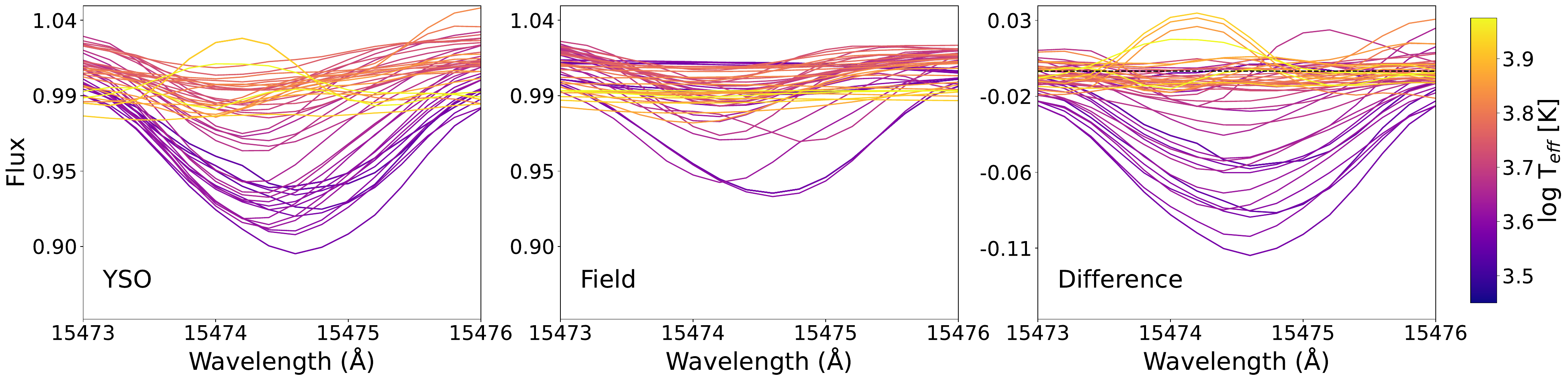}
\figsetgrpend
\figsetgrpnote{Comparison of the mean spectra of younger and older stars showing youth-sensitive features.}
\figsetend

\figsetgrpstart
\figsetgrpnum{\ref{fig:diff}.7}
\figsetgrptitle{15498--15502}
\figsetplot{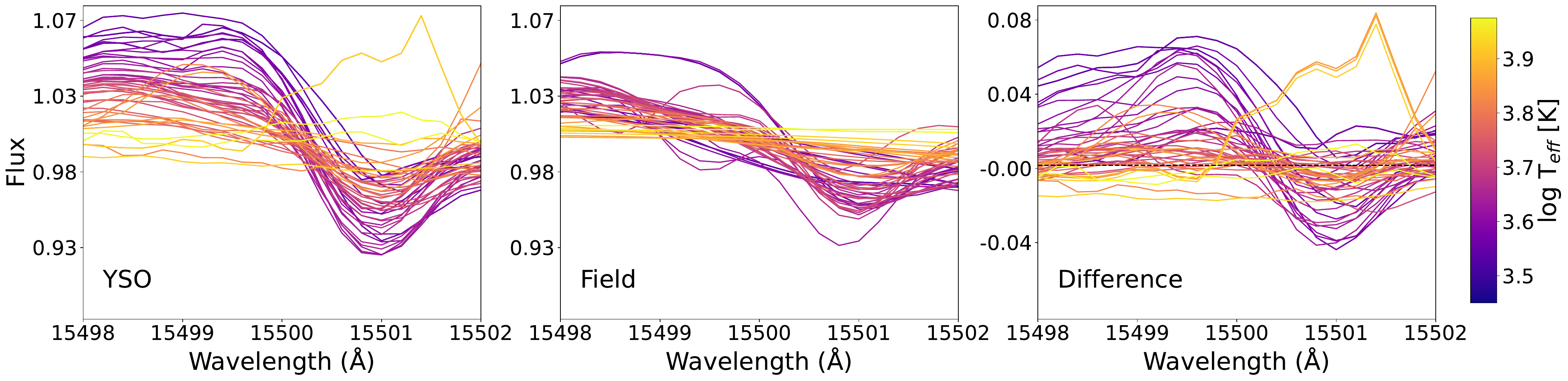}
\figsetgrpend
\figsetgrpnote{Comparison of the mean spectra of younger and older stars showing youth-sensitive features.}
\figsetend

\figsetgrpstart
\figsetgrpnum{\ref{fig:diff}.8}
\figsetgrptitle{15507--15513}
\figsetplot{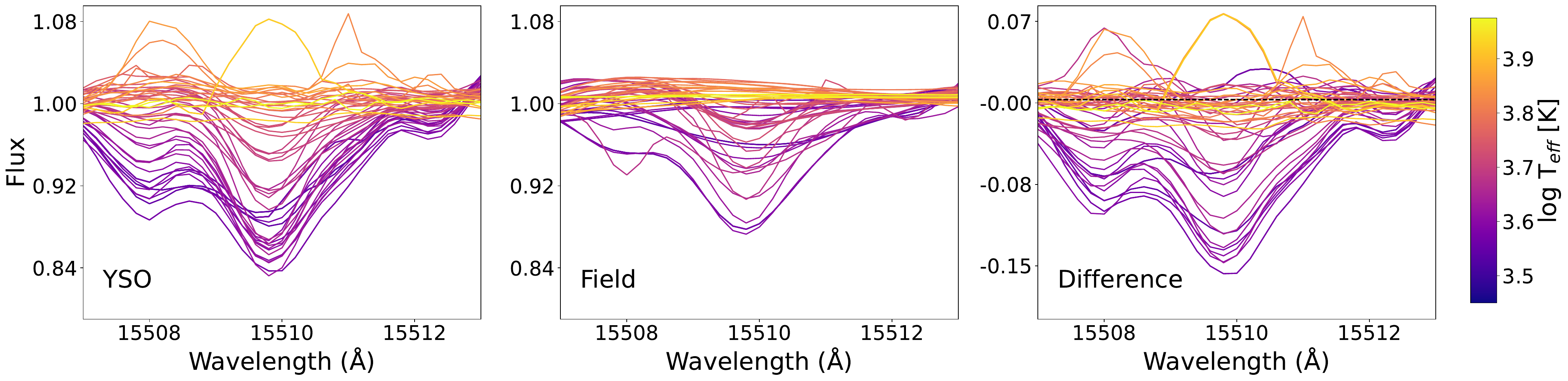}
\figsetgrpend
\figsetgrpnote{Comparison of the mean spectra of younger and older stars showing youth-sensitive features.}
\figsetend

\figsetgrpstart
\figsetgrpnum{\ref{fig:diff}.9}
\figsetgrptitle{15571--15578}
\figsetplot{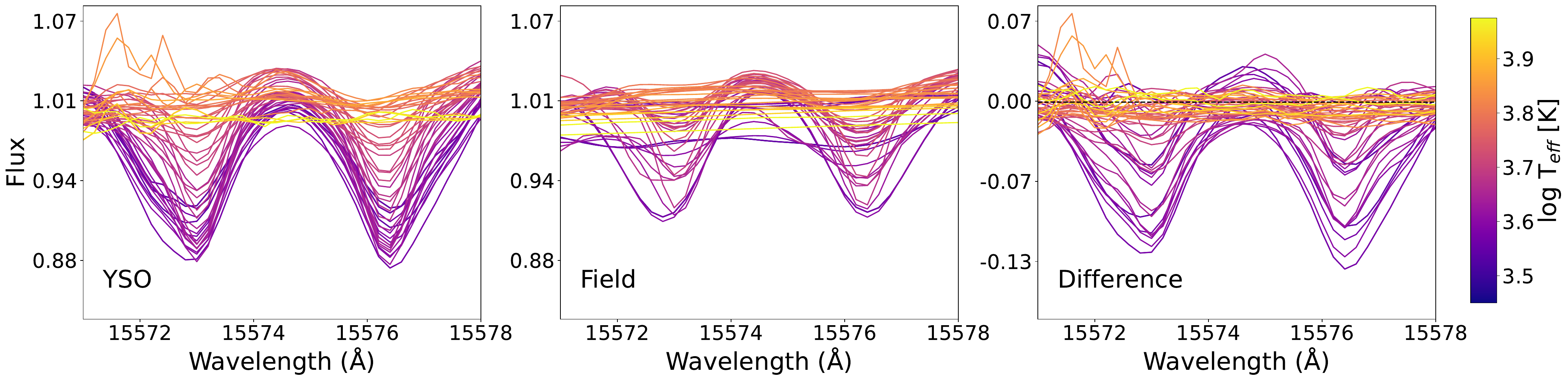}
\figsetgrpend
\figsetgrpnote{Comparison of the mean spectra of younger and older stars showing youth-sensitive features.}
\figsetend

\figsetgrpstart
\figsetgrpnum{\ref{fig:diff}.10}
\figsetgrptitle{15606--15610}
\figsetplot{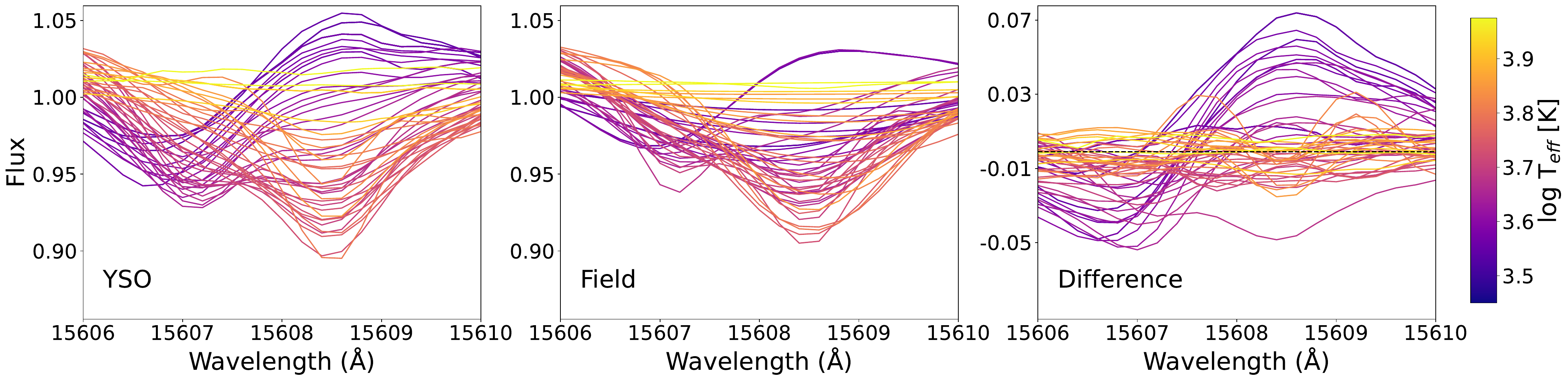}
\figsetgrpend
\figsetgrpnote{Comparison of the mean spectra of younger and older stars showing youth-sensitive features.}
\figsetend

\figsetgrpstart
\figsetgrpnum{\ref{fig:diff}.11}
\figsetgrptitle{15655--15660}
\figsetplot{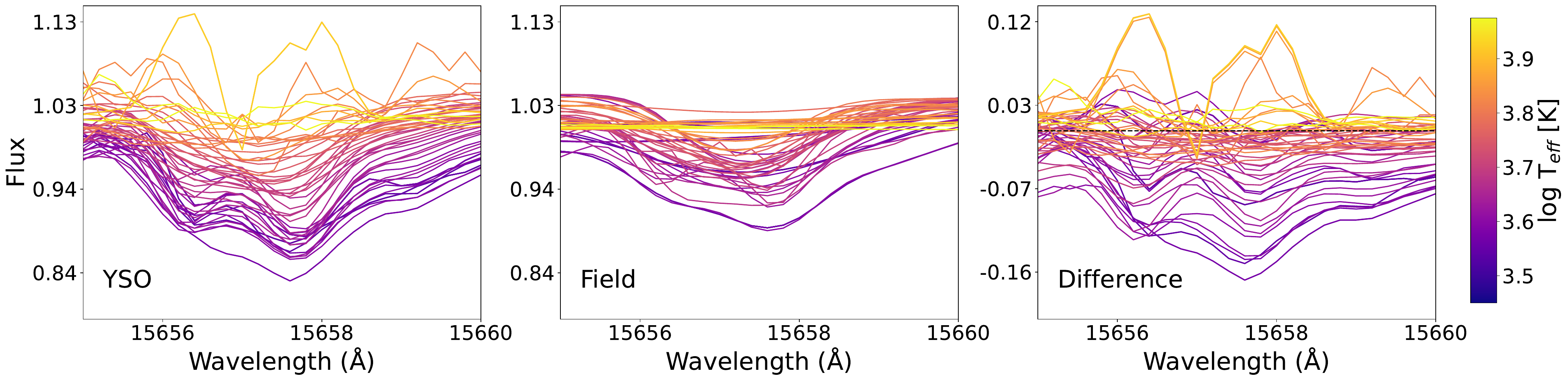}
\figsetgrpend
\figsetgrpnote{Comparison of the mean spectra of younger and older stars showing youth-sensitive features.}
\figsetend

\figsetgrpstart
\figsetgrpnum{\ref{fig:diff}.12}
\figsetgrptitle{15743--15756}
\figsetplot{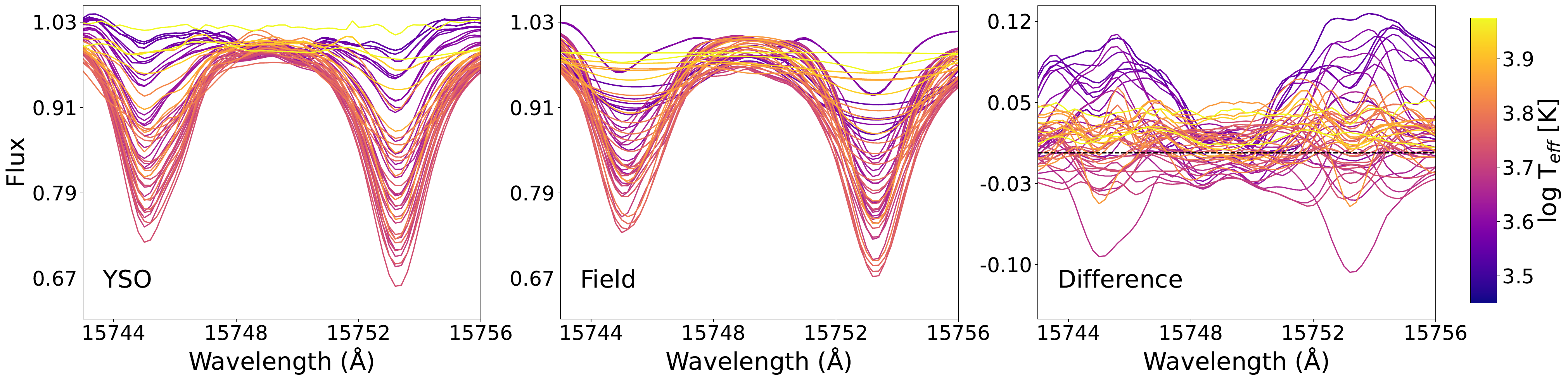}
\figsetgrpend
\figsetgrpnote{Comparison of the mean spectra of younger and older stars showing youth-sensitive features.}
\figsetend

\figsetgrpstart
\figsetgrpnum{\ref{fig:diff}.13}
\figsetgrptitle{15779--15787}
\figsetplot{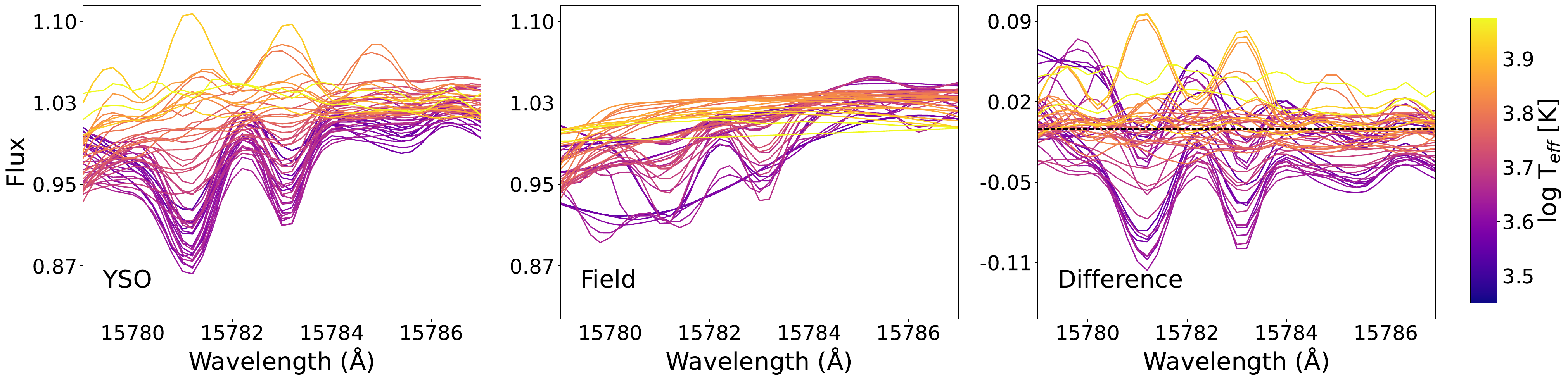}
\figsetgrpend
\figsetgrpnote{Comparison of the mean spectra of younger and older stars showing youth-sensitive features.}
\figsetend

\figsetgrpstart
\figsetgrpnum{\ref{fig:diff}.14}
\figsetgrptitle{15891--15899}
\figsetplot{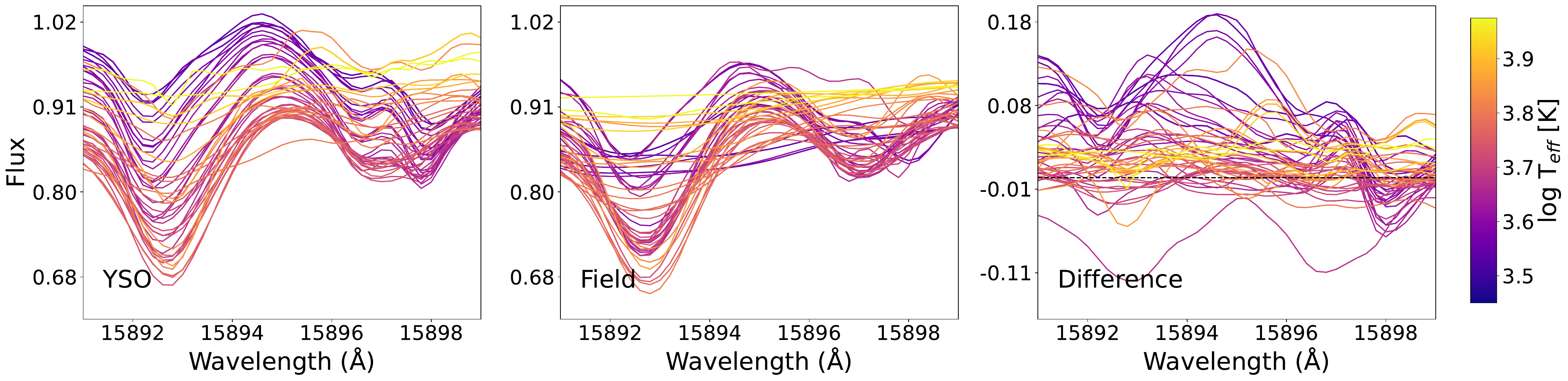}
\figsetgrpend
\figsetgrpnote{Comparison of the mean spectra of younger and older stars showing youth-sensitive features.}
\figsetend

\figsetgrpstart
\figsetgrpnum{\ref{fig:diff}.15}
\figsetgrptitle{15962--15967}
\figsetplot{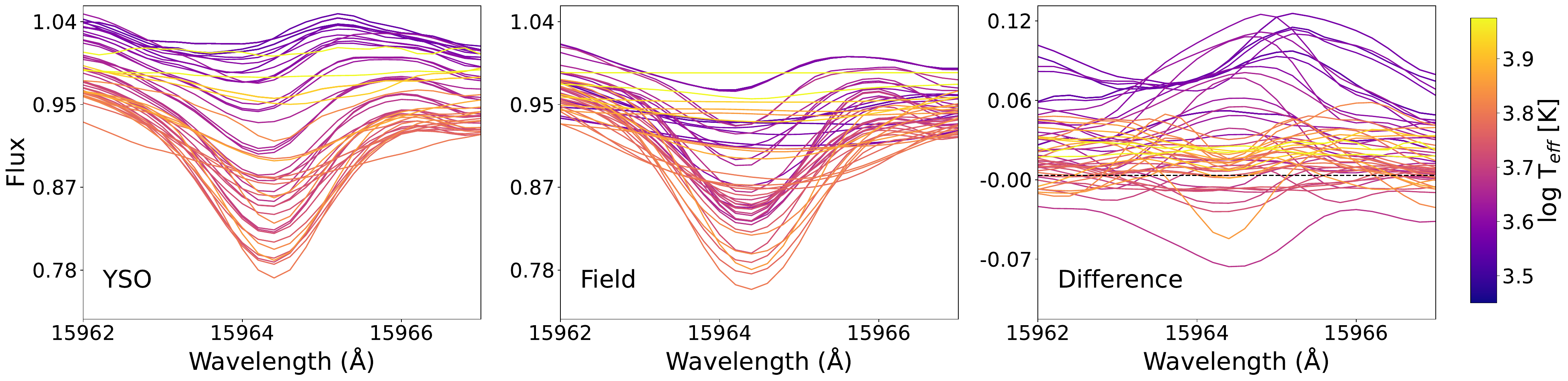}
\figsetgrpend
\figsetgrpnote{Comparison of the mean spectra of younger and older stars showing youth-sensitive features.}
\figsetend

\figsetgrpstart
\figsetgrpnum{\ref{fig:diff}.16}
\figsetgrptitle{16026--16036}
\figsetplot{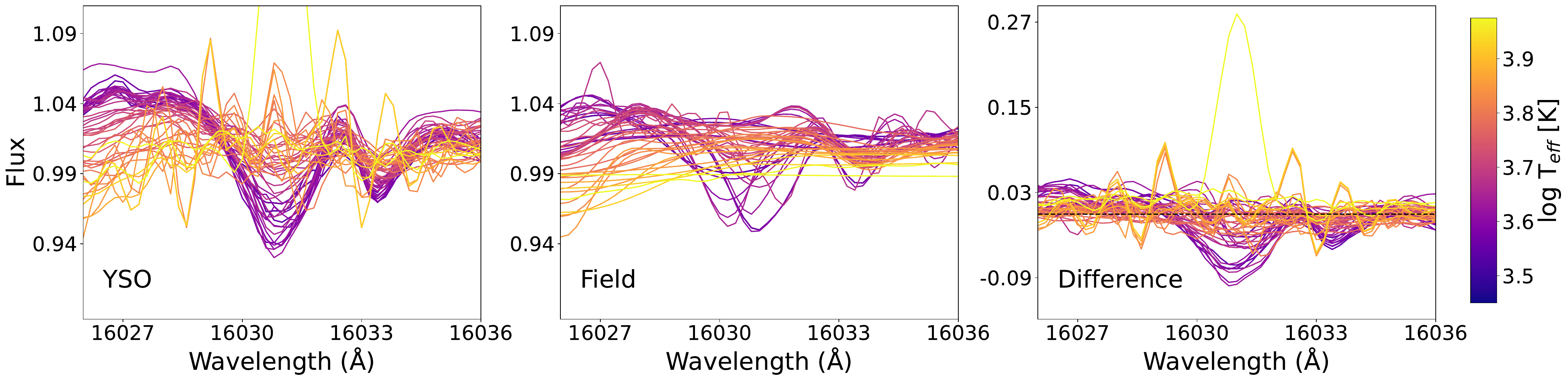}
\figsetgrpend
\figsetgrpnote{Comparison of the mean spectra of younger and older stars showing youth-sensitive features.}
\figsetend

\figsetgrpstart
\figsetgrpnum{\ref{fig:diff}.17}
\figsetgrptitle{16040--16044}
\figsetplot{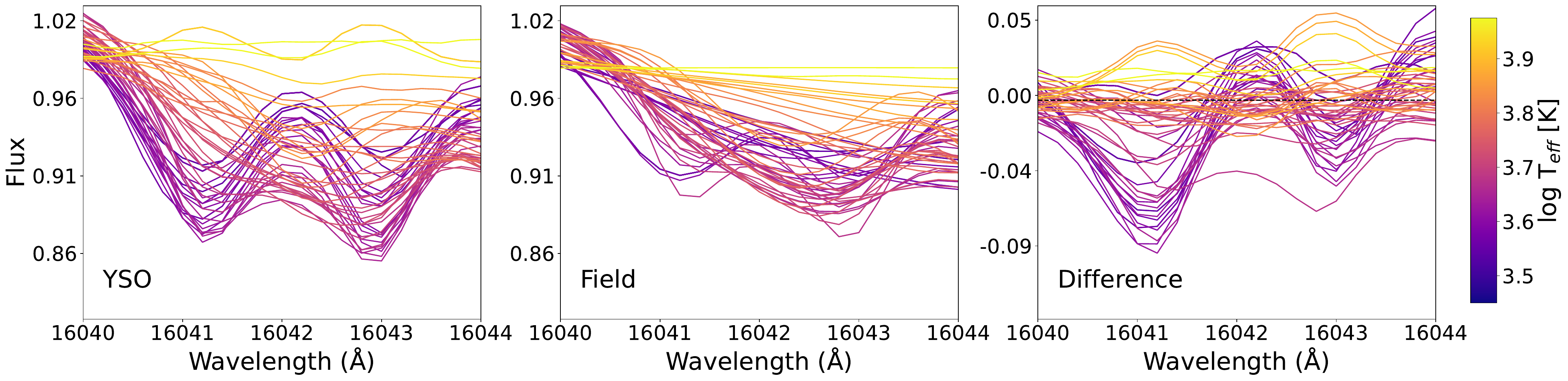}
\figsetgrpend
\figsetgrpnote{Comparison of the mean spectra of younger and older stars showing youth-sensitive features.}
\figsetend

\figsetgrpstart
\figsetgrpnum{\ref{fig:diff}.18}
\figsetgrptitle{16056--16061}
\figsetplot{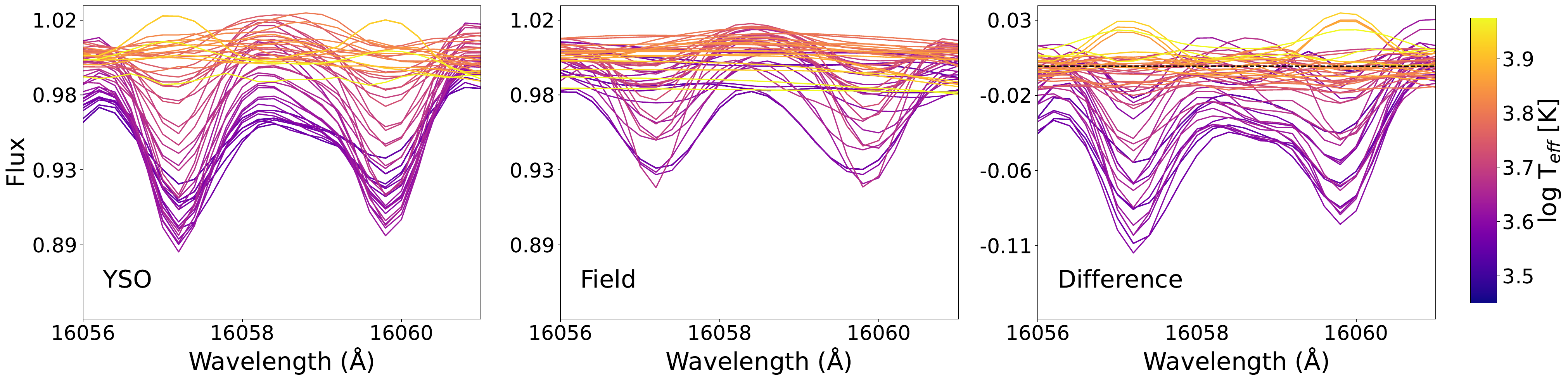}
\figsetgrpend
\figsetgrpnote{Comparison of the mean spectra of younger and older stars showing youth-sensitive features.}
\figsetend

\figsetgrpstart
\figsetgrpnum{\ref{fig:diff}.19}
\figsetgrptitle{16063--16067}
\figsetplot{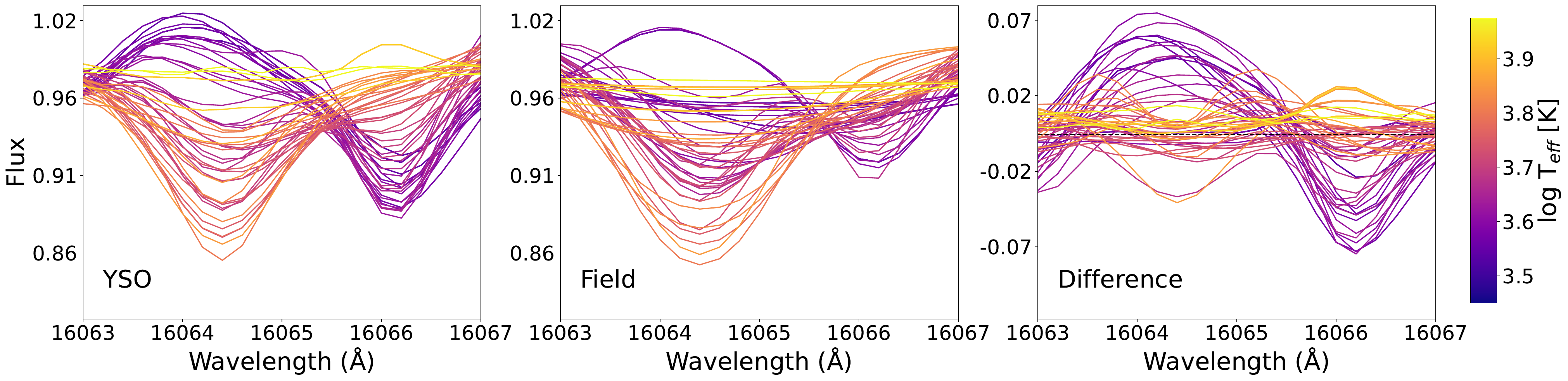}
\figsetgrpend
\figsetgrpnote{Comparison of the mean spectra of younger and older stars showing youth-sensitive features.}
\figsetend

\figsetgrpstart
\figsetgrpnum{\ref{fig:diff}.20}
\figsetgrptitle{16076--16083}
\figsetplot{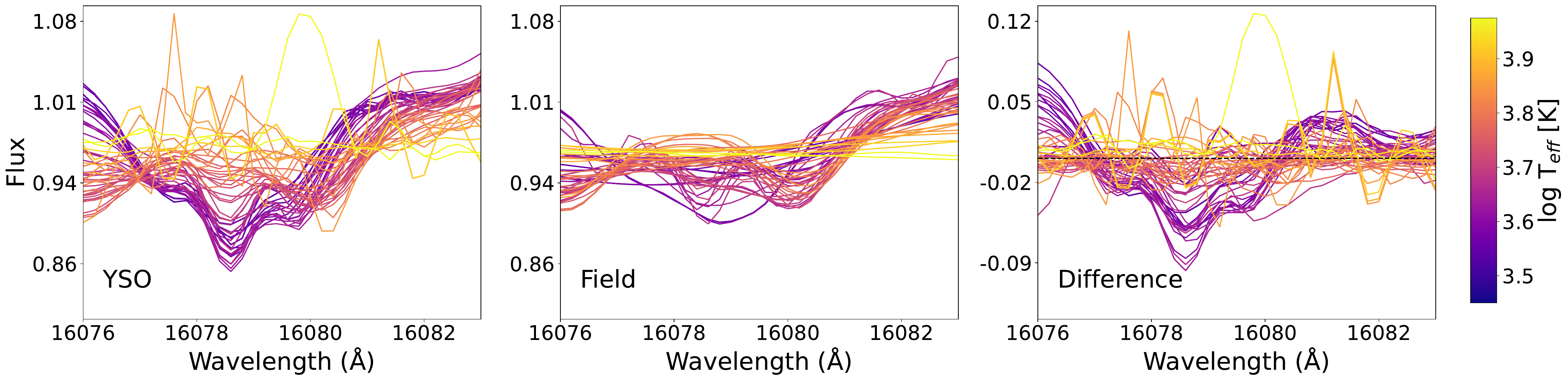}
\figsetgrpend
\figsetgrpnote{Comparison of the mean spectra of younger and older stars showing youth-sensitive features.}
\figsetend

\figsetgrpstart
\figsetgrpnum{\ref{fig:diff}.21}
\figsetgrptitle{16153--16157}
\figsetplot{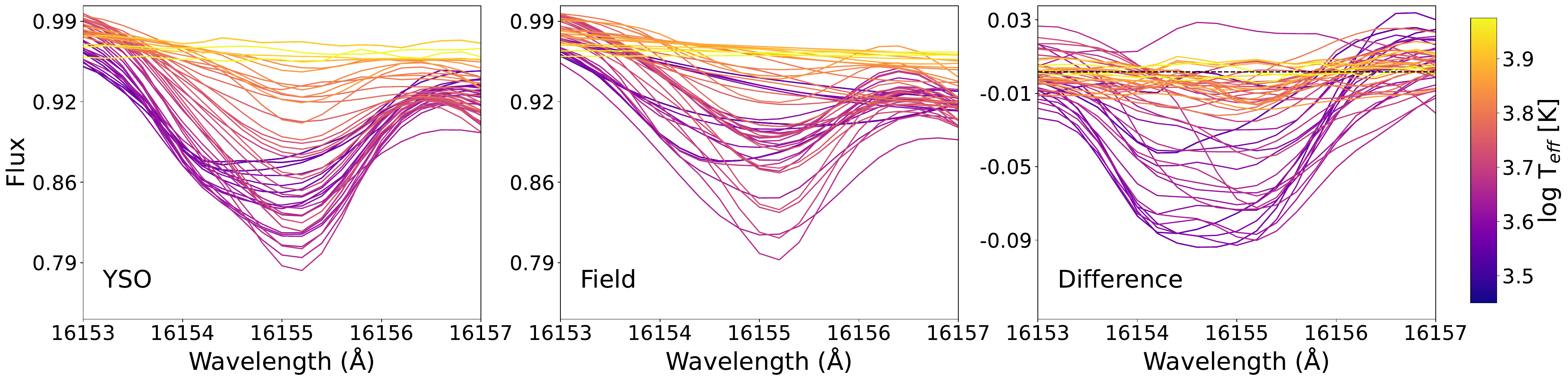}
\figsetgrpend
\figsetgrpnote{Comparison of the mean spectra of younger and older stars showing youth-sensitive features.}
\figsetend

\figsetgrpstart
\figsetgrpnum{\ref{fig:diff}.22}
\figsetgrptitle{16207--16213}
\figsetplot{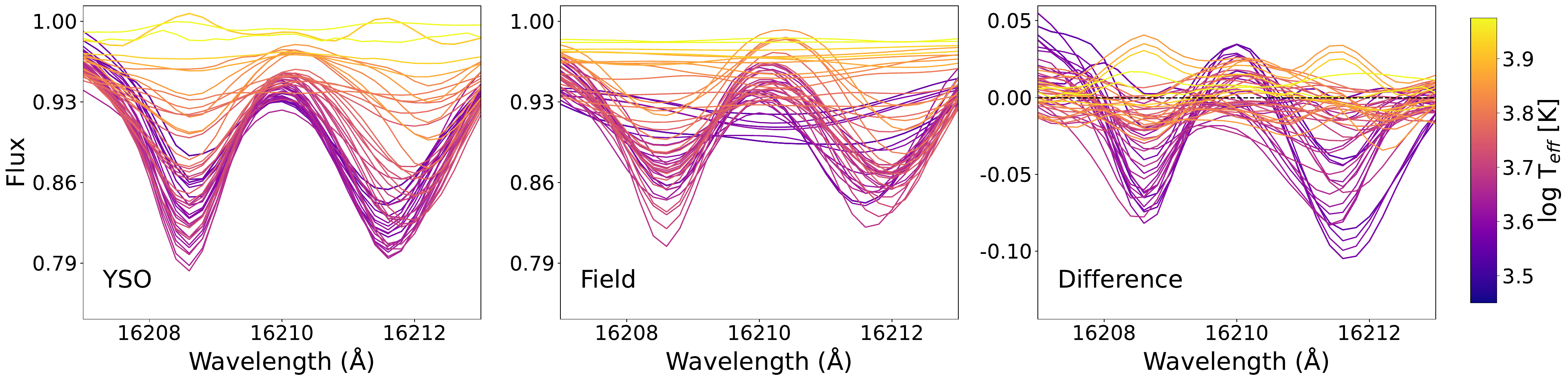}
\figsetgrpend
\figsetgrpnote{Comparison of the mean spectra of younger and older stars showing youth-sensitive features.}
\figsetend

\figsetgrpstart
\figsetgrpnum{\ref{fig:diff}.23}
\figsetgrptitle{16314--16322}
\figsetplot{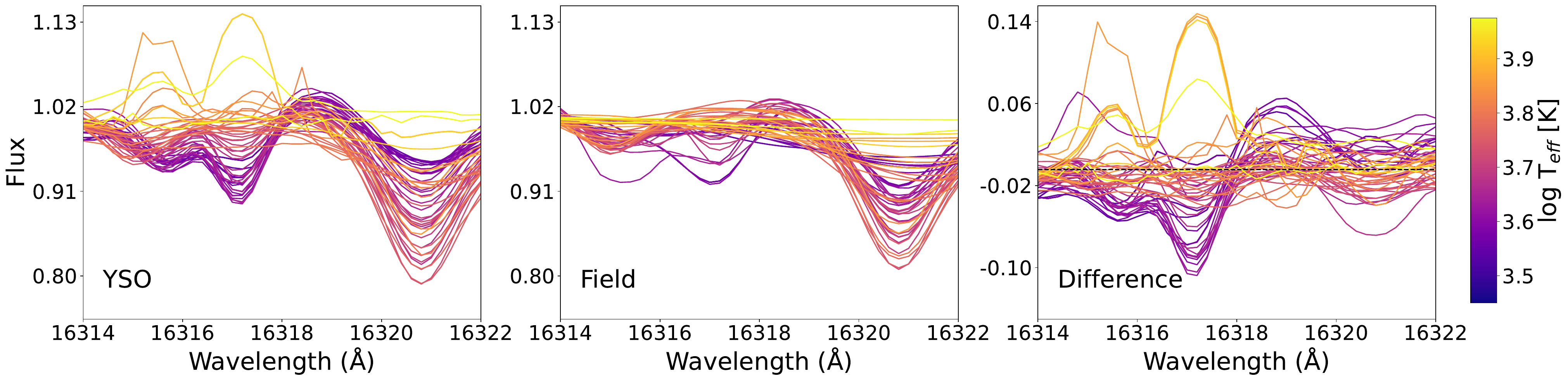}
\figsetgrpend
\figsetgrpnote{Comparison of the mean spectra of younger and older stars showing youth-sensitive features.}
\figsetend

\figsetgrpstart
\figsetgrpnum{\ref{fig:diff}.24}
\figsetgrptitle{16367--16374}
\figsetplot{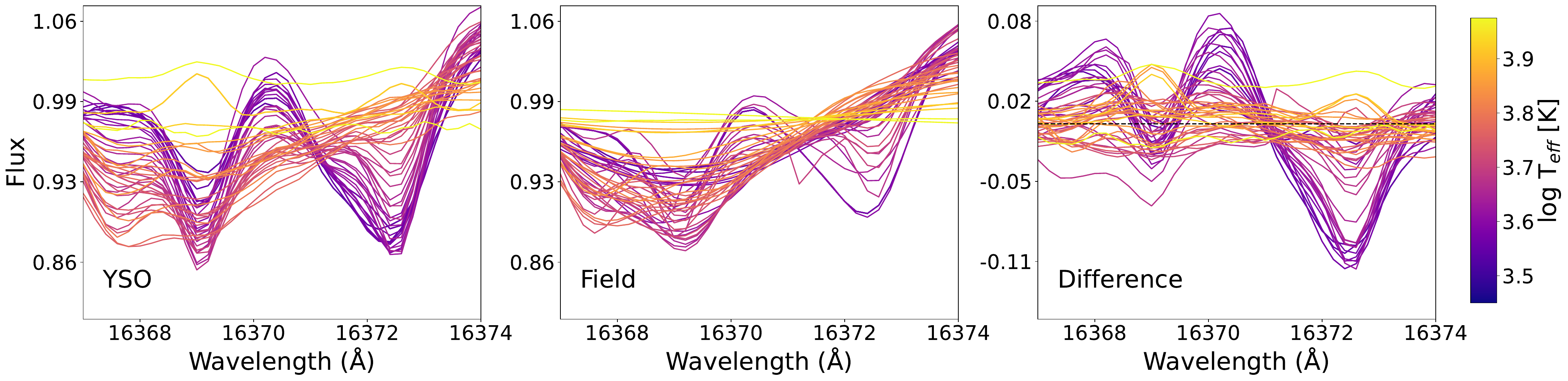}
\figsetgrpend
\figsetgrpnote{Comparison of the mean spectra of younger and older stars showing youth-sensitive features.}
\figsetend

\figsetgrpstart
\figsetgrpnum{\ref{fig:diff}.25}
\figsetgrptitle{16383--16390}
\figsetplot{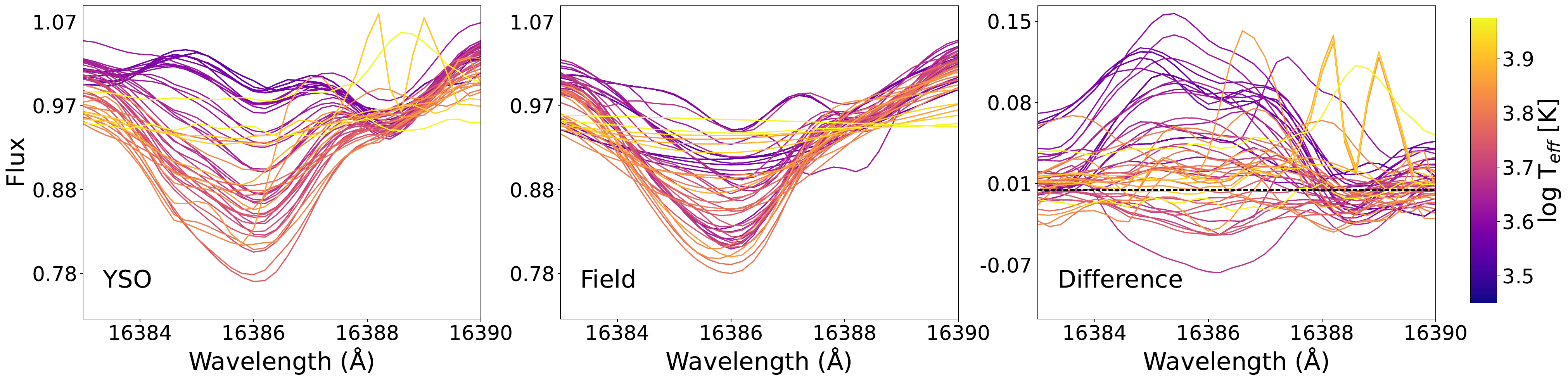}
\figsetgrpend
\figsetgrpnote{Comparison of the mean spectra of younger and older stars showing youth-sensitive features.}
\figsetend

\figsetgrpstart
\figsetgrpnum{\ref{fig:diff}.26}
\figsetgrptitle{16526--16533}
\figsetplot{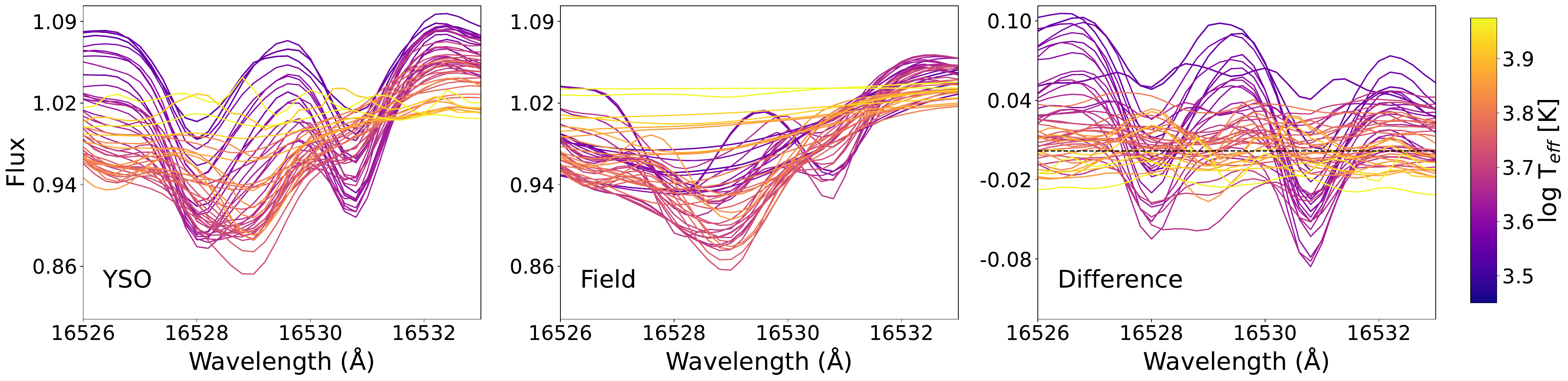}
\figsetgrpend
\figsetgrpnote{Comparison of the mean spectra of younger and older stars showing youth-sensitive features.}
\figsetend

\figsetgrpstart
\figsetgrpnum{\ref{fig:diff}.27}
\figsetgrptitle{16584--16590}
\figsetplot{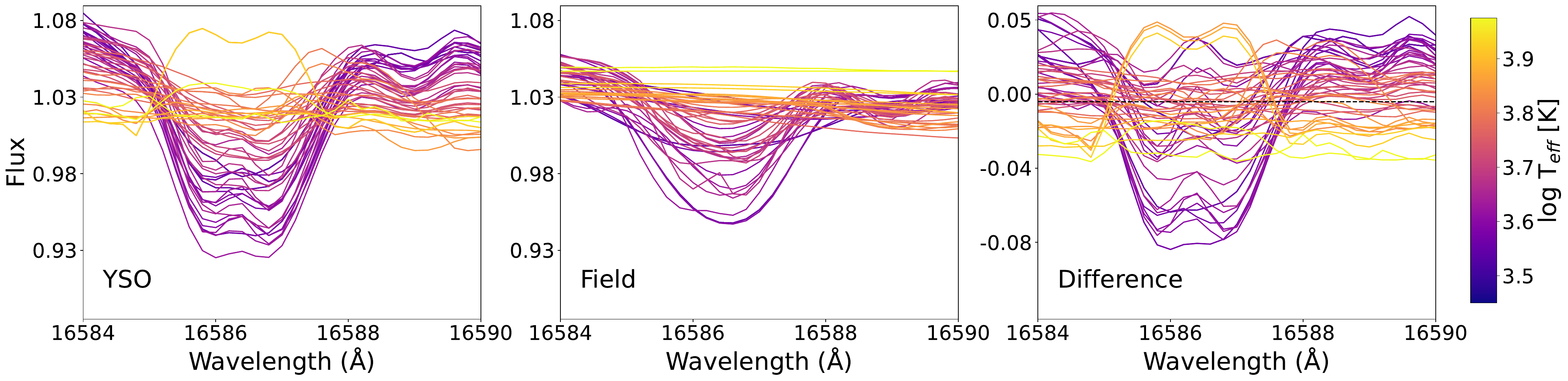}
\figsetgrpend
\figsetgrpnote{Comparison of the mean spectra of younger and older stars showing youth-sensitive features.}
\figsetend

\figsetgrpstart
\figsetgrpnum{\ref{fig:diff}.28}
\figsetgrptitle{16608--16614}
\figsetplot{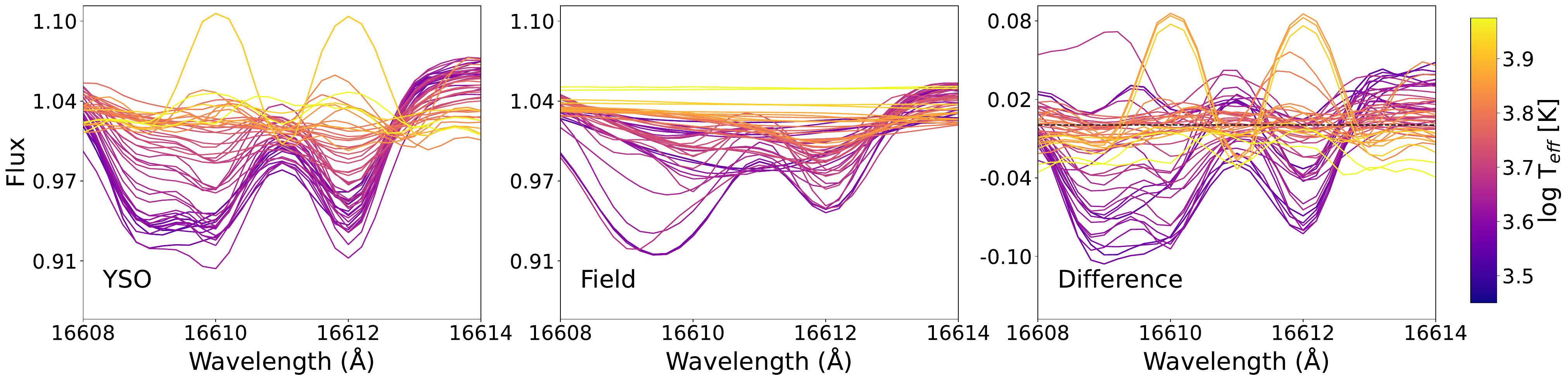}
\figsetgrpend
\figsetgrpnote{Comparison of the mean spectra of younger and older stars showing youth-sensitive features.}
\figsetend

\figsetgrpstart
\figsetgrpnum{\ref{fig:diff}.29}
\figsetgrptitle{16707--16714}
\figsetplot{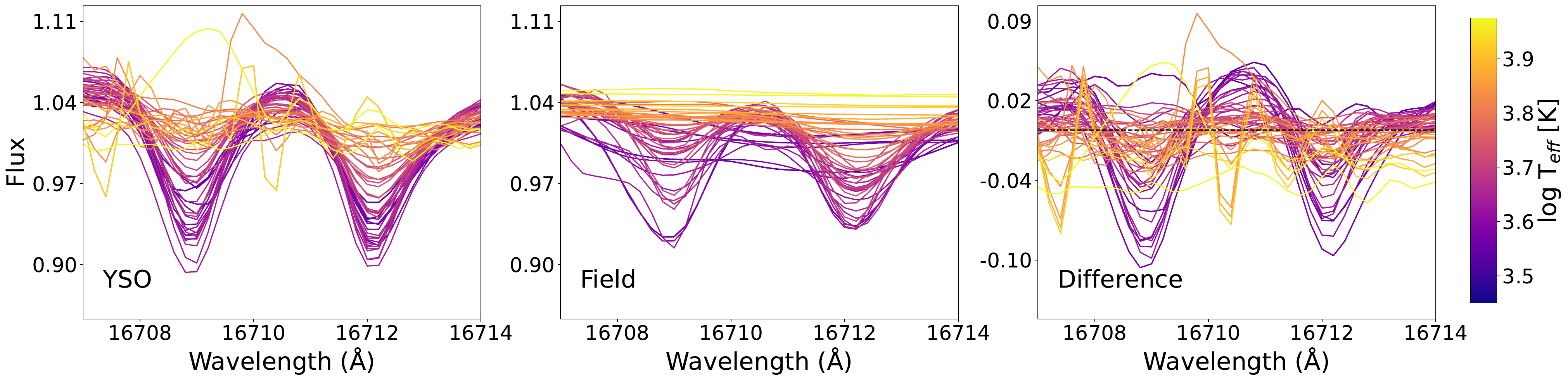}
\figsetgrpend
\figsetgrpnote{Comparison of the mean spectra of younger and older stars showing youth-sensitive features.}
\figsetend

\subsection{Evaluation of classification performance on the labeled data} \label{sec:eval}
Of the 19,498 sources labeled as YSOs in the training sample, the model recovered 13,663 sources (Figure \ref{fig:kiel}), yielding a recall (defined as a fraction of the number of true positives to the total number of YSOs in the labeled data) of 70\%---a fraction comparable to that observed in both the training set and the withheld dev set. However, because the labels themselves are not a definitive ground truth, the precision and recall values reported here should be interpreted as approximate measures of consistency with the adopted labeling scheme rather than absolute classification rates.

Unsurprisingly, given that the signatures of youth are easier to identify in low-mass stars, the recall strongly varies as a function of \teff\ (Figure \ref{fig:pr}), decreasing from almost 100\% for M dwarfs, down to $\sim$0\% for O-type stars. This decrease is, for the most part, smooth, although it does have a local minimum at $\sim$5000 K, most likely due to some residual contamination in the training set from the base of the red giant branch.

Additionally, the model identified 1,580 sources as likely young despite not being labeled as such in the training set, with a precision (defined as the ratio of true positives to all sources identified by the classifier as YSOs) of $\sim$90\%. Of these ``false positives,'' $\sim$45\% lie in regions of parameter space that were originally excluded on the basis of their \teff\ and \logg. However, because the model is not trained to predict \teff\ and \logg, but to identify candidate YSOs, they were nonetheless recovered by the model due to sufficient similarity to the youth signatures exhibited by the true positives. Indeed, many of these sources exhibit spatial and kinematic overdensities near known star-forming regions (e.g., Orion, Perseus, Ophiuchus) and were already present in the original training sample prior to the conservative \teff\, cuts. As such, the label of ``false positives'' refers merely to the sources that were originally excluded from the training set but recovered by the classifier, due to a significant fraction of these sources likely being bona fide YSOs.

The precision also varies with \teff; it is only 50\% among the coolest stars, primarily because many stars originally excluded by the isochrones are now recovered. Precision then reaches nearly 100\% at 3000 K, indicating that the original training sample is sufficient for identifying young stars, but it gradually decreases for hotter stars. It is difficult to determine to what extent the increase in ``false positives'' is driven by the difficulty of a priori identifying hotter young stars outside their membership in various moving groups and clusters, or by contamination, or by the intrinsic sparsity of the higher mass stars making them underrepresented in the training set.

These effects are also not uniform with age, which we estimate for the stars in the sample using Sagitta \citep{mcbride2021}. Sagitta is a neural network trained on the photometry of stars in nearby young clusters and moving groups with robust age measurements. This method allows the removal of systematic age-mass differences common in traditional isochrone fitting. Overall, we are able to recover stars as young as $\sim$40 Myr, but only among lower-mass stars, whereas the bulk of the sample has ages $<$10 Myr (Figure \ref{fig:age}).

The full catalog of sources identified by the classifier as likely YSOs is included in Table \ref{tab:catalog}.

\subsection{Inference on unlabeled and newly observed APOGEE spectra}\label{sec:unlabeled}

Since the work began, SDSS has continued its operations, significantly increasing the volume of the available spectra. Initially, we used spectra processed with IPL-3, which was equivalent to DR19 \citep{sdss-collaboration2025}. Since then, IPL-4 has become available. This approximately doubles the number of objects for which APOGEE spectra are available for analysis. This also provides a good test of the model's robustness on unlabeled data by ensuring that the resulting catalog of predictions appears to be qualitatively similar to the predictions obtained in the labeled data; that the sources predominantly appear to be pre-main sequence based on the bulk examination of their properties and their distribution on the sky, and that the catalog does not include a significant fraction of obvious contaminants.

In total, excluding sources that were present in the original dataset, we identify an additional 14,426 likely young stars in a catalog of $\sim1.3$ million stars (Figure \ref{fig:labels}). These candidates are included in Table \ref{tab:catalog}. Of them, 12,621 stars (87\%) have been targeted by ABYSS, indicating successful generalization of the model, as no targeting information was provided to the classifier; the only inputs are the raw spectra. Among the remaining 1,805 stars not originally targeted by ABYSS, we find strong clustering in the vicinity of Orion, NGC 2264, Cygnus, Camelopardalis, and a few other known star-forming regions.

\section{Discussion: Near-IR signatures of youth} \label{sec:discussion}

\begin{deluxetable*}{ccccccccc}[ht]
\tablecaption{Central wavelengths, widths, and closest known spectral line identifications of spectral features identified from large residuals between YSO coadded templates and artificially broadened field-star templates. The spectral line identifications were adapted from \textcite{tang2024} and \textcite{rayner2009}.
\label{tab:ave}}
\tabletypesize{\scriptsize}
\tablewidth{\linewidth}
\tablehead{
 \colhead{Center $\lambda$ (\AA)} &
 \colhead{Width (\AA)} &
 \colhead{Element} &
 \colhead{Center $\lambda$ (\AA)} &
 \colhead{Width (\AA)} &
 \colhead{Element} &
 \colhead{Center $\lambda$ (\AA)} &
 \colhead{Width (\AA)} &
 \colhead{Element}
}
\startdata
15282.7321 & 1.5 & ... & 15285.2305 & 2.5 & ... & 15296.9185 & 2.0 & ... \\
15332.5414 & 2.0 & ... & 15339.2336 & 1.5 & ... & 15340.2457 & 1.0 & ... \\
15395.4402 & 3.0 & ... & 15399.4488 & 3.0 & ... & 15411.2405 & 1.5 & ... \\
15413.4499 & 3.0 & ... & 15474.4463 & 3.0 & ... & 15477.2475 & 1.5 & ... \\
15500.9447 & 2.0 & ... & 15507.9480 & 2.0 & ... & 15509.9533 & 2.0 & ... \\
15572.9616 & 4.0 & ... & 15576.4582 & 3.0 & ... & 15606.8557 & 1.7 & ... \\
15608.6641 & 1.8 & Fe I & 15631.4657 & 3.0 & OH & 15656.9477 & 4.0 & Fe I\\
15745.0558 & 5.0 & Mg I & 15753.3819 & 5.3 & Mg I & 15781.1800 & 1.7 & \\
15783.1789 & 1.7 & ... & 15894.7265 & 4.7 & ... & 15964.4927 & 5.0 & Si I\\
16032.5098 & 5.0 & ... & 16042.2995 & 3.3 & Fe I & 16057.3061 & 3.3 & ... \\
16060.0093 & 2.0 & ... & 16064.5033 & 1.9 & Si I & 16066.2121 & 2.5 & ... \\
16078.7076 & 4.5 & ... & 16141.4170 & 3.2 & ... & 16143.7176 & 2.5 & ... \\
16155.2185 & 2.5 & Ca I & 16208.7192 & 2.5 & Fe I & 16211.7174 & 3.5 & ... \\
16317.2375 & 2.5 & ... & 16320.9273 & 3.0 & Fe I & 16369.2456 & 3.5 & ... \\
16372.5555 & 3.0 & ... & 16385.0274 & 2.0 & ... & 16387.2461 & 1.5 & ... \\
16388.5564 & 3.0 & ... & 16528.5564 & 3.0 & ... & 16530.8799 & 3.7 & ... \\
16551.2545 & 1.5 & ... & 16553.5596 & 3.0 & ... & 16586.5571 & 3.0 & ... \\
16609.5568 & 3.0 & ... & 16612.1637 & 1.7 & ... & 16709.0679 & 4.0 & ... \\
16712.2692 & 1.5 & ... & 16719.1679 & 2.3 & ... & 16723.5610 & 3.0 & Al I \\
\enddata
\end{deluxetable*}

While a number of spectroscopic signatures of youth are utilized in the literature for optical spectra, few robust counterparts have been established in the near-IR. While \logg\ can potentially be used as a criterion to select for youth for some stars, it quickly loses effectiveness for early K dwarfs and higher-mass stars, as they reach the main sequence quickly. The model we present in this work struggles to identify a number of bona fide YSOs, and, as described earlier, this effect is \teff-dependent. However, the classifier nonetheless retains some discriminatory power across the full \teff\ range, including in the range where \logg\ provides no discrimination. Thus, the question arises: which spectral features is the model using when deciding whether a particular spectrum belongs to a young star or to an older field source?

To answer this question, we used the empirical templates described in Section \ref{sec:templates} and compared representative spectra of sources identified as YSOs and field stars sharing the same \teff, \logg, and [Fe/H] bin. One of the signatures of youth becomes immediately apparent: spectra of YSOs are significantly more broadened than the spectra of field stars of the same type (Figure \ref{fig:rot}). When stars are born, they have high angular momentum, which they slowly lose as they continue to age \citep[e.g.,][]{serna2021,kounkel2022a,serna2024}. Thus, although the measurement of the rotational velocity \vsini\ does depend on the inclination, assuming an isotropic distribution, it is going to be on average higher for YSOs than it is for field (typically $>1$ Gyr old) stars. We also note that in addition to rotation, spectroscopic lines of young stars can be broadened magnetically due to Zeeman effects \citep{han2023a}.

Given the prominence of the effect of \vsini\ across the entire spectrum, we next attempted to isolate additional signatures of youth beyond rotational broadening alone. To do so, we artificially broadened the empirical templates of the field stars by convolving them with a rotational kernel, ensuring a good global fit to the template of the young stars. We find that, for stars with \teff $< 6300$ K, the typical best-fit \vsini\; is $\sim20$ km/s. We then compared the resulting templates and identified large residuals. In total, 57 features were identified (Table \ref{tab:ave}). This procedure should be regarded as a first-order correction rather than a complete physical model, because structured residuals may also be affected by imperfect broadening kernels, unresolved blends, or modest mismatches in the adopted atmospheric parameters. Even so, the persistence of these residuals across multiple temperature bins suggests that they are not purely stochastic.

We compare these features against the list of spectral lines that fall within the APOGEE bandpass \citep{shetrone2015}. Although the definitive identification of all of the lines is non-trivial, the residuals do not appear to be dominated by a single species, nor do they obviously point to a simple abundance anomaly associated with recent star formation. Rather, these features are more likely attributed to the heightened magnetic activity of young stars. Given that young stars tend to be strongly spotted \citep[often with $>$50\% of their photosphere covered by a cooler spot, e.g.,][]{gully-santiago2017}, the spectra of young stars consist not of a single distinct \teff\ profile, but of multiple. It is often possible to fit these multiple templates to the spectra of young stars \citep[e.g.,][]{perez-paolino2024}, and the effect from the cooler spots is particularly apparent in the near-IR. On the other hand, more evolved field stars would have significantly reduced magnetic activity, and they would have only weak spots that would not produce as pronounced a signature in the spectra \citep{morris2020,kounkel2022a}. We therefore regard enhanced magnetic activity resulting in spotted photospheres as the most plausible working interpretation for these near-IR residuals, particularly considering that some of these features are known to be strongly sensitive to \teff\ \citep{tang2024}. The presence of cool starspots would alter typical line ratios relative to single-temperature stars. This is because individual lines can have different strengths in the starspots or be completely absent, in which case the strength of the line in the observed spectrum would decrease \citep{2025ApJ...990..205P}. Hence, in most cases, the residuals would be relatively weak, and only some lines (which are stronger at lower temperatures) would be particularly strong \citep{cao2022a}.

More broadly, these results suggest that APOGEE spectra contain astrophysically meaningful signatures of youth beyond the classical optical diagnostics. Even if some of the individual lines remain unidentified, the combination of rotational broadening and spot-related residual structure appears to provide useful discriminatory power, particularly in regimes where \logg\ alone is insufficient. This interpretation also motivates future efforts to build more physically informed classifiers that jointly account for rotation, magnetic activity, and composite photospheres.

\section{Conclusions} \label{sec:conclusions}

In this work, we have developed a convolutional neural network classifier for APOGEE spectra to identify young stars and the signatures of youth used for this classification. Overall, the resulting classifier is effective at identifying young stars primarily among cooler stars, with the best performance being for cooler M and K dwarfs due to their pre-main-sequence evolution being the longest and their exhibiting the most prominent features of youth. The classifier's performance (both precision and recall) decreases with increasing temperature, missing bona fide young stars, and biasing the sample toward sources exhibiting the strongest youth indicators.

There are several features that the classifier likely uses to distinguish young stars from more evolved field stars, even when they have similar \teff\ and \logg. The first is line broadening due to high \vsini, which is common in young stars but less so in field stars, as stellar rotation slows down with age.

The second is that young stars commonly exhibit strong spots on their photospheres; rather than having a single temperature profile across the entire photosphere, their emergent spectra reflect a mixture of components at distinct temperatures. As such, there are several spectral lines which may be uncharacteristically strong for an evolved star of a particular \teff, because a cooler spot has contributed these lines. We assemble a list of the strongest magnetically sensitive features we have identified by comparing younger and more evolved stars with comparable spectroscopic parameters.

Overall, this work develops an effective classifier for identifying pre-main-sequence stars. Although the labels and the resulting predictions are not free of uncertainty, the presistence of the above features in the stacked spectra do demonstrate that the spectra of young stars are systematically distinct from the spectra of the more evolved stars, and it allows for the creation of a more robust catalog of young stars that minimizes contamination from more evolved sources. As SDSS continues operations, amassing a census of $\sim10^5$ candidate YSOs, such a classifier becomes particularly useful for developing a homogeneous catalog of spectroscopically selected likely young sources, providing a foundation for precise studies of star formation, stellar evolution, and the structure of young populations across the Galaxy.

\begin{appendix}
\section{BOSS classifier} \label{sec:appendix}

Previously, \citet{saad2024} developed a classifier to identify young stars in BOSS spectra; however, it used various equivalent widths and colors to make this classification. Since BOSS spectra are low-resolution, measuring equivalent widths reliably is more difficult, and the extraction process introduces an additional complication. Additionally, the spectrum as a whole contains much more information than a set of discrete lines. Furthermore, the inclusion of colors may introduce biases in characterizing spectral properties \citep{sprague2022,sizemore2024}. 

To combat these issues, we trained a new classifier. It operates similarly to the one for APOGEE described in this work, but ingests the entire BOSS spectrum instead. The labels used in training were also derived in the same manner as for APOGEE: using ABYSS targeting as a start, making cuts on the sample in \teff\ and \logg\ parameter space to exclude the most obvious field stars, manually evaluating ``false positives'' that exhibit strong clustering, and repeating the process multiple times until the final model achieved stable performance. 

In the final model, among 1.26 million sources, 23,154 were identified as likely YSOs (included in Table \ref{tab:catalog}). Of them, 2,096 were classified as ``false positives,'' for an overall precision of $\sim$91\%, i.e., comparable to APOGEE. However, there were also 24,288 false negatives, for a recall of 46\%, significantly lower than for APOGEE. Examining recall as a function of \teff, it appears to have a profile similar to that shown in Figure \ref{fig:pr}, but with a maximum recall of $\sim$80\% for cool stars, due to a reduced ability to distinguish older YSOs from the field.

This is likely due to two reasons. First, BOSS spectra are of lower resolution; therefore, rotational broadening cannot be reliably measured with this instrument. Second, these spectra are in the optical regime; therefore, the contribution from spots is weaker than in the NIR. As such, although optical spectra contain more recognizable features of youth, APOGEE spectra appear to be more efficient at recovering young stars than BOSS, at least within the framework adopted here.

\begin{figure*}    
\epsscale{1.0}
		\gridline{\fig{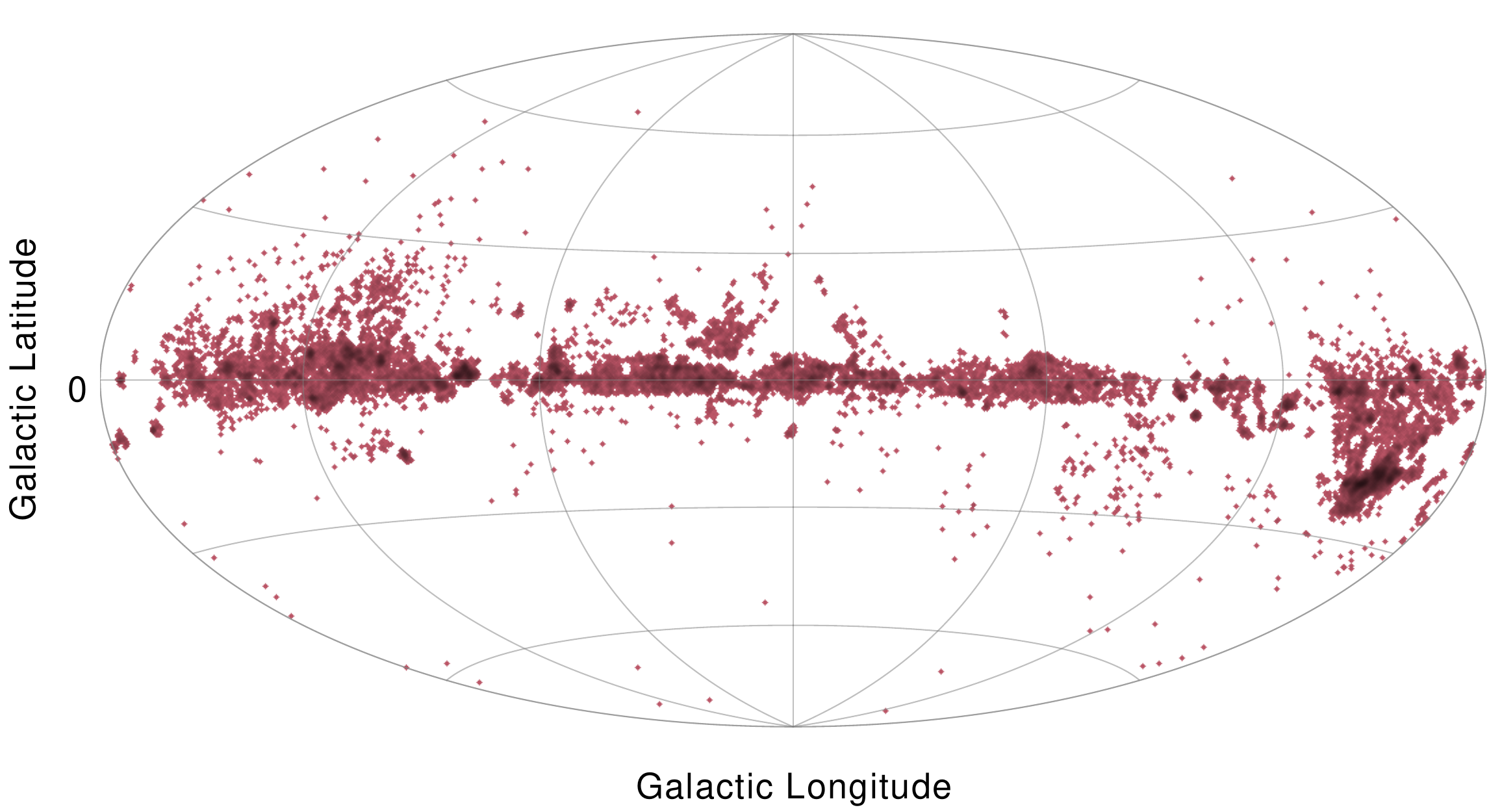}{1\textwidth}{}
        }\vspace{-1cm}
		\gridline{\fig{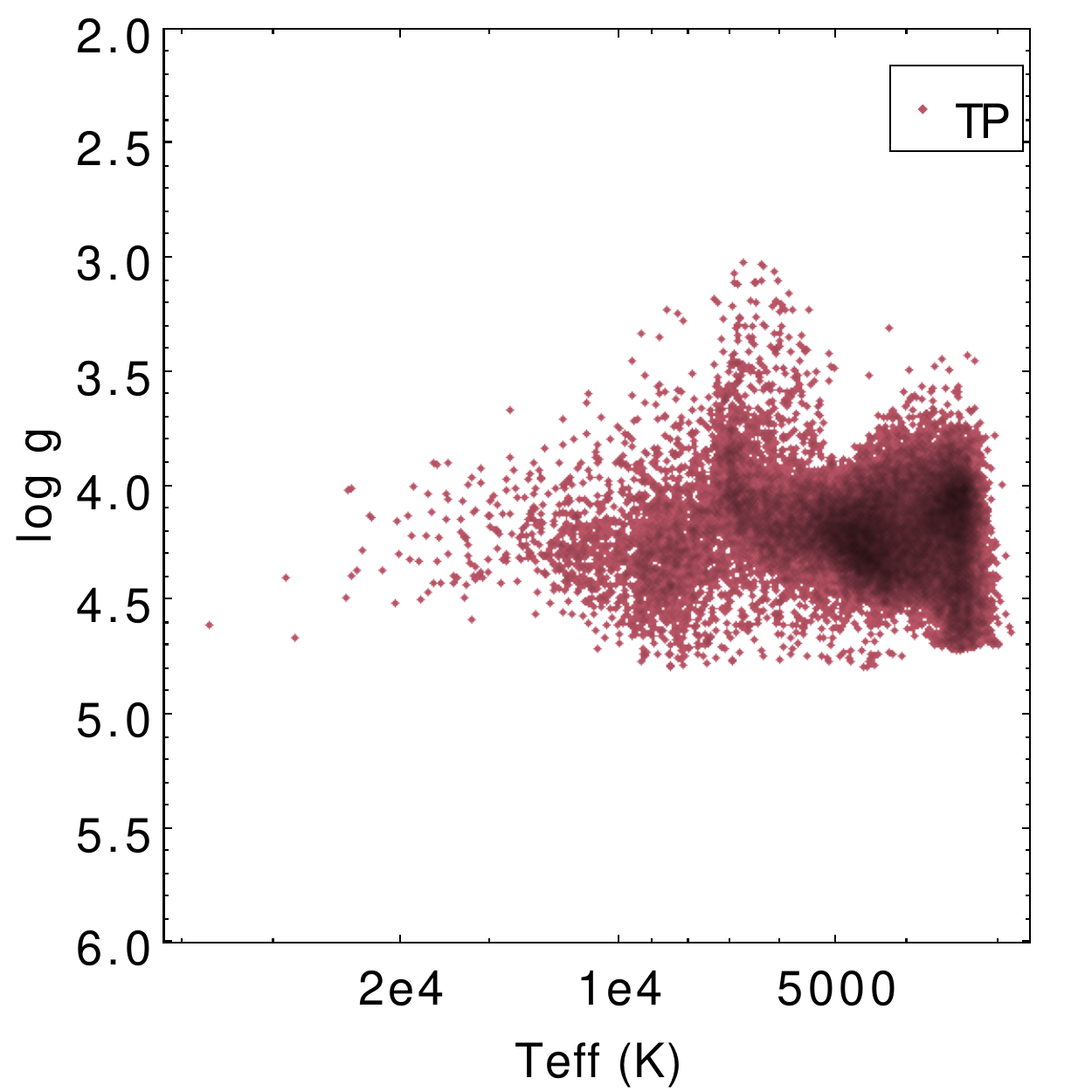}{0.33\textwidth}{}
		          \fig{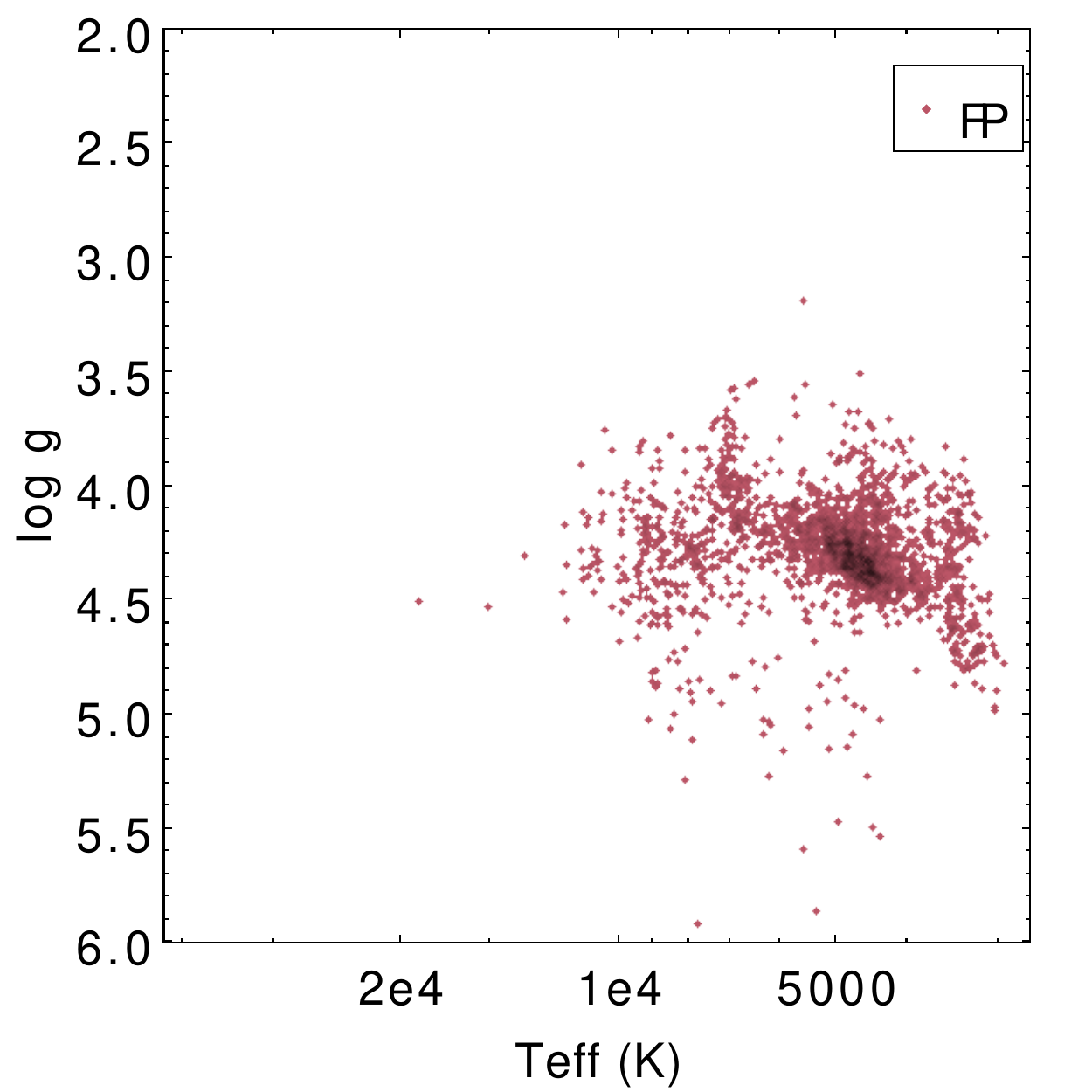}{0.33\textwidth}{}
                \fig{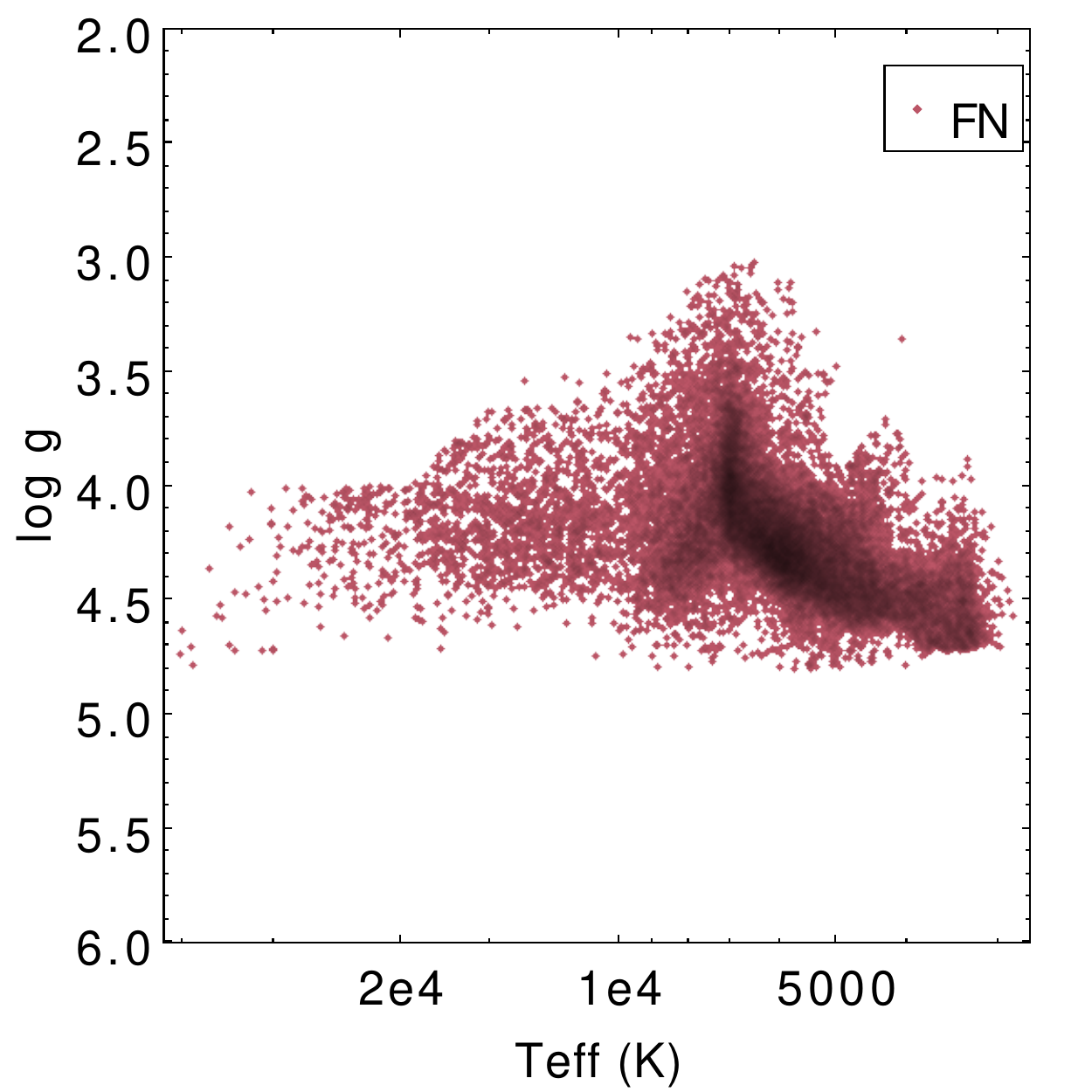}{0.33\textwidth}{}
        }\vspace{-1cm}
\caption{Sample of likely young stars identified by the classifier using BOSS spectra. The top panel shows the distribution of the identified objects in the plane of the sky, and the bottom three panels show the true positives, false positives, and false negatives obtained by comparing the predictions to the manually constructed labels.
\label{fig:boss}}
\end{figure*}







\end{appendix}




\software{TOPCAT \citep{topcat}, Tensorflow \citep{tensorflow}, YSO Classifier \citep{ysoclassifier}}

\begin{acknowledgments}

C. R-Z. acknowledges support from project UNAM-DGAPA-PAPIIT IN107226

Funding for the Sloan Digital Sky Survey V has been provided by the Alfred P. Sloan Foundation, the Heising-Simons Foundation, the National Science Foundation, and the Participating Institutions. SDSS acknowledges support and resources from the Center for High-Performance Computing at the University of Utah. SDSS telescopes are located at Apache Point Observatory, funded by the Astrophysical Research Consortium and operated by New Mexico State University, and at Las Campanas Observatory, operated by the Carnegie Institution for Science. The SDSS web site is \url{www.sdss.org}.

SDSS is managed by the Astrophysical Research Consortium for the Participating Institutions of the SDSS Collaboration, including the Carnegie Institution for Science, Chilean National Time Allocation Committee (CNTAC) ratified researchers, Caltech, the Gotham Participation Group, Harvard University, Heidelberg University, The Flatiron Institute, The Johns Hopkins University, L'Ecole polytechnique f\'{e}d\'{e}rale de Lausanne (EPFL), Leibniz-Institut f\"{u}r Astrophysik Potsdam (AIP), Max-Planck-Institut f\"{u}r Astronomie (MPIA Heidelberg), Max-Planck-Institut f\"{u}r Extraterrestrische Physik (MPE), Nanjing University, National Astronomical Observatories of China (NAOC), New Mexico State University, The Ohio State University, Pennsylvania State University, Smithsonian Astrophysical Observatory, Space Telescope Science Institute (STScI), the Stellar Astrophysics Participation Group, Universidad Nacional Aut\'{o}noma de M\'{e}xico, University of Arizona, University of Colorado Boulder, University of Illinois at Urbana-Champaign, University of Toronto, University of Utah, University of Virginia, Yale University, and Yunnan University. 

\end{acknowledgments}

\bibliographystyle{aasjournal.bst}
\bibliography{references.bib}

@article{2025ApJ...990..205P,
	adsurl = {https://ui.adsabs.harvard.edu/abs/2025ApJ...990..205P},
	archiveprefix = {arXiv},
	author = {{P{\'e}rez Paolino}, Facundo and {Bary}, Jeffrey S. and {Hillenbrand}, Lynne A. and {Horner}, Benjamin and {Carvalho}, Adolfo},
	doi = {10.3847/1538-4357/adf6ad},
	eid = {205},
	eprint = {2505.10837},
	journal = {\apj},
	month = sep,
	number = {2},
	pages = {205},
	primaryclass = {astro-ph.SR},
	title = {{Spectral Biases, Starspot Morphology, and Dynamo Transitions on the Pre-main Sequence: Insights from the X-Shooter WTTS Library}},
	volume = {990},
	year = 2025}

@software{ysoclassifier,
	author = {Bonilla Villalobos, Valentina and Kounkel, Marina},
	doi = {10.5281/zenodo.21225274},
	month = jul,
	publisher = {Zenodo},
	title = {Classifier of young stars in APOGEE \& BOSS spectra},
	url = {https://doi.org/10.5281/zenodo.21225274},
	year = 2026}

@article{tang2024,
	adsurl = {https://ui.adsabs.harvard.edu/abs/2024ApJ...973..124T},
	archiveprefix = {arXiv},
	author = {{Tang}, Shih-Yun and {Johns-Krull}, Christopher M. and {Prato}, L. and {Stahl}, Asa G.},
	doi = {10.3847/1538-4357/ad5e7f},
	eid = {124},
	eprint = {2406.13011},
	journal = {\apj},
	month = oct,
	number = {2},
	pages = {124},
	primaryclass = {astro-ph.SR},
	title = {{Measuring the Spot Variability of T Tauri Stars Using Near-infrared Atomic Fe and Molecular OH Lines}},
	volume = {973},
	year = 2024}

@article{morris2020,
	adsurl = {https://ui.adsabs.harvard.edu/abs/2020ApJ...893...67M},
	archiveprefix = {arXiv},
	author = {{Morris}, Brett M.},
	doi = {10.3847/1538-4357/ab79a0},
	eid = {67},
	eprint = {2002.09135},
	journal = {\apj},
	month = apr,
	number = {1},
	pages = {67},
	primaryclass = {astro-ph.SR},
	title = {{A Relationship between Stellar Age and Spot Coverage}},
	volume = {893},
	year = 2020}

@article{han2023a,
	adsurl = {https://ui.adsabs.harvard.edu/abs/2023AJ....166....4H},
	author = {{Han}, Eunkyu and {L{\'o}pez-Valdivia}, Ricardo and {Mace}, Gregory N. and {Jaffe}, Daniel T.},
	doi = {10.3847/1538-3881/acd2dd},
	eid = {4},
	journal = {\aj},
	month = jul,
	number = {1},
	pages = {4},
	title = {{Magnetic Field Measurements of Low-mass Stars from High-resolution Near-infrared IGRINS Spectra}},
	volume = {166},
	year = 2023}

@article{armstrong2025,
	adsurl = {https://ui.adsabs.harvard.edu/abs/2025MNRAS.543.2349A},
	archiveprefix = {arXiv},
	author = {{Armstrong}, Joseph J. and {Tan}, Jonathan C. and {Wright}, Nicholas J. and {Jeffries}, R.~D. and {Kos}, Janez and {Fiorellino}, E. and {Buder}, Sven and {Barrios L{\'o}pez}, D.},
	doi = {10.1093/mnras/staf1490},
	eprint = {2505.03716},
	journal = {\mnras},
	month = nov,
	number = {3},
	pages = {2349-2373},
	primaryclass = {astro-ph.SR},
	title = {{Investigating the Upper Scorpius OB association with HERMES ─ I. The spectroscopic sample and 6D kinematics}},
	volume = {543},
	year = 2025}

@article{serna2024,
	adsurl = {https://ui.adsabs.harvard.edu/abs/2024ApJ...968...68S},
	archiveprefix = {arXiv},
	author = {{Serna}, Javier and {Pinz{\'o}n}, Giovanni and {Hern{\'a}ndez}, Jes{\'u}s and {Manzo-Mart{\'\i}nez}, Ezequiel and {Mauco}, Karina and {Rom{\'a}n-Z{\'u}{\~n}iga}, Carlos G. and {Calvet}, Nuria and {Brice{\~n}o}, Cesar and {L{\'o}pez-Valdivia}, Ricardo and {Kounkel}, Marina and {Stringfellow}, Guy S. and {Stassun}, Keivan G. and {Pinsonneault}, Marc and {Adame}, Lucia and {Cao}, Lyra and {Covey}, Kevin and {Bayo}, Amelia and {Roman-Lopes}, Alexandre and {Nitschelm}, Christian and {Lane}, Richard R.},
	doi = {10.3847/1538-4357/ad3a6b},
	eid = {68},
	eprint = {2403.07505},
	journal = {\apj},
	month = jun,
	number = {2},
	pages = {68},
	primaryclass = {astro-ph.SR},
	title = {{Rotational Evolution of Classical T Tauri Stars: Models and Observations}},
	volume = {968},
	year = 2024}

@article{rayner2009,
	adsurl = {https://ui.adsabs.harvard.edu/abs/2009ApJS..185..289R},
	archiveprefix = {arXiv},
	author = {{Rayner}, John T. and {Cushing}, Michael C. and {Vacca}, William D.},
	doi = {10.1088/0067-0049/185/2/289},
	eprint = {0909.0818},
	journal = {\apjs},
	month = dec,
	number = {2},
	pages = {289-432},
	primaryclass = {astro-ph.SR},
	title = {{The Infrared Telescope Facility (IRTF) Spectral Library: Cool Stars}},
	volume = {185},
	year = 2009}

@article{savitzky1964,
	author = {Savitzky, Abraham. and Golay, M. J. E.},
	doi = {10.1021/ac60214a047},
	eprint = {https://doi.org/10.1021/ac60214a047},
	journal = {Analytical Chemistry},
	number = {8},
	pages = {1627-1639},
	title = {Smoothing and Differentiation of Data by Simplified Least Squares Procedures.},
	url = {https://doi.org/10.1021/ac60214a047},
	volume = {36},
	year = {1964}}

@misc{tensorflow,
	author = {Mart\'{i}n~Abadi and Ashish~Agarwal and Paul~Barham and Eugene~Brevdo and Zhifeng~Chen and Craig~Citro and Greg~S.~Corrado and Andy~Davis and Jeffrey~Dean and Matthieu~Devin and Sanjay~Ghemawat and Ian~Goodfellow and Andrew~Harp and Geoffrey~Irving and Michael~Isard and Yangqing Jia and Rafal~Jozefowicz and Lukasz~Kaiser and Manjunath~Kudlur and Josh~Levenberg and Dandelion~Man\'{e} and Rajat~Monga and Sherry~Moore and Derek~Murray and Chris~Olah and Mike~Schuster and Jonathon~Shlens and Benoit~Steiner and Ilya~Sutskever and Kunal~Talwar and Paul~Tucker and Vincent~Vanhoucke and Vijay~Vasudevan and Fernanda~Vi\'{e}gas and Oriol~Vinyals and Pete~Warden and Martin~Wattenberg and Martin~Wicke and Yuan~Yu and Xiaoqiang~Zheng},
	note = {Software available from tensorflow.org},
	title = {{TensorFlow}: Large-Scale Machine Learning on Heterogeneous Systems},
	url = {https://www.tensorflow.org/},
	year = {2015}}

@article{kollmeier2026,
	adsurl = {https://ui.adsabs.harvard.edu/abs/2026AJ....171...52K},
	archiveprefix = {arXiv},
	author = {{Kollmeier}, Juna A. and {Rix}, Hans-Walter and {Aerts}, Conny and {Aird}, James and {Vera Alfaro}, Pablo and {Almeida}, Andr{\'e}s and {Anderson}, Scott F. and {Arseneau}, Stefan M. and {Assef}, Roberto J. and {Aviram}, Shir and {Aydar}, Catarina and {Badenes}, Carles and {Bandyopadhyay}, Avrajit and {Barger}, Kat and {Barkhouser}, Robert H. and {Bauer}, Franz E. and {Behmard}, Aida and {Bender}, Chad and {Besser}, Felipe and {Bhattarai}, Binod and {Bilgi}, Pavaman and {Bird}, Jonathan and {Bizyaev}, Dmitry and {Blanc}, Guillermo A. and {Blanton}, Michael R. and {Bochanski}, John and {Bovy}, Jo and {Brandon}, Christopher and {Brandt}, William Nielsen and {Brownstein}, Joel R. and {Buchner}, Johannes and {Burchett}, Joseph N. and {Carlberg}, Joleen and {Casey}, Andrew R. and {Castaneda-Carlos}, Lesly and {Chakraborty}, Priyanka and {Chanam{\'e}}, Julio and {Chandra}, Vedant and {Cherinka}, Brian and {Chilingarian}, Igor and {Comparat}, Johan and {Cosens}, Maren and {Covey}, Kevin and {Crane}, Jeffrey D. and {Crumpler}, Nicole R. and {Cruz-Gonzalez}, Irene and {Cunha}, Katia and {Cunningham}, Tim and {Dai}, Xinyu and {Darling}, Jeremy and {Davidson}, Jr., James W. and {Davis}, Megan C. and {De Lee}, Nathan and {Deacon}, Niall and {M{\'e}ndez Delgado}, Jos{\'e} Eduardo and {Demasi}, Sebastian and {Demianenko}, Mariia and {Derwent}, Mark and {D'Onghia}, Elena and {Di Mille}, Francesco and {Dias}, Bruno and {Donor}, John and {Dow}, Peter N. and {Drory}, Niv and {Dwelly}, Tom and {Egorov}, Oleg and {Egorova}, Evgeniya and {El-Badry}, Kareem and {Engelman}, Mike and {Eracleous}, Mike and {Fan}, Xiaohui and {Farr}, Emily and {Fries}, Logan and {Frinchaboy}, Peter and {Froning}, Cynthia S. and {G{\"a}nsicke}, Boris T. and {Garc{\'\i}a}, Pablo and {Gelfand}, Joseph and {Gentile Fusillo}, Nicola Pietro and {Glover}, Simon and {Grabowski}, Katie and {Grebel}, Eva K. and {Green}, Paul J. and {Grier}, Catherine and {Gupta}, Pramod and {Gray}, Aidan C. and {H{\"a}berle}, Maximilian and {Hall}, Patrick B. and {Hammond}, Randolph P. and {Hawkins}, Keith and {Harding}, Albert C. and {Heged{\H{u}}s}, Viola and {Herbst}, Tom and {Hermes}, J.~J. and {Rodr{\'\i}guez Hidalgo}, Paola and {Hilder}, Thomas and {Hogg}, David W. and {Holtzman}, Jon A. and {Horta}, Danny and {Huang}, Yang and {Hwang}, Hsiang-Chih and {Ibarra-Medel}, Hector Javier and {Imig}, Julie and {Inight}, Keith and {Jana}, Arghajit and {Ji}, Alexander P. and {Jim{\'e}nez-Arranz}, {\'O}scar and {Jofre}, Paula and {Johns}, Matt and {Johnson}, Jennifer and {Johnson}, James W. and {Johnston}, Evelyn J. and {Jones}, Amy M. and {Katkov}, Ivan and {Knapp}, Gillian R. and {Koekemoer}, Anton M. and {Kounkel}, Marina and {Kreckel}, Kathryn and {Krishnarao}, Dhanesh and {Krumpe}, Mirko and {Kumari}, Nimisha and {Kupfer}, Thomas and {Lacerna}, Ivan and {Laporte}, Chervin and {Lepine}, Sebastien and {Li}, Jing and {Liu}, Xin and {Loebman}, Sarah and {Long}, Knox and {Roman-Lopes}, Alexandre and {Lu}, Yuxi and {Majewski}, Steven Raymond and {Maoz}, Dan and {McKinnon}, Kevin A. and {Medan}, Ilija and {Merloni}, Andrea and {Minniti}, Dante and {Morrison}, Sean and {Myers}, Natalie and {M{\'e}sz{\'a}ros}, Szabolcs and {Nandra}, Kirpal and {Nayak}, Prasanta K. and {Ness}, Melissa K. and {Nidever}, David L. and {O'Brien}, Thomas and {Oeur}, Micah and {Oravetz}, Audrey and {Oravetz}, Daniel and {Otto}, Jonah and {Pallathadka}, Gautham Adamane and {Palunas}, Povilas and {Pan}, Kaike and {Pappalardo}, Daniel and {Pandey}, Rakesh and {Negrete Pe{\~n}aloza}, Castalia Alenka and {Pinsonneault}, Marc H. and {Pogge}, Richard W. and {Taghizadeh Popp}, Manuchehr and {Price-Whelan}, Adrian M. and {Pulatova}, Nadiia and {Qiu}, Dan and {Ramirez}, Solange and {Rankine}, Amy and {Ricci}, Claudio and {Runnoe}, Jessie C. and {Sanchez}, Sebastian and {Salvato}, Mara and {Sarbadhicary}, Sumit K. and {Sattler}, Natascha and {Saydjari}, Andrew K. and {Sayres}, Conor and {Schinnerer}, Eva and {Schlaufman}, Kevin C. and {Schneider}, Donald P. and {Schreiber}, Matthias R. and {Schwope}, Axel and {Serna}, Javier and {Shen}, Yue and {Sif{\'o}n}, Crist{\'o}bal and {Singh}, Amrita and {Sinha}, Amaya and {Smee}, Stephen and {Song}, Ying-Yi and {Souto}, Diogo and {Stassun}, Keivan G. and {Steinmetz}, Matthias and {Stone-Martinez}, Alexander and {Stringfellow}, Guy and {Stutz}, Amelia and {S{\'a}nchez-Gallego}, Jos{\'e} and {Tan}, Jonathan C. and {Tayar}, Jamie and {Thai}, Riley and {Thakar}, Ani and {Ting}, Yuan-Sen and {Tkachenko}, Andrew and {Tovmassian}, Gagik and {Trakhtenbrot}, Benny and {Fern{\'a}ndez-Trincado}, Jos{\'e} G. and {Troup}, Nicholas},
	doi = {10.3847/1538-3881/ae0576},
	eid = {52},
	eprint = {2507.06989},
	journal = {\aj},
	month = jan,
	number = {1},
	pages = {52},
	primaryclass = {astro-ph.IM},
	title = {{Sloan Digital Sky Survey. V. Pioneering Panoptic Spectroscopy}},
	volume = {171},
	year = 2026}

@article{sizemore2024,
	adsurl = {https://ui.adsabs.harvard.edu/abs/2024AJ....167..173S},
	archiveprefix = {arXiv},
	author = {{Sizemore}, Logan and {Llanes}, Diego and {Kounkel}, Marina and {Hutchinson}, Brian and {Stassun}, Keivan G. and {Chandra}, Vedant},
	doi = {10.3847/1538-3881/ad291d},
	eid = {173},
	eprint = {2402.05184},
	journal = {\aj},
	month = apr,
	number = {4},
	pages = {173},
	primaryclass = {astro-ph.SR},
	title = {{A Self-consistent Data-driven Model for Determining Stellar Parameters from Optical and Near-infrared Spectra}},
	volume = {167},
	year = 2024}

@article{cao2022a,
	adsurl = {https://ui.adsabs.harvard.edu/abs/2022MNRAS.517.2165C},
	archiveprefix = {arXiv},
	author = {{Cao}, Lyra and {Pinsonneault}, Marc H.},
	doi = {10.1093/mnras/stac2706},
	eprint = {2209.10549},
	journal = {\mnras},
	month = dec,
	number = {2},
	pages = {2165-2189},
	primaryclass = {astro-ph.SR},
	title = {{Star-spots and magnetism: testing the activity paradigm in the Pleiades and M67}},
	volume = {517},
	year = 2022}

@article{sdss-collaboration2025,
	adsurl = {https://ui.adsabs.harvard.edu/abs/2025arXiv250707093S},
	archiveprefix = {arXiv},
	author = {{SDSS Collaboration} and {Adamane Pallathadka}, Gautham and {Aghakhanloo}, Mojgan and {Aird}, James and {Almeida}, Andr{\'e}s and {Amrita}, Singh and {Anders}, Friedrich and {Anderson}, Scott F. and {Arseneau}, Stefan and {Gonz{\'a}lez Avila}, Consuelo and {Aviram}, Shir and {Aydar}, Catarina and {Badenes}, Carles and {Barrera-Ballesteros}, Jorge K. and {Bauer}, Franz E. and {Behmard}, Aida and {Berg}, Michelle and {Besser}, F. and {Moni Bidin}, Christian and {Bizyaev}, Dmitry and {Blanc}, Guillermo and {Blanton}, Michael R. and {Bovy}, Jo and {Brandt}, William Nielsen and {Brownstein}, Joel R. and {Buchner}, Johannes and {Bulbul}, Esra and {Burchett}, Joseph N. and {Carigi}, Leticia and {Carlberg}, Joleen K. and {Casey}, Andrew R. and {Chakraborty}, Priyanka and {Chanam{\'e}}, Julio and {Chandra}, Vedant and {Chiappini}, Cristina and {Chilingarian}, Igor and {Comparat}, Johan and {Covey}, Kevin and {Crumpler}, Nicole and {Cunha}, Katia and {D'Onghia}, Elena and {Dai}, Xinyu and {Darling}, Jeremy and {Davis}, Megan and {De Lee}, Nathan and {Deacon}, Niall and {M{\'e}ndez Delgado}, Jos{\'e} Eduardo and {Demasi}, Sebastian and {Demianenko}, Mariia and {Demke}, Delvin and {Donor}, John and {Drory}, Niv and {Villa Durango}, Monica Alejandra and {Dwelly}, Tom and {Egorov}, Oleg and {Egorova}, Evgeniya and {El-Badry}, Kareem and {Eracleous}, Mike and {Fan}, Xiaohui and {Farr}, Emily and {Finkbeiner}, Douglas P. and {Fries}, Logan and {Frinchaboy}, Peter and {Gentile Fusillo}, Nicola Pietro and {Serrano F{\'e}lix}, Luis Daniel and {Gaensicke}, Boris and {Galligan}, Emma and {Garc{\'\i}a}, Pablo and {Gelfand}, Joseph and {Grabowski}, Katie and {Grebel}, Eva and {Green}, Paul J and {Greve}, Hannah and {Grier}, Catherine and {Griffith}, Emily and {Guetzoyan}, Paloma and {Gupta}, Pramod and {Hackshaw}, Zoe and {Hall}, Patrick B. and {Hawkins}, Keith and {Heged{\H{u}}s}, Viola and {Hekker}, Saskia and {Herbst}, T.~M. and {Hermes}, J.~J. and {Hern{\'a}ndez-Garc{\'\i}a}, Lorena and {Hiremath}, Pranavi and {Hogg}, David W and {Holtzman}, Jon and {Horne}, Keith and {Horta}, Danny and {Huang}, Yang and {Hutchinson}, Brian and {H{\"a}berle}, Maximilian and {Ibarra-Medel}, Hector Javier and {Ji}, Alexander P. and {Jofre}, Paula and {Johnson}, James W. and {Johnson}, Jennifer and {Johnston}, Evelyn J. and {Kaldor}, Mary and {Katkov}, Ivan and {Khalatyan}, Arman and {Khoperskov}, Sergey and {Klessen}, Ralf and {Kluge}, Matthias and {Koekemoer}, Anton M. and {Kollmeier}, Juna A. and {Kounkel}, Marina and {Kreckel}, Kathryn and {Krishnarao}, Dhanesh and {Krumpe}, Mirko and {Lacerna}, Ivan and {Laporte}, Chervin and {Lepine}, Sebastien and {Li}, Jing and {Liang}, Fu-Heng and {Limberg}, Guilherme and {Liu}, Xin and {Loebman}, Sarah and {Long}, Knox and {Lu}, Yuxi and {Lucey}, Madeline and {Lugo-Aranda}, Alejandra Z. and {Mart{\'\i}nez Martinez-Aldama}, Mary Loli and {McKinnon}, Kevin and {Medan}, Ilija and {Merloni}, Andrea and {Morrison}, Sean and {Myers}, Natalie and {M{\'e}sz{\'a}ros}, Szabolcs and {M{\"u}ller-Horn}, Johanna and {Nepal}, Samir and {Ness}, Melissa and {Nidever}, David and {Nitschelm}, Christian and {Oravetz}, Audrey and {Otto}, Jonah and {Pan}, Kaike and {P{\'e}rez Paolino}, Facundo and {Negrete Pe{\~n}aloza}, Castalia Alenka and {Pinsonneault}, Marc and {Taghizadeh Popp}, Manuchehr and {Price-Whelan}, Adrian and {Pulatova}, Nadiia and {Queiroz}, Anna Barbara and {Raddick}, Jordan and {Rankine}, Amy and {Rix}, Hans-Walter and {Rom{\'a}n-Z{\'u}{\~n}iga}, Carlos and {Fern{\'a}ndez Rosso}, Daniela and {Runnoe}, Jessie and {Mahmud Saad}, Serat and {Salvato}, Mara and {Sanchez}, Sebastian F. and {Sattler}, Natascha and {Saydjari}, Andrew and {Sayres}, Conor and {Schlaufman}, Kevin and {Schneider}, Donald P. and {Schwope}, Axel and {Seaton}, Lucas M. and {Seeburger}, Rhys and {Serna}, Javier and {Sharma}, Sanjib and {Shen}, Yue and {Sinha}, Amaya and {Sizemore}, Brian and {Sniegowska}, Marzena and {Song}, Yingyi and {Souto}, Diogo and {Stassun}, Keivan and {Steinmetz}, Matthias and {Stone}, Zachary and {Stone-Martinez}, Alexander and {Stringfellow}, Guy S. and {Mata S{\'a}nchez}, Aurora and {S{\'a}nchez-Gallego}, Jos{\'e} and {Tan}, Jonathan and {Tayar}, Jamie and {Thai}, Riley and {Thakar}, Ani and {Thibodeaux}, Pierre and {Ting}, Yuan-Sen and {Tkachenko}, Andrew and {Trakhtenbrot}, Benny and {Fernandez Trincado}, Jose G. and {Troup}, Nicholas and {Trump}, Jonathan R. and {Ulloa}, Natalie and {Van der Marel}, Roeland P. and {Vera}, Pablo and {Villanova}, Sandro and {Villase{\~n}or}, Jaime and {Wang}, Ji and {Way}, Zachary and {Weijmans}, Anne-Marie and {Wheeler}, Adam and {Wilson}, John C. and {Wofford}, Aida and {Wong}, Tony},
	eid = {arXiv:2507.07093},
	eprint = {2507.07093},
	journal = {arXiv e-prints},
	month = jul,
	pages = {arXiv:2507.07093},
	primaryclass = {astro-ph.GA},
	title = {{The Nineteenth Data Release of the Sloan Digital Sky Survey}},
	year = 2025}

@article{perez-paolino2024,
	adsurl = {https://ui.adsabs.harvard.edu/abs/2024ApJ...967...45P},
	archiveprefix = {arXiv},
	author = {{P{\'e}rez Paolino}, Facundo and {Bary}, Jeffrey S. and {Hillenbrand}, Lynne A. and {Markham}, Madison},
	doi = {10.3847/1538-4357/ad393b},
	eid = {45},
	eprint = {2403.20255},
	journal = {\apj},
	month = may,
	number = {1},
	pages = {45},
	primaryclass = {astro-ph.SR},
	title = {{The Effect of Starspots on Spectroscopic Age and Mass Estimates of Nonaccreting T Tauri Stars in the Taurus{\textendash}Auriga Star-forming Region}},
	volume = {967},
	year = 2024}

@article{gagne2024,
	adsurl = {https://ui.adsabs.harvard.edu/abs/2024arXiv240512860G},
	archiveprefix = {arXiv},
	author = {{Gagn{\'e}}, Jonathan},
	doi = {10.48550/arXiv.2405.12860},
	eid = {arXiv:2405.12860},
	eprint = {2405.12860},
	journal = {arXiv e-prints},
	month = may,
	pages = {arXiv:2405.12860},
	primaryclass = {astro-ph.SR},
	title = {{A Quick Guide to Nearby Young Association}},
	year = 2024}

@article{saad2024,
	adsurl = {https://ui.adsabs.harvard.edu/abs/2024AJ....167..125S},
	archiveprefix = {arXiv},
	author = {{Saad}, Serat and {Lane}, Kaitlyn and {Kounkel}, Marina and {Stassun}, Keivan G. and {L{\'o}pez-Valdivia}, Ricardo and {Kim}, Jinyoung Serena and {Pe{\~n}a Ram{\'\i}rez}, Karla and {Stringfellow}, Guy S. and {Rom{\'a}n-Z{\'u}{\~n}iga}, Carlos G. and {Hern{\'a}ndez}, Jes{\'u}s and {Wolk}, Scott J. and {Hillenbrand}, Lynne A.},
	doi = {10.3847/1538-3881/ad2001},
	eid = {125},
	eprint = {2401.01932},
	journal = {\aj},
	month = mar,
	number = {3},
	pages = {125},
	primaryclass = {astro-ph.SR},
	title = {{ABYSS. II. Identification of Young Stars in Optical SDSS Spectra and Their Properties}},
	volume = {167},
	year = 2024}

@article{wright2023,
	adsurl = {https://ui.adsabs.harvard.edu/abs/2023arXiv231108358W},
	archiveprefix = {arXiv},
	author = {{Wright}, Nicholas J. and {Jeffries}, R.~D. and {Jackson}, R.~J. and {Sacco}, G.~G. and {Arnold}, Becky and {Franciosini}, E. and {Gilmore}, G. and {Gonneau}, A. and {Morbidelli}, L. and {Prisinzano}, L. and {Randich}, S. and {Worley}, Clare C.},
	eid = {arXiv:2311.08358},
	eprint = {2311.08358},
	journal = {arXiv e-prints},
	month = nov,
	pages = {arXiv:2311.08358},
	primaryclass = {astro-ph.GA},
	title = {{The Gaia-ESO Survey: 3D dynamics of young groups and clusters from GES and Gaia EDR3}},
	year = 2023}

@article{kounkel2023,
	adsurl = {https://ui.adsabs.harvard.edu/abs/2023ApJS..266...10K},
	archiveprefix = {arXiv},
	author = {{Kounkel}, Marina and {Zari}, Eleonora and {Covey}, Kevin and {Tkachenko}, Andrew and {Z{\'u}{\~n}iga}, Carlos Rom{\'a}n and {Stassun}, Keivan and {Stutz}, Amelia M. and {Stringfellow}, Guy and {Roman-Lopes}, Alexandre and {Hern{\'a}ndez}, Jes{\'u}s and {Pe{\~n}a Ram{\'\i}rez}, Karla and {Bayo}, Amelia and {Kim}, Jinyoung Serena and {Cao}, Lyra and {Wolk}, Scott J. and {Kollmeier}, Juna and {L{\'o}pez-Valdivia}, Ricardo and {Rojas-Ayala}, B{\'a}rbara},
	doi = {10.3847/1538-4365/acc106},
	eid = {10},
	eprint = {2301.07186},
	journal = {\apjs},
	month = may,
	number = {1},
	pages = {10},
	primaryclass = {astro-ph.GA},
	title = {{ABYSS. I. Targeting Strategy for the APOGEE and BOSS Young Star Survey in SDSS-V}},
	volume = {266},
	year = 2023}

@article{kang2023,
	adsurl = {https://ui.adsabs.harvard.edu/abs/2023arXiv230408398K},
	archiveprefix = {arXiv},
	author = {{Kang}, Da Eun and {Ksoll}, Victor F. and {Itrich}, Dominika and {Testi}, Leonardo and {Klessen}, Ralf S. and {Hennebelle}, Patrick and {Molinari}, Sergio},
	doi = {10.48550/arXiv.2304.08398},
	eid = {arXiv:2304.08398},
	eprint = {2304.08398},
	journal = {arXiv e-prints},
	month = apr,
	pages = {arXiv:2304.08398},
	primaryclass = {astro-ph.SR},
	title = {{Spectral classification of young stars using conditional invertible neural networks I. Introducing and validating the method}},
	year = 2023}

@article{roman-zuniga2023,
	adsurl = {https://ui.adsabs.harvard.edu/abs/2023AJ....165...51R},
	archiveprefix = {arXiv},
	author = {{Rom{\'a}n-Z{\'u}{\~n}iga}, Carlos G. and {Kounkel}, Marina and {Hern{\'a}ndez}, Jes{\'u}s and {Pe{\~n}a Ram{\'\i}rez}, Karla and {L{\'o}pez-Valdivia}, Ricardo and {Covey}, Kevin R. and {Stutz}, Amelia M. and {Roman-Lopes}, Alexandre and {Campbell}, Hunter and {Khilfeh}, Elliott and {Tapia}, Mauricio and {Stringfellow}, Guy S. and {Downes}, Juan Jos{\'e} and {Stassun}, Keivan G. and {Minniti}, Dante and {Bayo}, Amelia and {Kim}, Jinyoung Serena and {Su{\'a}rez}, Genaro and {Ybarra}, Jason E. and {Fern{\'a}ndez-Trincado}, Jos{\'e} G. and {Longa-Pe{\~n}a}, Pen{\'e}lope and {Ram{\'\i}rez-Preciado}, Valeria and {Serna}, Javier and {Lane}, Richard R. and {Garc{\'\i}a-Hern{\'a}ndez}, D.~A. and {Beaton}, Rachael L. and {Bizyaev}, Dmitry and {Pan}, Kaike},
	doi = {10.3847/1538-3881/aca3a4},
	eid = {51},
	eprint = {2211.09217},
	journal = {\aj},
	month = feb,
	number = {2},
	pages = {51},
	primaryclass = {astro-ph.GA},
	title = {{Stellar Properties for a Comprehensive Collection of Star-forming Regions in the SDSS APOGEE-2 Survey}},
	volume = {165},
	year = 2023}

@article{kounkel2022a,
	adsurl = {https://ui.adsabs.harvard.edu/abs/2022AJ....164..137K},
	archiveprefix = {arXiv},
	author = {{Kounkel}, Marina and {Stassun}, Keivan G. and {Bouma}, Luke G. and {Covey}, Kevin and {Hillenbrand}, Lynne A. and {Curtis}, Jason Lee},
	doi = {10.3847/1538-3881/ac866d},
	eid = {137},
	eprint = {2206.13545},
	journal = {\aj},
	month = oct,
	number = {4},
	pages = {137},
	primaryclass = {astro-ph.SR},
	title = {{Untangling the Galaxy. IV. Empirical Constraints on Angular Momentum Evolution and Gyrochronology for Young Stars in the Field}},
	volume = {164},
	year = 2022}

@article{smee2013,
	adsurl = {https://ui.adsabs.harvard.edu/abs/2013AJ....146...32S},
	archiveprefix = {arXiv},
	author = {{Smee}, Stephen A. and {Gunn}, James E. and {Uomoto}, Alan and {Roe}, Natalie and {Schlegel}, David and {Rockosi}, Constance M. and {Carr}, Michael A. and {Leger}, French and {Dawson}, Kyle S. and {Olmstead}, Matthew D. and {Brinkmann}, Jon and {Owen}, Russell and {Barkhouser}, Robert H. and {Honscheid}, Klaus and {Harding}, Paul and {Long}, Dan and {Lupton}, Robert H. and {Loomis}, Craig and {Anderson}, Lauren and {Annis}, James and {Bernardi}, Mariangela and {Bhardwaj}, Vaishali and {Bizyaev}, Dmitry and {Bolton}, Adam S. and {Brewington}, Howard and {Briggs}, John W. and {Burles}, Scott and {Burns}, James G. and {Castander}, Francisco Javier and {Connolly}, Andrew and {Davenport}, James R.~A. and {Ebelke}, Garrett and {Epps}, Harland and {Feldman}, Paul D. and {Friedman}, Scott D. and {Frieman}, Joshua and {Heckman}, Timothy and {Hull}, Charles L. and {Knapp}, Gillian R. and {Lawrence}, David M. and {Loveday}, Jon and {Mannery}, Edward J. and {Malanushenko}, Elena and {Malanushenko}, Viktor and {Merrelli}, Aronne James and {Muna}, Demitri and {Newman}, Peter R. and {Nichol}, Robert C. and {Oravetz}, Daniel and {Pan}, Kaike and {Pope}, Adrian C. and {Ricketts}, Paul G. and {Shelden}, Alaina and {Sandford}, Dale and {Siegmund}, Walter and {Simmons}, Audrey and {Smith}, D. Shane and {Snedden}, Stephanie and {Schneider}, Donald P. and {SubbaRao}, Mark and {Tremonti}, Christy and {Waddell}, Patrick and {York}, Donald G.},
	doi = {10.1088/0004-6256/146/2/32},
	eid = {32},
	eprint = {1208.2233},
	journal = {\aj},
	month = aug,
	number = {2},
	pages = {32},
	primaryclass = {astro-ph.IM},
	title = {{The Multi-object, Fiber-fed Spectrographs for the Sloan Digital Sky Survey and the Baryon Oscillation Spectroscopic Survey}},
	volume = {146},
	year = 2013}

@article{sprague2022,
	adsurl = {https://ui.adsabs.harvard.edu/abs/2022AJ....163..152S},
	archiveprefix = {arXiv},
	author = {{Sprague}, Dani and {Culhane}, Connor and {Kounkel}, Marina and {Olney}, Richard and {Covey}, K.~R. and {Hutchinson}, Brian and {Lingg}, Ryan and {Stassun}, Keivan G. and {Rom{\'a}n-Z{\'u}{\~n}iga}, Carlos G. and {Roman-Lopes}, Alexandre and {Nidever}, David and {Beaton}, Rachael L. and {Borissova}, Jura and {Stutz}, Amelia and {Stringfellow}, Guy S. and {Ram{\'\i}rez}, Karla Pe{\~n}a and {Ram{\'\i}rez-Preciado}, Valeria and {Hern{\'a}ndez}, Jes{\'u}s and {Kim}, Jinyoung Serena and {Lane}, Richard R.},
	doi = {10.3847/1538-3881/ac4de7},
	eid = {152},
	eprint = {2201.03661},
	journal = {\aj},
	month = apr,
	number = {4},
	pages = {152},
	primaryclass = {astro-ph.GA},
	title = {{APOGEE Net: An Expanded Spectral Model of Both Low-mass and High-mass Stars}},
	volume = {163},
	year = 2022}

@article{serna2021,
	adsurl = {https://ui.adsabs.harvard.edu/abs/2021ApJ...923..177S},
	archiveprefix = {arXiv},
	author = {{Serna}, Javier and {Hernandez}, Jesus and {Kounkel}, Marina and {Manzo-Mart{\'\i}nez}, Ezequiel and {Roman-Lopes}, Alexandre and {Rom{\'a}n-Z{\'u}{\~n}iga}, Carlos G. and {Batista}, Maria Gracia and {Pinz{\'o}n}, Giovanni and {Calvet}, Nuria and {Brice{\~n}o}, Cesar and {Tapia}, Mauricio and {Su{\'a}rez}, Genaro and {Ram{\'\i}rez}, Karla Pe{\~n}a and {G. Stassun}, Keivan and {Covey}, Kevin and {Vargas-Gonz{\'a}lez}, J. and {Fern{\'a}ndez-Trincado}, Jos{\'e} G.},
	doi = {10.3847/1538-4357/ac300a},
	eid = {177},
	eprint = {2110.06431},
	journal = {\apj},
	month = dec,
	number = {2},
	pages = {177},
	primaryclass = {astro-ph.SR},
	title = {{Stellar Rotation of T Tauri Stars in the Orion Star-forming Complex}},
	volume = {923},
	year = 2021}

@article{gully-santiago2017,
	adsurl = {https://ui.adsabs.harvard.edu/abs/2017ApJ...836..200G},
	archiveprefix = {arXiv},
	author = {{Gully-Santiago}, Michael A. and {Herczeg}, Gregory J. and {Czekala}, Ian and {Somers}, Garrett and {Grankin}, Konstantin and {Covey}, Kevin R. and {Donati}, J.~F. and {Alencar}, Silvia H.~P. and {Hussain}, Gaitee A.~J. and {Shappee}, Benjamin J. and {Mace}, Gregory N. and {Lee}, Jae-Joon and {Holoien}, T.~W. -S. and {Jose}, Jessy and {Liu}, Chun-Fan},
	doi = {10.3847/1538-4357/836/2/200},
	eid = {200},
	eprint = {1701.06703},
	journal = {\apj},
	month = feb,
	number = {2},
	pages = {200},
	primaryclass = {astro-ph.SR},
	title = {{Placing the Spotted T Tauri Star LkCa 4 on an HR Diagram}},
	volume = {836},
	year = 2017}

@article{mcbride2021,
	adsurl = {https://ui.adsabs.harvard.edu/abs/2021AJ....162..282M},
	archiveprefix = {arXiv},
	author = {{McBride}, Aidan and {Lingg}, Ryan and {Kounkel}, Marina and {Covey}, Kevin and {Hutchinson}, Brian},
	doi = {10.3847/1538-3881/ac2432},
	eid = {282},
	eprint = {2012.10463},
	journal = {\aj},
	month = dec,
	number = {6},
	pages = {282},
	primaryclass = {astro-ph.SR},
	title = {{Untangling the Galaxy. III. Photometric Search for Pre-main-sequence Stars with Deep Learning}},
	volume = {162},
	year = 2021}

@article{bowen1973,
	adsurl = {https://ui.adsabs.harvard.edu/abs/1973ApOpt..12.1430B},
	author = {{Bowen}, I.~S. and {Vaughan}, A.~H., Jr.},
	doi = {10.1364/AO.12.001430},
	journal = {\ao},
	month = jan,
	pages = {1430-1434},
	title = {{The optical design of the 40-in. telescope and of the Ir{\'e}n{\'e}e DuPont telescope at Las Campanas Observatory, Chile.}},
	volume = {12},
	year = 1973}

@article{wilson2019,
	adsurl = {https://ui.adsabs.harvard.edu/abs/2019PASP..131e5001W},
	archiveprefix = {arXiv},
	author = {{Wilson}, J.~C. and {Hearty}, F.~R. and {Skrutskie}, M.~F. and {Majewski}, S.~R. and {Holtzman}, J.~A. and {Eisenstein}, D. and {Gunn}, J. and {Blank}, B. and {Henderson}, C. and {Smee}, S. and {Nelson}, M. and {Nidever}, D. and {Arns}, J. and {Barkhouser}, R. and {Barr}, J. and {Beland}, S. and {Bershady}, M.~A. and {Blanton}, M.~R. and {Brunner}, S. and {Burton}, A. and {Carey}, L. and {Carr}, M. and {Colque}, J.~P. and {Crane}, J. and {Damke}, G.~J. and {Davidson}, J.~W., Jr. and {Dean}, J. and {Di Mille}, F. and {Don}, K.~W. and {Ebelke}, G. and {Evans}, M. and {Fitzgerald}, G. and {Gillespie}, B. and {Hall}, M. and {Harding}, A. and {Harding}, P. and {Hammond}, R. and {Hancock}, D. and {Harrison}, C. and {Hope}, S. and {Horne}, T. and {Karakla}, J. and {Lam}, C. and {Leger}, F. and {MacDonald}, N. and {Maseman}, P. and {Matsunari}, J. and {Melton}, S. and {Mitcheltree}, T. and {O'Brien}, T. and {O'Connell}, R.~W. and {Patten}, A. and {Richardson}, W. and {Rieke}, G. and {Rieke}, M. and {Roman-Lopes}, A. and {Schiavon}, R.~P. and {Sobeck}, J.~S. and {Stolberg}, T. and {Stoll}, R. and {Tembe}, M. and {Trujillo}, J.~D. and {Uomoto}, A. and {Vernieri}, M. and {Walker}, E. and {Weinberg}, D.~H. and {Young}, E. and {Anthony-Brumfield}, B. and {Bizyaev}, D. and {Breslauer}, B. and {De Lee}, N. and {Downey}, J. and {Halverson}, S. and {Huehnerhoff}, J. and {Klaene}, M. and {Leon}, E. and {Long}, D. and {Mahadevan}, S. and {Malanushenko}, E. and {Nguyen}, D.~C. and {Owen}, R. and {S{\'a}nchez-Gallego}, J.~R. and {Sayres}, C. and {Shane}, N. and {Shectman}, S.~A. and {Shetrone}, M. and {Skinner}, D. and {Stauffer}, F. and {Zhao}, B.},
	doi = {10.1088/1538-3873/ab0075},
	eprint = {1902.00928},
	journal = {\pasp},
	month = may,
	number = {999},
	pages = {055001},
	primaryclass = {astro-ph.IM},
	title = {{The Apache Point Observatory Galactic Evolution Experiment (APOGEE) Spectrographs}},
	volume = {131},
	year = 2019}

@article{zerjal2019,
	adsurl = {https://ui.adsabs.harvard.edu/abs/2019MNRAS.484.4591Z},
	archiveprefix = {arXiv},
	author = {{{\v{Z}}erjal}, M. and {Ireland}, M.~J. and {Nordlander}, T. and {Lin}, J. and {Buder}, S. and {Casagrande}, L. and {{\v{C}}otar}, K. and {de Silva}, G. and {Horner}, J. and {Martell}, S. and {Traven}, G. and {Zwitter}, T. and {Galah Collaboration}},
	doi = {10.1093/mnras/stz296},
	eprint = {1810.10435},
	journal = {\mnras},
	month = apr,
	number = {4},
	pages = {4591-4600},
	primaryclass = {astro-ph.SR},
	title = {{The GALAH Survey: lithium-strong KM dwarfs}},
	volume = {484},
	year = 2019}

@article{kounkel2018a,
	adsurl = {https://ui.adsabs.harvard.edu/abs/2018AJ....156...84K},
	archiveprefix = {arXiv},
	author = {{Kounkel}, Marina and {Covey}, Kevin and {Su{\'a}rez}, Genaro and {Rom{\'a}n-Z{\'u}{\~n}iga}, Carlos and {Hernandez}, Jesus and {Stassun}, Keivan and {Jaehnig}, Karl O. and {Feigelson}, Eric D. and {Pe{\~n}a Ram{\'\i}rez}, Karla and {Roman-Lopes}, Alexandre and {Da Rio}, Nicola and {Stringfellow}, Guy S. and {Kim}, J. Serena and {Borissova}, Jura and {Fern{\'a}ndez-Trincado}, Jos{\'e} G. and {Burgasser}, Adam and {Garc{\'\i}a-Hern{\'a}ndez}, D.~A. and {Zamora}, Olga and {Pan}, Kaike and {Nitschelm}, Christian},
	doi = {10.3847/1538-3881/aad1f1},
	eid = {84},
	eprint = {1805.04649},
	journal = {\aj},
	month = {Sep},
	number = {3},
	pages = {84},
	primaryclass = {astro-ph.SR},
	title = {{The APOGEE-2 Survey of the Orion Star-forming Complex. II. Six-dimensional Structure}},
	volume = {156},
	year = {2018}}

@article{choi2016,
	adsurl = {https://ui.adsabs.harvard.edu/abs/2016ApJ...823..102C},
	archiveprefix = {arXiv},
	author = {{Choi}, Jieun and {Dotter}, Aaron and {Conroy}, Charlie and {Cantiello}, Matteo and {Paxton}, Bill and {Johnson}, Benjamin D.},
	doi = {10.3847/0004-637X/823/2/102},
	eid = {102},
	eprint = {1604.08592},
	journal = {\apj},
	month = {Jun},
	number = {2},
	pages = {102},
	primaryclass = {astro-ph.SR},
	title = {{Mesa Isochrones and Stellar Tracks (MIST). I. Solar-scaled Models}},
	volume = {823},
	year = {2016}}

@article{haisch2001,
	adsurl = {https://ui.adsabs.harvard.edu/abs/2001ApJ...553L.153H},
	archiveprefix = {arXiv},
	author = {{Haisch}, Karl E., Jr. and {Lada}, Elizabeth A. and {Lada}, Charles J.},
	doi = {10.1086/320685},
	eprint = {astro-ph/0104347},
	journal = {\apjl},
	month = {Jun},
	number = {2},
	pages = {L153-L156},
	primaryclass = {astro-ph},
	title = {{Disk Frequencies and Lifetimes in Young Clusters}},
	volume = {553},
	year = {2001}}

@article{gunn2006,
	adsurl = {http://adsabs.harvard.edu/abs/2006AJ....131.2332G},
	arxivurl = {http://arxiv.org/abs/astro-ph/0602326},
	author = {{Gunn}, J.~E. and {Siegmund}, W.~A. and {Mannery}, E.~J. and {Owen}, R.~E. and {Hull}, C.~L. and {Leger}, R.~F. and {Carey}, L.~N. and {Knapp}, G.~R. and {York}, D.~G. and {Boroski}, W.~N. and {Kent}, S.~M. and {Lupton}, R.~H. and {Rockosi}, C.~M. and {Evans}, M.~L. and {Waddell}, P. and {Anderson}, J.~E. and {Annis}, J. and {Barentine}, J.~C. and {Bartoszek}, L.~M. and {Bastian}, S. and {Bracker}, S.~B. and {Brewington}, H.~J. and {Briegel}, C.~I. and {Brinkmann}, J. and {Brown}, Y.~J. and {Carr}, M.~A. and {Czarapata}, P.~C. and {Drennan}, C.~C. and {Dombeck}, T. and {Federwitz}, G.~R. and {Gillespie}, B.~A. and {Gonzales}, C. and {Hansen}, S.~U. and {Harvanek}, M. and {Hayes}, J. and {Jordan}, W. and {Kinney}, E. and {Klaene}, M. and {Kleinman}, S.~J. and {Kron}, R.~G. and {Kresinski}, J. and {Lee}, G. and {Limmongkol}, S. and {Lindenmeyer}, C.~W. and {Long}, D.~C. and {Loomis}, C.~L. and {McGehee}, P.~M. and {Mantsch}, P.~M. and {Neilsen}, Jr., E.~H. and {Neswold}, R.~M. and {Newman}, P.~R. and {Nitta}, A. and {Peoples}, Jr., J. and {Pier}, J.~R. and {Prieto}, P.~S. and {Prosapio}, A. and {Rivetta}, C. and {Schneider}, D.~P. and {Snedden}, S. and {Wang}, S.-i.},
	doi = {10.1086/500975},
	eprint = {astro-ph/0602326},
	journal = {\aj},
	month = apr,
	pages = {2332-2359},
	title = {{The 2.5 m Telescope of the Sloan Digital Sky Survey}},
	volume = 131,
	year = 2006}

@article{shetrone2015,
	adsurl = {http://adsabs.harvard.edu/abs/2015ApJS..221...24S},
	archiveprefix = {arXiv},
	arxivurl = {http://arxiv.org/abs/1502.04080},
	author = {{Shetrone}, M. and {Bizyaev}, D. and {Lawler}, J.~E. and {Allende Prieto}, C. and {Johnson}, J.~A. and {Smith}, V.~V. and {Cunha}, K. and {Holtzman}, J. and {Garc{\'{\i}}a P{\'e}rez}, A.~E. and {M{\'e}sz{\'a}ros}, S. and {Sobeck}, J. and {Zamora}, O. and {Garc{\'{\i}}a-Hern{\'a}ndez}, D.~A. and {Souto}, D. and {Chojnowski}, D. and {Koesterke}, L. and {Majewski}, S. and {Zasowski}, G.},
	doi = {10.1088/0067-0049/221/2/24},
	eid = {24},
	eprint = {1502.04080},
	journal = {\apjs},
	month = dec,
	pages = {24},
	primaryclass = {astro-ph.IM},
	title = {{The SDSS-III APOGEE Spectral Line List for H-band Spectroscopy}},
	volume = 221,
	year = 2015}

@article{cottaar2015,
	adsurl = {http://adsabs.harvard.edu/abs/2015ApJ...807...27C},
	archiveprefix = {arXiv},
	arxivurl = {http://arxiv.org/abs/1505.07504},
	author = {{Cottaar}, M. and {Covey}, K.~R. and {Foster}, J.~B. and {Meyer}, M.~R. and {Tan}, J.~C. and {Nidever}, D.~L. and {Chojnowski}, S.~D. and {da Rio}, N. and {Flaherty}, K.~M. and {Frinchaboy}, P.~M. and {Majewski}, S. and {Skrutskie}, M.~F. and {Wilson}, J.~C. and {Zasowski}, G.},
	doi = {10.1088/0004-637X/807/1/27},
	eid = {27},
	eprint = {1505.07504},
	journal = {\apj},
	month = jul,
	pages = {27},
	primaryclass = {astro-ph.SR},
	title = {{IN-SYNC. III. The Dynamical State of IC 348 - A Super-virial Velocity Dispersion and a Puzzling Sign of Convergence}},
	volume = 807,
	year = 2015}

@article{foster2015,
	adsurl = {http://adsabs.harvard.edu/abs/2015ApJ...799..136F},
	archiveprefix = {arXiv},
	arxivurl = {http://arxiv.org/abs/1411.6013},
	author = {{Foster}, J.~B. and {Cottaar}, M. and {Covey}, K.~R. and {Arce}, H.~G. and {Meyer}, M.~R. and {Nidever}, D.~L. and {Stassun}, K.~G. and {Tan}, J.~C. and {Chojnowski}, S.~D. and {da Rio}, N. and {Flaherty}, K.~M. and {Rebull}, L. and {Frinchaboy}, P.~M. and {Majewski}, S.~R. and {Skrutskie}, M. and {Wilson}, J.~C. and {Zasowski}, G.},
	doi = {10.1088/0004-637X/799/2/136},
	eid = {136},
	eprint = {1411.6013},
	journal = {\apj},
	month = feb,
	pages = {136},
	title = {{IN-SYNC. II. Virial Stars from Subvirial Cores - the Velocity Dispersion of Embedded Pre-main-sequence Stars in NGC 1333}},
	volume = 799,
	year = 2015}

@article{cottaar2014,
	adsurl = {http://adsabs.harvard.edu/abs/2014ApJ...794..125C},
	archiveprefix = {arXiv},
	arxivurl = {http://arxiv.org/abs/1408.7113},
	author = {{Cottaar}, M. and {Covey}, K.~R. and {Meyer}, M.~R. and {Nidever}, D.~L. and {Stassun}, K.~G. and {Foster}, J.~B. and {Tan}, J.~C. and {Chojnowski}, S.~D. and {da Rio}, N. and {Flaherty}, K.~M. and {Frinchaboy}, P.~M. and {Skrutskie}, M. and {Majewski}, S.~R. and {Wilson}, J.~C. and {Zasowski}, G.},
	doi = {10.1088/0004-637X/794/2/125},
	eid = {125},
	eprint = {1408.7113},
	journal = {\apj},
	month = oct,
	pages = {125},
	primaryclass = {astro-ph.SR},
	title = {{IN-SYNC I: Homogeneous Stellar Parameters from High-resolution APOGEE Spectra for Thousands of Pre-main Sequence Stars}},
	volume = 794,
	year = 2014}

@article{jeffries2017,
	adsurl = {http://adsabs.harvard.edu/abs/2017MNRAS.464.1456J},
	archiveprefix = {arXiv},
	arxivurl = {http://arxiv.org/abs/1609.07150},
	author = {{Jeffries}, R.~D. and {Jackson}, R.~J. and {Franciosini}, E. and {Randich}, S. and {Barrado}, D. and {Frasca}, A. and {Klutsch}, A. and {Lanzafame}, A.~C. and {Prisinzano}, L. and {Sacco}, G.~G. and {Gilmore}, G. and {Vallenari}, A. and {Alfaro}, E.~J. and {Koposov}, S.~E. and {Pancino}, E. and {Bayo}, A. and {Casey}, A.~R. and {Costado}, M.~T. and {Damiani}, F. and {Hourihane}, A. and {Lewis}, J. and {Jofre}, P. and {Magrini}, L. and {Monaco}, L. and {Morbidelli}, L. and {Worley}, C.~C. and {Zaggia}, S. and {Zwitter}, T.},
	doi = {10.1093/mnras/stw2458},
	eprint = {1609.07150},
	journal = {\mnras},
	month = jan,
	pages = {1456-1465},
	primaryclass = {astro-ph.SR},
	title = {{The Gaia-ESO Survey: lithium depletion in the Gamma Velorum cluster and inflated radii in low-mass pre-main-sequence stars}},
	volume = 464,
	year = 2017}

@inproceedings{topcat,
	adsurl = {http://adsabs.harvard.edu/abs/2005ASPC..347...29T},
	author = {{Taylor}, M.~B.},
	booktitle = {Astronomical Data Analysis Software and Systems XIV},
	editor = {{Shopbell}, P. and {Britton}, M. and {Ebert}, R.},
	month = dec,
	pages = {29},
	series = {Astronomical Society of the Pacific Conference Series},
	title = {{TOPCAT {\amp} STIL: Starlink Table/VOTable Processing Software}},
	volume = 347,
	year = 2005}

@article{lada2003,
	adsurl = {http://adsabs.harvard.edu/abs/2003ARA%26A..41...57L},
	arxivurl = {http://arxiv.org/abs/astro-ph/0301540},
	author = {{Lada}, C.~J. and {Lada}, E.~A.},
	doi = {10.1146/annurev.astro.41.011802.094844},
	eprint = {astro-ph/0301540},
	journal = {\araa},
	pages = {57-115},
	title = {{Embedded Clusters in Molecular Clouds}},
	volume = 41,
	year = 2003}

@article{white2003,
	adsurl = {http://adsabs.harvard.edu/abs/2003ApJ...582.1109W},
	arxivurl = {http://arxiv.org/abs/astro-ph/0209164},
	author = {{White}, R.~J. and {Basri}, G.},
	doi = {10.1086/344673},
	eprint = {astro-ph/0209164},
	journal = {\apj},
	month = jan,
	pages = {1109-1122},
	title = {{Very Low Mass Stars and Brown Dwarfs in Taurus-Auriga}},
	volume = 582,
	year = 2003}

@article{baraffe2015,
	adsurl = {http://adsabs.harvard.edu/abs/2015A%26A...577A..42B},
	archiveprefix = {arXiv},
	arxivurl = {http://arxiv.org/abs/1503.04107},
	author = {{Baraffe}, I. and {Homeier}, D. and {Allard}, F. and {Chabrier}, G.},
	doi = {10.1051/0004-6361/201425481},
	eid = {A42},
	eprint = {1503.04107},
	journal = {\aap},
	month = may,
	pages = {A42},
	primaryclass = {astro-ph.SR},
	title = {{New evolutionary models for pre-main sequence and main sequence low-mass stars down to the hydrogen-burning limit}},
	volume = 577,
	year = 2015}

@article{da-rio2016,
	adsurl = {http://adsabs.harvard.edu/abs/2016ApJ...818...59D},
	archiveprefix = {arXiv},
	arxivurl = {http://arxiv.org/abs/1511.04147},
	author = {{Da Rio}, N. and {Tan}, J.~C. and {Covey}, K.~R. and {Cottaar}, M. and {Foster}, J.~B. and {Cullen}, N.~C. and {Tobin}, J.~J. and {Kim}, J.~S. and {Meyer}, M.~R. and {Nidever}, D.~L. and {Stassun}, K.~G. and {Chojnowski}, S.~D. and {Flaherty}, K.~M. and {Majewski}, S. and {Skrutskie}, M.~F. and {Zasowski}, G. and {Pan}, K.},
	doi = {10.3847/0004-637X/818/1/59},
	eid = {59},
	eprint = {1511.04147},
	journal = {\apj},
	month = feb,
	pages = {59},
	title = {{IN-SYNC. IV. The Young Stellar Population in the Orion A Molecular Cloud}},
	volume = 818,
	year = 2016}

@article{briceno1997,
	adsurl = {http://adsabs.harvard.edu/abs/1997AJ....113..740B},
	author = {{Briceno}, C. and {Hartmann}, L.~W. and {Stauffer}, J.~R. and {Gagne}, M. and {Stern}, R.~A. and {Caillault}, J.-P.},
	doi = {10.1086/118293},
	journal = {\aj},
	month = feb,
	pages = {740-752},
	title = {{X-Rays Surveys and the Post-T Tauri Problem}},
	volume = 113,
	year = 1997}

\end{document}